\documentclass[prd,superscriptaddress,nofootinbib,11pt, tightenlines]{revtex4-2} 
\usepackage{epsfig}  
\usepackage{bbold}
\usepackage{amsmath} 
\usepackage{amsfonts}  
\usepackage{xfrac}
\usepackage{float}    
\usepackage{amssymb}    
\usepackage{slashed} 
\usepackage{color}      
\usepackage{pbox}
\usepackage{bbding}
\usepackage{subcaption}
\usepackage{ragged2e}
\usepackage{array,multirow,makecell}
\usepackage[colorlinks,citecolor=blue, linkcolor = blue, urlcolor = blue]{hyperref}
\usepackage{cleveref}
\usepackage{titlesec} 
\usepackage{scrextend}
\usepackage{adjustbox}
\usepackage{natbib} 
\usepackage{lipsum}
\usepackage[normalem]{ulem} 
\usepackage{tikz,xcolor,hyperref}
\usepackage{cancel}
\usepackage{marginnote}
\usepackage{tikz}
\usepackage[compat=1.1.0]{tikz-feynman} 
\usepackage{contour}
\usetikzlibrary{shapes.geometric}
\usepackage{ulem}
\usepackage{orcidlink}

\newcommand{\GeV}{{\, \rm GeV}}
\newcommand{\TeV}{{\, \rm TeV}}
\newcommand{\eps}{\epsilon}

\newcommand{\gmcon}{\gamma_\mu}

\newcommand{\sigsubmn}{\sigma_{\mu\nu}}

\newcommand{\be}{\begin{equation}} 
\newcommand{\ee}{\end{equation}} 
\newcommand{\bea}{\begin{eqnarray}}
\newcommand{\eea}{\end{eqnarray}}
\newcommand{\bee}{\begin{eqnarray*}}
\newcommand{\eee}{\end{eqnarray*}}

\newcommand{\cO}{{\mathcal O}} 
\newcommand{\cL}{{\mathcal L}}

\newcommand{\CRmue}{{\rm CR} (\mu\rightarrow e)}
\newcommand{\KLemu}{K^0_L\to e^{\pm}\mu^{\mp}}

\newcommand{\BZmue}{B^0\to \mu^{\mp}e^{\pm}}

\def\Re{\mbox{Re}\,}

\def\ie{\hbox{\it i.e.}{}}
\def\eg{\hbox{\it e.g.}{}}
\def\etc{\hbox{\it etc}{}}

\renewcommand{\bar}{\overline}

\newcommand{\BR}{\text{BR}}

\newcommand{\vp}{\varphi}

\def\lQ1{\mathcal C_{\ell q}^{(1)}}
\def\lq3{\mathcal C_{\ell q}^{(3)}}
\def\qe{\mathcal C_{qe}}
\def\eu{\mathcal C_{eu}}

\def\Phil1{\mathcal C_{\vp \ell}^{(1)}}
\def\phil3{\mathcal C_{\vp \ell}^{(3)}}
\def\lu{\mathcal C_{\ell u}}
\def\ld{\mathcal C_{\ell d}}
\def\ed{\mathcal C_{ed}}

\def\ledq{\mathcal C_{\ell edq}}

\def\leQu1{\mathcal C_{\ell equ}^{(1)}}
\def\lequ3{\mathcal C_{\ell equ}^{(3)}}
\def\mesonV{\mathcal V}
\def\mesonP{\mathcal P}

\tikzfeynmanset{
  every fermion={black},
  every photon={blue},
  every blob={/tikz/fill=black!30,/tikz/inner sep=1pt},
}

\begin{document}

\date{\today}
\title{Constraining lepton flavor violating SMEFT $2q 2\ell$  operators from low-energy cLFV processes}

\author{Utpal Chattopadhyay\orcidlink{0000-0002-6824-9465}}
\email{tpuc@iacs.res.in}
\affiliation{School of Physical Sciences, Indian Association for the Cultivation of Science, Jadavpur, Kolkata 700032, India}

\author{Debottam Das}
\email{debottam@iopb.res.in}
\affiliation{Institute of Physics, Sachivalaya Marg, Bhubaneswar, 751005, India}
\affiliation{Homi Bhabha National Institute, Training School Complex, Anushakti Nagar, Mumbai 400094, India}

\author{Rahul Puri\orcidlink{0000-0002-8051-4313}}
\email{rahul.puri@iopb.res.in}
\affiliation{Institute of Physics, Sachivalaya Marg, Bhubaneswar, 751005, India}
\affiliation{Homi Bhabha National Institute, Training School Complex, Anushakti Nagar, Mumbai 400094, India}

\author{Joydeep Roy\orcidlink{0000-0001-7406-0127}}
\email{joydeeproy@acharya.ac.in}
\email{joyroy.phy@gmail.com}
\affiliation{School of Physical Sciences, Indian Association for the Cultivation of Science, Jadavpur, Kolkata 700032, India}
\affiliation{Department of Physics, Acharya Institute of Technology, Bengaluru, 560107, Karnataka, India}

\begin{abstract}
Charged lepton flavour-violating (cLFV) processes, which are definite proof of new physics beyond the Standard Model, have remained experimentally elusive to date. 
Effective Field Theory (EFT) has been very useful in providing information about such new physics through the higher-dimensional operators. 
These operators respect SM gauge invariance, and they are suppressed by appropriate powers of the energy scale $\Lambda$. 
In regard to lepton flavour violating (LFV) processes, the Standard Model Effective Field Theory (SMEFT) is shown to be a useful tool for estimating any new physics effect at the scale $\Lambda$. 
It is worth noting that a large class of cLFV processes involves both quarks and leptons. 
Thus, low-energy observables play a significant role in providing bounds on lepton-flavour-violating 2-quark-2-lepton ($2q2\ell$) operators. In this work, we explore several low-energy cLFV processes that can be addressed within the SMEFT framework, relating them to the specific operators of relevance. 
Keeping in mind the correlation that exists among the SMEFT operators, we want to extract the strongest constraints on these $2q2\ell$ operators.
\end{abstract}

\maketitle
\newpage


\section{Introduction}
\label{sec:intro}

The Standard Model (SM)~\cite{Peskin:1995ev} of particle physics has been extraordinarily successful in elucidating the fundamental interactions between constituent particles.
It has made precise predictions that have been verified by experiments at accelerators such as LEP~\cite{Myers:1990sk}, Tevatron~\cite{Wilson:1977nk}, and the LHC~\cite{Evans:2008zzb}. The discovery of the Higgs boson at the LHC in 2012~\cite{ATLAS:2012yve} was the last missing piece of the SM puzzle, and it firmly established the SM as the appropriate theory for the energy range we have explored. However, SM is still far from becoming a comprehensive account of particle physics~\cite{Crivellin:2023zui}. There are several experimental facts and theoretical questions that cannot be addressed by staying within the SM framework. These include the gauge hierarchy problem, the mass of neutrinos, the lack of a particle dark matter (DM) candidate, and the baryon asymmetry of the universe, among other significant problems that drive our investigation of physics beyond the SM (BSM) scenarios. Direct and indirect avenues exist for exploring the potential existence of New Physics (NP). Direct approaches involve detecting new particles through ongoing and upcoming collider experiments, while indirect probes rely on evidence gathered from various low-energy processes. Among many potential BSM signals, charged lepton flavour violation stands out as an intriguing and promising candidate for investigating NP scenarios. It is known that flavour-changing neutral current (FCNC) processes can be substantially large in many BSM scenarios.
In contrast, they are heavily suppressed in the SM by small Cabibbo-Kobayashi-Maskawa (CKM)  matrix elements, loop effects, etc. Probing NP models with FCNC effects can thus be quite useful. Such effects can be seen in several cLFV processes, which are tabulated in Table-\ref{Tab:LFVlimits} with their current experimental upper limits.

\begin{table}[ht!]
\centering
\resizebox{\textwidth}{!}{
\renewcommand{\arraystretch}{0.95}
\begin{tabular}{|l|cl|cl|c|}
\hline
Observables of cLFV modes. & \multicolumn{2}{c}{Present bounds} & \multicolumn{2}{|c|}{Expected future limits} & \sf Flavio \\
\hline\hline
BR$(\tau\to e\gamma)$ & $3.3\times10^{-8}$ &
BaBar(2010)~\cite{BaBar:2009hkt} & 
&
&
$\checkmark$ \\
BR$(\tau\to \mu\gamma)$ & $4.2\times10^{-8}$ &
Belle(2021)~\cite{Belle:2021ysv} & 
&
&
$\checkmark$ \\\hline
BR$(\tau\to e\pi^0)$ & $8.0\times10^{-8}$ &
Belle(2007)~\cite{Ohshima:2007zz} & $7.3\times10^{-10}$  &
Belle-II\cite{Belle-II:2022cgf} &
$\checkmark$ \\
BR$(\tau\to \mu\pi^0)$ & $1.1\times10^{-7}$ &
BaBar(2006)~\cite{BaBar:2006jhm} &
$(5\times 10^{-10})$ $7.1\times10^{-10}$ &
Belle-II\cite{Belle-II:2018jsg, Belle-II:2022cgf} &
$\checkmark$ \\
\hline
BR$(\tau\to e\rho^0)$ & $2.2\times10^{-8}$ &
Belle(2023)~\cite{Belle:2023ziz} & $3.8\times10^{-10}$ &
Belle-II\cite{Belle-II:2018jsg} & $\checkmark$ \\
BR$(\tau\to \mu\rho^0)$ &
$1.7\times10^{-8}$ &
Belle(2023)~\cite{Belle:2023ziz} & $5.5\times10^{-10}$ &
Belle-II\cite{Belle-II:2018jsg} &
$\checkmark$ \\ \hline 
BR$(\tau \to e \phi)$ &
$2.0\times10^{-8}$&
Belle(2023)~\cite{Belle:2023ziz} &
$7.4\times10^{-10}$ &
Belle-II\cite{Belle-II:2018jsg} &
$\checkmark$ \\
BR$(\tau \to \mu \phi)$ &
$2.3\times10^{-8}$ &
Belle(2023)~\cite{Belle:2023ziz} &
$8.4\times10^{-10}$ & 
Belle-II\cite{Belle-II:2018jsg} &
$\checkmark$ \\
\hline
BR$(\tau \to e(\mu) K_S^0)$   & $0.8(1.2)\times 10^{-8}$ & Belle-II(2025)~\cite{Belle-II:2025yiq} & & & $\checkmark\checkmark$ \\
\hline\hline
CR$(\mu\to e,{\rm Au})$ &
$7.0\times10^{-13}$ &
SINDRUMII(2006)~\cite{SINDRUMII:2006dvw}& --\hspace{.7cm} & -- & $\checkmark$ \\
CR$(\mu\to e,{\rm Ti})$ &
$6.1\times10^{-13}$ &
SINDRUMII(1998)\cite{Wintz:1998rp}&
--\hspace{.7cm}  & -- & $\checkmark$ \\
\multirow{2}{*}{CR$(\mu\to e,{\rm Al})$} & - & - & $10^{-17}$ & Mu2e\cite{Mu2e:2014fns} &  \\
 & - & -& $10^{-15}$ (Phase I) \& $10^{-17}$ (Phase II) & COMET at J-PARC\cite{Kuno:2013mha,COMET:2018auw} & $\checkmark$ \\ 
BR$(\mu\to e\gamma)$ &
$3.1\times10^{-13}$ &
MEG-II(2024)~\cite{MEGII:2023ltw}& --\hspace{.7cm} & -- & $\checkmark$ \\
 \hline\hline
BR$(J/\Psi \to e^{\pm}\mu^{\mp})$ &
$4.5\times 10^{-9}$ &
BESIII(2022)~\cite{BESIII:2022exh} & & & $\checkmark$ \\
BR$(J/\Psi \to e \tau)$ & $7.5\times10^{-8}$ &
BESIII(2021)~\cite{BESIII:2021slj} & & & $\checkmark$ \\
BR$(J/\Psi \to\mu \tau)$ & $2.0\times10^{-6}$ &
BES(2004)~\cite{BES:2004jiw} & & & $\checkmark$ \\ \hline
BR$(\Upsilon{(1S)} \to e \mu)$ &
$3.9\times10^{-7}$ &
Belle(2022)~\cite{Belle:2022cce}& & & $\checkmark$ \\
BR$(\Upsilon{(1S)} \to e \tau)$ &
$2.7\times10^{-6}$ &
Belle(2022)~\cite{Belle:2022cce}&
& & $\checkmark$ \\
BR$(\Upsilon{(1S)} \to \mu \tau)$ & $2.7\times10^{-6}$& Belle(2022)~\cite{Belle:2022cce}& & & $\checkmark$ \\
BR$(\Upsilon{(2S)} \to e \tau)$ &
$1.12\times10^{-6}$&
Belle(2024)~\cite{Belle:2023iln}& & & $\checkmark$ \\
BR$(\Upsilon{(2S)} \to \mu \tau)$ & $2.3\times10^{-7}$&
Belle(2024)~\cite{Belle:2023iln}& & & $\checkmark$ \\
BR$(\Upsilon(3S) \to e \mu)$ & $3.6\times 10^{-7}$ &
BaBar(2022)~\cite{BaBar:2021loj}& & & $\checkmark$ \\
BR$(\Upsilon(3S) \to e \tau)$ & $4.2\times 10^{-6}$ &  BaBar(2010)~\cite{BaBar:2010vxb}& & & $\checkmark$ \\
BR$(\Upsilon(3S) \to \mu \tau)$ &
$3.1\times 10^{-6}$ &
BaBar(2010)~\cite{BaBar:2010vxb}&
& & $\checkmark$ \\ \hline\hline
BR$(\pi^0\to e^{+}\mu^{-})$ &
$3.2\times10^{-10}$ &
NA62(2021)~\cite{NA62:2021zxl} &
- & - & $\checkmark$$\checkmark$ \\
BR$(\pi^0\to e^{-}\mu^{+})$ & $3.8\times10^{-10}$ &
SPEC(2000)~\cite{Appel:2000wg} & - & - & $\checkmark$ \\ 
BR$(\pi^0\to e^{\pm}\mu^{\mp})$ &
$3.6\times10^{-10}$ & KTeV(2007)~\cite{KTeV:2007cvy} & - & - & \bf{$\checkmark$$\checkmark$} \\ \hline
BR$(\eta^0\to e\mu)$ & $6.0\times10^{-6}$ &
SPES2(1995)~\cite{White:1995jc}& & & $\checkmark$  \\
BR$(\eta^{0\prime}\to e\mu)$ &  $4.7\times10^{-4}$ & 
CLEO II(1999)~\cite{CLEO:1999nsy} &
JEF~\cite{JEF}, REDTOP~\cite{REDTOP:2022slw} &
& $\checkmark$$\checkmark$ \\ \hline
BR$(K_L^0 \to e^{\pm} \mu^{\mp}$)   & $4.7 \times 10^{-12}$ & BNL E871(1998)~\cite{BNL:1998apv} &
& & $\boldsymbol{\ast\ast}$\protect\footnote{\scriptsize Although BR$(K_L\to e^\pm\mu^\mp)$ is already available in {\sf Flavio-2.4.2}, the implementation contains an error. Therefore, we have implemented this observable by ourselves for this analysis.} \\
BR$(K_L \to \pi^0 e \mu)$ & 
$7.6\times 10^{-11}$  &  KTeV(2008)~\cite{KTeV:2007cvy} & - & - & $\checkmark$$\checkmark$\\
BR$(K^+\to \pi^+ \mu^- e^+)$  & $6.6\times 10^{-11}$ & NA62(2021)~\cite{NA62:2021zxl} &  & & $\checkmark$$\checkmark$ \\
BR$(K^+\to \pi^+ \mu^+ e^-)$  & $1.3\times 10^{-11}$ & E865(2005)~\cite{Sher:2005sp} & & & $\checkmark$$\checkmark$ \\ \hline
BR$(D^0 \to e^{\pm} \mu^{\mp}$) &
$1.3\times 10^{-8}$ & LHCb(2016)~\cite{Zyla:2020zbs} & & & $\checkmark$$\checkmark$ \\
BR$(D^0 \to\pi^0 e^{\pm} \mu^{\mp}$) &
$8\times 10^{-7}$ & BaBar(2020)~\cite{BaBar:2020faa} & & & $\times$ \\
BR$(D^0 \to\rho^0(\phi) e^{\pm} \mu^{\mp}$) &
$5(5.1)\times 10^{-7}$ &
BaBar(2020)~\cite{BaBar:2020faa} & & & $\times$ \\
BR$(D^0 \to\omega e^{\pm} \mu^{\mp}$) &
$17.1\times 10^{-7}$ & BaBar(2020)~\cite{BaBar:2020faa} & & & $\times$ \\
BR$(D^0 \to\eta e^{\pm} \mu^{\mp}$) &
$22.5\times 10^{-7}$ &
BaBar(2020)~\cite{BaBar:2020faa} &
& & $\times$ \\
BR$(D^+ \to \pi^+ e^{+(-)} \mu^{-(+)})$ &
$2.1\,(2.2)\times10^{-7}$ & LHCb(2020)~\cite{LHCb:2020car} & & & $\checkmark$$\checkmark$  \\ 
BR$(D_s^+ \to K^+ e^{+(-)} \mu^{-(+)}) $ & $7.9(5.6)\times10^{-7}$ & 
LHCb(2020)~\cite{LHCb:2020car} & & &  $\checkmark$$\checkmark$  \\ \hline
BR$(B^+\to K^+e^{+(-)}\mu^{-(+)})$ &
$ 7.0\,(6.4)\times 10^{-9}$ & LHCb(2019)~\cite{LHCb:2019bix} & $-$ & $-$ &  $\checkmark$ \\
BR$(B^0\to K^{*0}e^{+(-)}\mu^{-(+)})$ &
$ 6.8\,(5.7)\times 10^{-9}$ &
LHCb(2022)~\cite{LHCb:2022lrd} & $-$ & - & $\checkmark$ \\
BR$(B^{0}\to \pi^{0}e^{\pm}\mu^{\mp})$ &
$ 1.4\times 10^{-7}$ & BaBar(2007)~\cite{BaBar:2007xeb} & $-$ & 
$-$ & $\checkmark$ \\
BR$(B^{+}\to \pi^{+}e^{\pm}\mu^{\mp})$ &
$ 1.7\times 10^{-7}$ & BaBar(2007)~\cite{BaBar:2007xeb} & $-$ & 
- & $\checkmark$ \\
BR$(B_s^0\to \phi\mu^{\pm}e^{\mp})$ & $ 16\times 10^{-9}$ & LHCb(2022)~\cite{LHCb:2022lrd} & $-$ & $-$ & $\checkmark$ \\
BR$(B_{s}^0\to \mu^{\mp}e^{\pm})$ & $5.4\times 10^{-9}$ & LHCb(2018)~\cite{LHCb:2017hag} & $3 \times 10^{-10}$ & LHCb-II~\cite{LHCb:2018roe} & $\checkmark$ \\
BR$(B^0\to \mu^{\mp}e^{\pm})$ & $1.0\times10^{-9}$ &
LHCb(2018)~\cite{LHCb:2017hag} & $3 \times 10^{-10}$ & LHCb-II~\cite{LHCb:2018roe} & $\checkmark$ \\
\hline
BR$(B^+\to K^+\mu^{-(+)}\tau^{+(-)})$ & $ 0.59\,(2.45)\times 10^{-5}$ & Belle(2022)~\cite{Belle:2022pcr} & $3.3 \times 10^{-6}$ & Belle-II~\cite{Belle-II:2018jsg} & $\checkmark$ \\
BR$(B^+\to K^{+}e^{+(-)}\tau^{-(+)})$ &
$ 1.53 \,(1.51) \times 10^{-5}$ &
Belle(2022)~\cite{Belle:2022pcr} &
$2.1 \times 10^{-6}$ &
Belle-II~\cite{Belle-II:2018jsg} &
$\checkmark$ \\
BR$(B^0\to K^{*0}\mu^{+(-)}\tau^{-(+)})$ &
$8.2\,(10)\times10^{-6}$&
LHCb(2022)~\cite{LHCb:2022wrs} & $-$ & $-$ &
$\checkmark$ \\
BR$(B^+\to \pi^+e^{+(-)}\tau^{-(+)})$ & 
$7.4\,(2.0)\times10^{-5}$ &
BaBar(2012)~\cite{BaBar:2012azg} & $-$ & $-$ &
$\checkmark$ \\
BR$(B^+\to \pi^+\mu^{+(-)}\tau^{-(+)})$ &
$6.2\,(4.5)\times10^{-5}$ &
BaBar(2012)~\cite{BaBar:2012azg}&  $-$ & $-$ & 
$\checkmark$ \\
 BR$(B_s^0\to \phi\mu^{\pm}\tau^{\mp})$ &
$ 1.0\times 10^{-5}$ &
 LHCb(2024)~\cite{LHCb:2024wve} & $-$ & $-$ &
$\checkmark$ \\
BR$(B^0\to \tau^{\pm}e^{\mp})$ &
$ 1.6\times 10^{-5}$ &
Belle(2021)~\cite{Belle:2021rod} & $-$ & $-$ &
$\checkmark$ \\
BR$(B^0\to \tau^{\pm}\mu^{\mp})$ &
$1.2\times10^{-5}$ &
LHCb(2019)~\cite{LHCb:2019ujz} &
$1.3\times 10^{-6}$ & Belle-II~\cite{Belle-II:2018jsg} &
$\checkmark$ \\
BR$(B_s^0\to \tau^{\pm}e^{\mp})$ &
$ 1.4\times 10^{-3}$ &
Belle(2023)~\cite{Belle:2023jwr} & $-$ & $-$ &
$\checkmark$ \\
BR$(B_s^0\to \tau^{\pm}\mu^{\mp})$ &
$ 3.4\times 10^{-5}$ &
LHCb(2019)~\cite{LHCb:2019ujz} & $-$ & $-$ &
$\checkmark$ \\
\hline\hline
\end{tabular}}

\caption{\justifying\footnotesize Present upper bounds (with $90\%\,\text{CL}$), and future expected sensitivities of branching ratios for the set of low-energy cLFV transitions relevant for our analysis. $K_L^0 \to (e/\mu)^{\pm} \tau^{\mp}$ decay modes are forbidden by phase space. Similar for the mode $D^0 \to \mu^{\pm} \tau^{\mp}$~\cite{Hazard:2017udp}. The last column represents whether the {\sf Python} codes for the corresponding LFV processes are available in the \textsf{Flavio} package or not. A regular `$\checkmark$' means it is already implemented, and a bold `\textbf{$\checkmark$$\checkmark$}' means it is implemented by us for the first time for this analysis. `$\times$' sign represents an observable which is neither available in \textsf{Flavio} nor implemented by us.\protect\footnote{\scriptsize We plan to implement a few other LFV processes, such as $ \tau^{\pm} \to (e/\mu)^{\pm} \omega$ or $D^0\to(\omega/\rho^0) e^{\pm}\mu^{\mp}~\etc.$, that are not implemented in flavio, in our future analysis.}}
\label{Tab:LFVlimits}
\end{table}

In the model-dependent category, various popular models have been used to accommodate cLFV processes, but to date, searches at the LHC have not yielded any direct evidence of a new particle. Strong arguments in support of BSM physics and these null results, can become the motivation for considering an Effective Field Theory (EFT) approach \cite{Georgi:1993mps,Buchmuller:1985jz,Bechtle:2022tck} to estimate the level of unknown physics interactions at a given scale. In contrast to considering a BSM model that is associated with a top-down approach to an EFT framework, one can adopt a bottom-up approach~\cite{Bechtle:2022tck} and here, this refers to the model-independent investigation within the Standard Model Effective Field Theory (SMEFT)~\cite{Bechtle:2022tck,Buchmuller:1985jz,Grzadkowski:2010es} where the energy scale $\Lambda$ of effective interactions can be above the reach of current experiments.  In SMEFT, one considers higher-dimensional effective local operators out of SM fields only. The operators respect SM gauge invariance, and they are suppressed by appropriate powers of $\Lambda$. In regard to LFV processes, SMEFT is shown to be a useful tool for estimating 
any new physics effect at the scale $\Lambda$~\cite{Davidson:2018kud,Crivellin:2017rmk,Cirigliano:2017azj,Davidson:2017nrp,Davidson:2020ord,Davidson:2020hkf,Cirigliano:2021img,Kumar:2021yod}.

 For example, SMEFT has been used to study general LFV processes \cite{Crivellin:2013hpa,Pruna:2014asa,Crivellin:2017rmk}, LFV $B$-meson decays (LFVBDs) \cite{Alonso:2014csa, Aebischer:2015fzz, Crivellin:2015era, Becirevic:2016zri, Descotes-Genon:2023pen, Ali:2023kua}, Z boson LFV decays (ZLFVDs) \cite{Calibbi:2021pyh}, Higgs boson LFV decays (HLFVDs) \cite{Cullen:2020zof} or LFV Quarkonium decays (LFVQDs) \cite{Calibbi:2022ddo}. Although these references are hardly an exhaustive list of such works, it has been shown in all of them that the most dominant contributions to LFV processes come from dimension-6 operators, and the operators contributing to leptonic or semi-leptonic processes are either of the Higgs-fermion or four-fermion type. 
 For our present purpose, we focus our attention on the inputs from low-energy precision experiments that may be particularly sensitive to NP. In this regard, a subset of dimension-6 four-fermion operators involving two quarks and two leptons is particularly relevant. Specifically, these operators with different lepton flavours contribute to a wide range of low-energy LFV observables that are experimentally accessible. We shall refer to them as LFV $2q2\ell$ operators hereafter in this work and estimate the bounds on them by comparing low-energy observables dependent on them. A comprehensive list of these operators within the SMEFT framework, along with their chirality structures, is provided in Table \ref{tab:Dim-6 2q2l operators1}. A similar list of effective operators appropriate for low-energy (LEFT) is provided in Table-\ref{tab:JMSop}.

Several studies have compiled constraints on four-fermionic operators \cite{Zarnecki:1999je,Ibarra:2004pe,Carpentier:2010ue,Garosi:2023yxg,Fernandez-Martinez:2024bxg,Ali:2025xkw}. 
However, in all of these works, the analyses were either limited to a subset of the $2q2\ell$ operators or conducted in terms of the WCs with broad Lorentz structures--vector, axial-vector, tensor, or pseudoscalar--without distinguishing between specific flavour indices.
In this analysis, we try to impose constraints only on LFV $2q2\ell$ operators with specific quark and lepton indices, which are responsible for specific LFV processes. Based on the involvement of quarks and leptons in each process and corresponding EFT analysis, Table-\ref{tab:LFV Oprtr} presents a set of LFV $2q2\ell$ operators that contribute dominantly to those processes.\footnote{We do not consider the high-energy LFV processes involving bosons (Higgs or $Z$) or $3$-body decays of leptons in this analysis or the processes where these operators can be produced in the loop, e.g. LFVZ decays.} Furthermore, in Table-\ref{tab:mesons} we tabulate all mesons considered in this analysis and their quark compositions, whereas Tables-\ref{tab:SVG}--\ref{tab:effco} mention the relevant data used.

The said $2q2\ell$ operators contribute to the LFV processes in two possible scenarios. First, where processes involve different quark-flavour transitions, like decays of pseudoscalar mesons $(B,K,D)$, as described in Table-\ref{Tab:Operator_different-quark_indices}. Second, where the system involves the same quark flavour like quarkonium $(q\bar{q})$ or $\tau$ decaying to vector and pseudoscalar mesons, as described in Tables-\ref{Tab:Operator_same-quark_indices} and \ref{Tab:Operator_indices_tau}. Moreover, these operators have important phenomenological consequences because several bosonic mediators, such as a second Higgs
doublet, vector bosons, various low-energy scalar or leptoquark models~\cite{deBlas:2017xtg,Plakias:2023esq,Chattopadhyay:2019ycs,De:2021crr,Gherardi:2020det,Buras:2020xsm,Bobeth:2017ecx,Dekens:2018bci,Aebischer:2025qhh} can induce them. 

In most SMEFT analyses, a common simplification is to consider only one operator at a time, with the others being zero. Although this single-operator setup is not ideal for a realistic low-energy description of any UV-completion scenario, such analysis probes the new physics scale $\Lambda$ appropriate to the operator under discussion. Moreover, when several observables, all of which are dependent on a specific operator, are considered, the single-operator method provides a way to perform the sensitivity test among those observables. As we shall see, the information from such a test helps to identify the observables that are among the most relevant ones to constrain a particular operator. The third important aspect of this so-called `1-d analysis' comes from the fact that several of these LFV $2q2\ell$ operators ($\mathcal{C}_{\ell q}^{(1)}, \mathcal{C}_{\ell q}^{(3)}$ and $\mathcal{C}_{qe}$) are important in imposing constraints on other LFVs such as ZLFV or 3-body leptonic decays~\cite{Calibbi:2021pyh} through the Higgs-lepton operators $(\mathcal{C}_{\phi\ell}^{(1)},\mathcal{C}_{\phi\ell}^{(3)},\mathcal{C}_{\phi e})$, as they are strongly related via RGEs. Moreover, several of these LFV operators can be related to other non-LFV processes as well and thus signifying the phenomenological relevance of the constraints obtained from this analysis.

We also performed analysis beyond this simplistic approach by considering the simultaneous presence of two operators at the UV scale and presented some of those results. In sec-\ref{sec: EFT framework} we provide a general effective field theory framework and a comprehensive list of dim-6 operators responsible for LFV processes. There, we also discuss the procedure and methods used to match the SMEFT and LEFT operator bases. All relevant expressions for our analysis are presented in terms of SMEFT operators in sec-\ref{sec:observables}. In this analysis, for the first time, we have implemented several LFV decay modes that were not hitherto available in \textsf{Flavio}. Form factors and other necessary elements needed to write the associated codes are presented in Tables-\ref{tab:effco}--\ref{tab:decay constants} in sec-\ref{sec:Calculation method}. Results of this analysis are presented in \cref{sec:results}, and finally we conclude in sec-\ref{sec:conclusion}.


\section{Standard Model Effective Field Theory Framework}
\label{sec: EFT framework}

New physics involving heavy particles may be present at scales ($\Lambda$) beyond the current experimental reach. In a model-independent way, SMEFT incorporates higher-dimensional non-renormalizable and gauge-invariant operators involving SM fields to describe a
BSM effect on observables of interest. The operators are suppressed by appropriate powers of $\Lambda~(\gg m_W)$ and corresponding Wilson coefficients (WCs) parameterise the low-energy behaviour of such high-energy theory through the running of the renormalisation group equations. (RGEs) of masses and coupling parameters of the theory. The SMEFT Lagrangian is given by \cite{Bechtle:2022tck,Grzadkowski:2010es}
\be \label{eq:SMEFT Lagrng} 
\cL_{\rm SMEFT} = \cL_{\rm SM} + \frac{1}{\Lambda} \sum_i \mathcal{C}^{(5)}_i \cO^{(5)}_i + \frac{1}{\Lambda^2} \sum_i \mathcal{C}_i^{(6)} \cO_i^{(6)} + \cO \left(\frac{1}{\Lambda^3}\right) + ...\hspace{1 em}\,,
\ee
where $\cL_{\rm SM}$ is the usual renormalizable SM Lagrangian, $\cO^{(5)}$ represents the gauge-invariant mass dimension-5 operators, known as neutrino mass generating Weinberg operator, $\mathcal{C}^{(5)}$ is the corresponding WCs. Similarly, $\cO_n^{(6)}$ and $\mathcal{C}_n^{(6)}$ represent mass dimension-6 operators and corresponding WCs respectively. 
In this work, we will not consider any more terms with suppression level greater than $1/\Lambda^2$.

\begin{table}[h!]
\centering
\renewcommand{\arraystretch}{1.4}
\setlength{\tabcolsep}{15pt}
\begin{tabular}{|c|ccc|} 
\hline
\rule{0pt}{2.2em}\shortstack{Lorentz\\Structure} & \shortstack{WC\\$(\mathcal C_i)$} & \shortstack{Operator\\$(\mathcal O_i)$} & \shortstack{Chirality\\Structure}
\\
\hline\hline
\bm{}& \bm{$[\mathcal C_{\ell q}^{(1)}]_{\alpha\beta ij}$} & \bm{$(\bar L_{\alpha} \gamma_\mu L_{\beta})(\bar Q _i\gamma^\mu Q_j)$} & $(\bar{L}L)(\bar{L}L)$ \\
& \bm{$[\mathcal C_{\ell q}^{(3)}]_{\alpha\beta ij}$} & \bm{$(\bar L_{\alpha} \gamma_\mu \tau^I L_{\beta})(\bar Q_i \gamma^\mu \tau^I Q_j)$} & $(\bar{L}L)(\bar{L}L)$ \\
& \bm{$[\mathcal C_{qe}]_{ij \alpha\beta }$}  & \bm{$(\bar Q_i \gamma_\mu Q_j)(\bar E_{\alpha} \gamma^\mu E_{\beta})$} & $(\bar{L}L)(\bar{R}R)$ \\
Vector & \bm{$[\mathcal C_{\ell d}]_{\alpha\beta ij}$} & \bm{$(\bar L_{\alpha} \gamma_\mu L_{\beta})(\bar D_i \gamma^\mu D_j)$} & $(\bar{L}L)(\bar{R}R)$ \\
& \bm{$[\mathcal C_{\ell u}]_{\alpha\beta ij}$} & \bm{$(\bar L_{\alpha} \gamma_\mu L_{\beta})(\bar U_i \gamma^\mu U_j)$} & $(\bar{L}L)(\bar{R}R)$ \\
& \bm{$[\mathcal C_{ed}]_{\alpha\beta ij}$} & \bm{$(\bar E_{\alpha} \gamma_\mu E_{\beta})(\bar D_i\gamma^\mu D_j)$} & $(\bar{R}R)(\bar{R}R)$\\
& \bm{$[\mathcal C_{eu}]_{\alpha\beta ij}$} & \bm{$(\bar E_{\alpha} \gamma_\mu E_{\beta})(\bar U_i \gamma^\mu U_j)$} & $(\bar{R}R)(\bar{R}R)$ \\
\hline\hline
Scalar & \bm{$[\mathcal C_{\ell edq}]_{\alpha\beta ij}$} & \bm{$(\bar L^a_{\alpha} E_{\beta})(\bar D_i Q^a_j)$} & $(\bar{L}R)(\bar{R}L)$\\
& \bm{$[\mathcal C_{\ell equ}^{(1)}]_{\alpha\beta ij}$} &  \bm{$(\bar L^a_{\alpha} E_{\beta}) \eps_{ab} (\bar Q^b_i U_j)$} & $(\bar{L}R)(\bar{L}R)$\\
 \hline\hline
Tensor & \bm{$[\mathcal C_{\ell equ}^{(3)}]_{\alpha\beta ij}$} &  \bm{$(\bar L^a_{\alpha} \sigma_{\mu\nu} E_{\beta}) \eps_{ab} (\bar Q^b_i \sigma^{\mu\nu} U_j)$} & $(\bar{L}R)(\bar{L}R)$ \\
\hline
\end{tabular}
\caption{\justifying A comprehensive list of dimension-6 $2q2\ell$ SMEFT operators, with the corresponding WCs, that remain invariant under the SM gauge group and contribute to LFV observables. In these expressions, $Q$ and $L$ represent left-handed quark and lepton $SU(2)$ doublets, respectively. $U,\,D$ and $E$ denote right-handed up, down quark and lepton singlets. The last column provides the chirality of fermionic currents of these operators.} 
\label{tab:Dim-6 2q2l operators1}
\end{table}

In general studies of low-energy observables in SMEFT, there are three energy scales. A high energy scale $\Lambda$, an intermediate energy scale of electroweak symmetry-breaking $m_{Z/W}$ and a low-energy scale like $\sim m_b$ or $m_{\tau / \mu}$. Therefore for probing the level of contributions of the higher dimensional operators for LFV studies that may be consistent with experimental constraints, a general method of ``match and run" of RGEs is described below. At the first step, the SMEFT 1-loop RGEs \cite{Jenkins:2013zja, Jenkins:2013wua, Alonso:2013hga} of relevant WCs would be initialised at the scale $\Lambda(\sim\TeV)$ and run down to the electroweak scale $\sim m_{Z/W}$. The WCs under study are given a non-vanishing value like unity, while other WCs are set to zero at the scale $\Lambda$ and RGE evolution is completed till the electroweak scale $m_{Z/W}$. Of course, the WCs are hardly expected to remain at their initial values, including also the ones that were vanishing at the high scale $\Lambda$.
 This level of evolution is adequate for the LFV decays of the top-quark, $Z$ or Higgs bosons but not enough for processes referring to energies below the electroweak scale. Further down the scale, in the second step, as in a top-down approach of EFT, the heavy particles of the theory ($W^{\pm}$, $Z$, the Higgs boson and the top quark) are integrated out and the operators invariant under the QCD$\otimes$QED gauge groups and consisting of fields of light charged fermions $(u, d, c, s, b, e, \mu, \tau)$, neutral fermions $(\nu_e, \nu_{\mu}, \nu_{\tau})$ and massless gauge bosons 
describe the effective interactions. These operators are known as Low-energy Effective Field Theory (LEFT) operators that contribute to the total LEFT Lagrangian containing dimension three and higher dimensional $(d>4)$ operators. The most general interaction Lagrangian involving dimension-5 and dimension-6 operators can be written as:
\begin{align}
    \cL_{\rm LEFT}^{(\rm int)} = -\mathcal{H}_{\rm eff} \supset \sum_{d=5,6}\left[\sum_{O_i^{(d)} = O_i^{(d)\dagger}} \frac{1}{v^{d-4}}C_i^{(d)}\,O_i^{(d)} + \sum_{O_i^{(d)} \neq O_i^{(d)\dagger}} \frac{1}{v^{d-4}}(C_i^{(d)}\,O_i^{(d)} + \text{h.c.})\right],
    \label{eq:LEFT eff Ham}
\end{align}
where, $v=(\sqrt{2}G_F)^{-1/2}$, $G_F$ being the Fermi-constant. For our purpose the most relevant dimension-6  LEFT operators are those containing four fermions, with one spinor bilinear involving charged leptons of different flavours and the other involving quarks~\cite{Grinstein:1988me}. Schematically, these operators take the form 
\begin{align}
\label{eq:LEFT oprtr}
{O}^{S,AB}_{eq} = (\bar e_\alpha \Gamma_S P_A e_\beta)(\bar q_i \Gamma^S P_B q_j),~~~q\in\{u,d\}
\end{align}
where $e,u$ and $d$, represent charged leptons, up-type quarks and down-type quarks, respectively. The indices $(\alpha\beta)$ and $(ij)$ specify the generations of the leptons and quarks, respectively. $P_{A,B}$ are the left and right projection operators and $\Gamma_S \in \{\mathbb{1}, \gmcon,  \sigsubmn\}$ are associated with scalar, vector and tensor operators, respectively. All such operators, along with the leptonic dipole operators, with the corresponding WCs, are listed in Table-\ref{tab:JMSop}.

\begin{table}[h!]
    \centering
    \renewcommand{\arraystretch}{1.7}
    \begin{tabular}{|c|c|c|c|}
        \hline
        Dimensions& Lorentz Structure& Wilson Coefficient $(C)$ & Operator $(O)$ \\\hline\hline
        \multirow{6}{*}{$d=6$}&\multirow{2}{*}{Vector} 
        & $(C^{V,XY}_{eu})_{\alpha\beta ij}$ & $(\bar e_\alpha\gamma^\mu P_X e_\beta)(\bar u_i\gamma_\mu P_Y u_j)$ \\
        && $(C^{V,XY}_{ed})_{\alpha\beta ij}$ & $(\bar e_\alpha\gamma^\mu P_X e_\beta)(\bar d_i\gamma_\mu P_Y d_j)$ \\\cline{2-4}
        
        &\multirow{2}{*}{Scalar}
        & $(C^{S,XY}_{eu})_{\alpha\beta ij}$ & $(\bar e_\alpha P_X e_\beta)(\bar u_i P_Y u_j)$ \\
        && $(C^{S,XY}_{ed})_{\alpha\beta ij}$ & $(\bar e_\alpha P_X e_\beta)(\bar d_i P_Y d_j)$ \\\cline{2-4}
        
        &\multirow{2}{*}{Tensor}
        & $(C^{T,RR}_{eu})_{\alpha\beta ij}$ & $(\bar e_\alpha \sigma^{\mu\nu} P_R e_\beta)(\bar u_i \sigma_{\mu\nu} P_R u_j)$ \\
        && $(C^{T,RR}_{ed})_{\alpha\beta ij}$ & $(\bar e_\alpha \sigma^{\mu\nu} P_R e_\beta)(\bar d_i \sigma_{\mu\nu} P_R d_j)$ \\\hline\hline

         $d=5$&Dipole
         & $(C_{e\gamma})_{\alpha\beta}$ & $(\bar e_\alpha \sigma^{\mu\nu} P_Re_\beta)F_{\mu\nu}$ \\\hline
    \end{tabular}
    \caption{\justifying 
    Relevant Wilson coefficients and corresponding LEFT operators. Here, $P_{X,Y}$ denote the chiral projection operators with $X,Y \in \{L,R\}$, and the indices $\alpha,\beta,i,j \in \{1,2,3\}$ label fermion generations excluding the top-quark.}
    \label{tab:JMSop}
\end{table}

Following the ``\textit{running and matching}" procedure, the LEFT WCs are matched at tree level to the SMEFT operators at the scale $m_Z$, which are given in Appendix~\ref{appndx:LEFT SMEFT matching} (for details, see Refs.~\cite{Jenkins:2017jig, Jenkins:2017dyc}). In the third and final step, these LEFT coefficients are evolved down to the relevant low-energy scale of the process under consideration ($m_\tau, m_\mu$ or $m_b$) to evaluate the desired experimental observables. We implement all these procedures in our numerical analysis with the help of the {\sf Wilson}~\cite{Aebischer:2018bkb} and {\sf Flavio}~\cite{Straub:2018kue} packages.


\subsubsection*{Change of Basis in LEFT}

Instead of chiral operators, the LEFT can also be expressed in terms of operators constructed from fermion currents classified by their Lorentz structure (LS), namely scalar $(S)$, pseudoscalar $(P)$, vector $(V)$, axial-vector $(A)$, tensor $(T)$, and pseudotensor $(T5)$. 
The leptonic dipole operators can also be written as tensor-dipole ($D$) and pseudotensor-dipole ($D5$) as: 
\begin{align}
    \cL^{(\rm int)}_{\rm LEFT} &\supset \frac{1}{4} \sum_{J,K} \frac{1}{v^2}C_{JK}^{\alpha\beta ij}\, \left(\bar e_\alpha \hat{G}_J e_\beta\right)\left(\bar q_i \hat{G}_K q_j\right) 
    + \frac{1}{4} \sum_{M,N}\frac{1}{v^2} C_{MN}^{\alpha\beta ij}\, \left(\bar e_\alpha \hat{G}^\mu_M e_\beta\right)\left(\bar q_i {(\hat{G}_N)}_\mu q_j\right)\nonumber\\
    &~~~~~~~~~+\frac{1}{2v^2}C_T^{\alpha\beta ij}\left(\bar e_\alpha\sigma^{\mu\nu}e_\beta\right)\left(\bar q_i\sigma_{\mu\nu}q_j\right)+\frac{1}{2v^2}C_{T5}^{\alpha\beta ij}\left(\bar e_\alpha\sigma^{\mu\nu}e_\beta\right)\left(\bar q_i\sigma_{\mu\nu}\gamma_5q_j\right)\nonumber\\
    &~~~~~~~~~+\frac{1}{2v}C_D^{\alpha\beta}\left(\bar e_\alpha\sigma^{\mu\nu}e_\beta\right)F_{\mu\nu}+\frac{1}{2v}C_{D5}^{\alpha\beta}\left(\bar e_\alpha\sigma^{\mu\nu}\gamma_5e_\beta\right)F_{\mu\nu}\,,
    \label{eq: master lagrangian}
\end{align}
where $J, K \in \{S, P\}$ and $M, N \in \{V, A\}$.  The matrices $\hat{G}^{(\mu)}$ are defined as:
\begin{align}
    \hat{G}_S &= \mathbb{1}, &     \hat{G}_P &= \gamma_5, & 
    \hat{G}^\mu_V &= \gamma^\mu, & 
    \hat{G}^\mu_A &= \gamma^\mu \gamma_5 \,.
\end{align}

The transformation from the chiral WCs (Table-\ref{tab:JMSop}) to the LS WCs is given by:
\begin{align}
\begin{bmatrix}
C_{VV} \\ C_{AV} \\ C_{VA} \\ C_{AA}
\end{bmatrix}
&=
\begin{bmatrix}
\phantom{-}1 &\phantom{-} 1 & \phantom{-}1 & \phantom{-}1 \\
-1 & -1 & \phantom{-}1 & \phantom{-}1 \\
-1 & \phantom{-}1 & -1 & \phantom{-}1 \\
\phantom{-}1 & -1 & -1 & \phantom{-}1
\end{bmatrix}
\begin{bmatrix}
C^{V,LL} \\
C^{V,LR} \\
C^{V,RL} \\
C^{V,RR}
\end{bmatrix}, &
\begin{bmatrix}
C_{SS} \\ C_{SP} \\ C_{PS} \\ C_{PP}
\end{bmatrix}
&=
\begin{bmatrix}
\phantom{-}1 & \phantom{-}1 & \phantom{-}1 & \phantom{-}1 \\
-1 & -1 & \phantom{-}1 & \phantom{-}1 \\
-1 & \phantom{-}1 & -1 & \phantom{-}1 \\
\phantom{-}1 & -1 & -1 & \phantom{-}1
\end{bmatrix}
\begin{bmatrix}
C^{S,LL} \\
C^{S,LR} \\
C^{S,RL} \\
C^{S,RR}
\end{bmatrix}\nonumber
\end{align}\\[-20pt]
\begin{align}
    \begin{bmatrix}
        C_T^{\alpha\beta ij}\\C_{T5}^{\alpha\beta ij}
    \end{bmatrix}&=
    \begin{bmatrix}
        \phantom{-}1 &\phantom{-}1\\-1&\phantom{-}1
    \end{bmatrix}
    \begin{bmatrix}
        (C^{T,RR})^\ast_{\beta\alpha ji}\\[4pt](C^{T,RR})_{\alpha\beta ij}
    \end{bmatrix} ,&\begin{bmatrix}
        C_D^{\alpha\beta}\\C_{D5}^{\alpha\beta}
    \end{bmatrix}&=
    \begin{bmatrix}
        \phantom{-}1 &\phantom{-}1\\-1&\phantom{-}1
    \end{bmatrix}
    \begin{bmatrix}
        C_{e\gamma}^{\beta\alpha\,\ast}\\C_{e\gamma}^{\alpha\beta}
    \end{bmatrix}
 \label{JMS WET basis tranform}   
\end{align}
where, the lepton flavour indices ($\alpha\beta$) and quark flavour indices ($ij$) as well as the class of WCs ($C_{eu}/C_{ed}$) are implicit unless specified.


\section{Observables \& operators relevant for different LFV  processes}
\label{sec:observables}

All the processes considered in this analysis can be broadly classified into three categories based on the quark FCNC involved. The majority of the processes feature interactions through a $d$-type FCNC, while some involve a $u$-type FCNC. The remaining processes include interactions through both types of FCNCs. This classification by quark FCNC, together with the meson-type (vector or pseudoscalar), is appropriate in identifying the class of operators relevant to each process. Table-\ref{tab:LFV Oprtr} associates the LFV processes that have experimental data along with the appropriate classes of operators.

\begin{table}[h!]
	\renewcommand{\arraystretch}{1.3}
	\begin{tabular}{|c|c|l|c|c|}\hline
		\multirow{2}{*}{\shortstack{Interacting\\ quark type}}&	\multirow{2}{*}{ \shortstack{Meson\\type}}&	\makecell[c]{\multirow{2}{*}{ Process}} & 	\multirow{2}{*}{\shortstack{Energy scale\\($\mu$)}} & 	\multirow{2}{*}{\parbox{7cm}{\centering Relevant classes for $2q2\ell$ Operators}}\\
		&&&& \\\hline\hline
	\multirow{11}{*}{\rotatebox{90}{ \shortstack{$d$-type \\$(d,s,b)$}}}	& \multirow{11}{*}{$\mathcal P$}  &  ~$K_L^0\to\ell_1\ell_2$    &    \multirow{3}{*}{500\,MeV}         &  \multirow{9}{*}{$\cO_{\ell q}^{(1)},\cO_{\ell q}^{(3)},\cO_{qe},\cO_{ed},\cO_{\ell d},\cO_{\ell edq}$}  \\
	 &    &    ~$K_L^0\to\pi^0\ell_1\ell_2$      &             &    \\
	&     &    ~$K^\pm\to\pi^\pm\ell_1\ell_2$      &              &    \\\cline{3-4}
	 &    &    ~$B^0\to\ell_1\ell_2$    &      \multirow{5}{*}{5\,GeV}        &    \\
	 &    &    ~$B^{0(\pm)}\to \pi^{0(\pm)}\ell_1\ell_2$    &              &    \\
	 &    &     ~$B^{0(\pm)}\to K^{\ast0(\pm)}\ell_1\ell_2$    &              &    \\
	&     &     ~$B^0_s\to\ell_1\ell_2$    &              &    \\
	&     &     ~$B^0_s\to\phi^0\ell_1\ell_2$    &              &    \\\cline{3-4}
	& 	&     ~$\tau^\pm\to K_S\ell^\pm$    &       2\,GeV       &    \\\cline{2-5}
	&  \multirow{2}{*}{$\mathcal V$}	&    ~$\tau^\pm\to \phi^0\ell^\pm$    &       2\,GeV         & \multirow{2}{*}{$\cO_{\ell q}^{(1)},\cO_{\ell q}^{(3)},\cO_{qe},\cO_{ed},\cO_{\ell d}$ }\\
		\cline{3-4}
	&	&     ~$\Upsilon\to\ell_1\ell_2$    &      10\,GeV        &    \\\hline
	\multirow{4}{*}{\rotatebox{90}{ \shortstack{$u$-type\\ $(u,c)$}}}	&  \multirow{3}{*}{$\mathcal P$}&	   ~$D^0\to\mu^\pm e^\mp$    &      \multirow{3}{*}{2\,GeV}        &   \multirow{3}{*}{$\cO_{\ell q}^{(1)},\cO_{\ell q}^{(3)},\cO_{qe},\cO_{eu},\cO_{\ell u},\cO_{\ell equ}^{(1)},\cO_{\ell equ}^{(3)}$}  \\
	&	&  ~$D^{0(\pm)}\to\pi^{0(\pm)}\mu^\pm e^\mp$  ~    &              &    \\
	&		&   ~$D_s^\pm\to K^\pm\mu^\pm e^\mp$      &              &    \\\cline{2-5}
	&	{$\mathcal V$}	&    ~$J/\psi\to\ell_1\ell_2$     &        3\,GeV      &  $\cO_{\ell q}^{(1)},\cO_{\ell q}^{(3)},\cO_{qe},\cO_{eu},\cO_{\ell u},\cO_{\ell equ}^{(3)}$  \\\hline
	\multirow{6}{*}{\rotatebox{90}{ \shortstack{both\\ $(u,d,s,c,b)$}}}&	--	&   ~$\mu^\pm N\to e^\pm N$      &     1\,GeV          &    \multirow{5}{*}{\shortstack{$\cO_{\ell q}^{(1)},\cO_{\ell q}^{(3)},\cO_{qe},\cO_{eu},\cO_{\ell u},\cO_{ed},\cO_{\ell d},$\\$\cO_{\ell edq},\cO_{\ell equ}^{(1)},\cO_{\ell equ}^{(3)}$}} \\\cline{2-4}
	&	\multirow{4}{*}{$\mathcal P$}	&    ~$\pi^0\to\mu^\pm e^\mp$     &     150\,GeV         &  \\\cline{3-4}
	&		&    ~$\eta\to\mu^\pm e^\mp$     &       550\,MeV       &  \\\cline{3-4}
	&		&    ~$\eta'\to\mu^\pm e^\mp$     &        1\,GeV      &  \\\cline{3-4}
	&		&     ~$\tau^\pm\to\pi^0\ell^\pm$    &        2\,GeV      &  \\\cline{2-5}
	&	$\mathcal V$	&     ~$\tau^\pm\to\rho^0\ell^\pm$    &      2\,GeV         & $\cO_{\ell q}^{(1)},\cO_{\ell q}^{(3)},\cO_{qe},\cO_{eu},\cO_{\ell u},\cO_{ed},\cO_{\ell d},\cO_{\ell equ}^{(3)}$ \\\hline
	\end{tabular}
\caption{\justifying List of different LFV processes of phenomenological interest and corresponding SMEFT operators belonging to only $2q2\ell$ operators contributing at the tree-level. We have refrained from mentioning any other possible SMEFT operators beyond $2q2\ell$ ones.}
\label{tab:LFV Oprtr}
\end{table}

The above observables can further be categorised based on the initial and final states of the processes, namely pure leptonic decays of pseudoscalar and vector mesons ($\mathcal P/\mathcal V\to\ell_i\ell_j$), semileptonic decays of pseudoscalars ($\mathcal P\to(\mathcal P'/\mathcal V)~\ell_i\ell_j $), $\tau$ decays to mesons ($\tau\to(\mathcal P/\mathcal V)~\ell$) and conversion of muon to electron at a nucleus ($\mu~N\to e~N$). Below, we discuss briefly the general procedure to analyse these processes and then provide relevant general expressions for all of them in terms of  LEFT WCs, which can be related to the SMEFT WCs through ``{\it running and matching}''. The same can be identified based on the flavours of leptons and quarks involved in the process (see Tables-\ref{Tab:Operator_different-quark_indices}--\ref{Tab:Operator_indices_tau}). 


\subsection{LFV semi-leptonic decays of pseudoscalar $(\mathcal P)$ mesons }

In this section, we shall discuss the LFV decays of pseudoscalar mesons, which are characterised by zero spin and odd parity. This class of mesons can further be categorised into two segments. One class is the meson \textit{nonet}, consisting of light quarks ($u, d$ and $s$) such as pions $(\pi)$, eta mesons $(\eta)$, and $K$ mesons. The other class of mesons include \textit{heavy} quarks, bottom ($B$ meson) and charm ($D$ meson). All of these mesons can potentially give rise to LFV decays in both leptonic and semileptonic modes. Here we first discuss the latter type of decays, which can further be classified into two categories. First one is $\mathcal P\to \mathcal P'\ell_i^-\ell_j^+$, where the decay product of a pseudoscalar meson $(\mathcal P)$ contains another pseudoscalar meson $\mathcal P^{\prime}(\pi^0,\pi^{\pm}, K^0,K^+ ~ \rm etc.)$. Fig.~\ref{Fig:Generic PtoPp LFV} represents a generic Feynman diagram corresponding to these processes. The quark composition of mesons considered for this analysis is given in Table-\ref{tab:mesons}.

\begin{figure}[h!]
\centering
\begin{tikzpicture}
\begin{feynman}
\vertex (a1) {\(q_i\)};
\vertex[blob,shape=circle,fill=black,minimum height=1.0cm,minimum width=0.5cm] at ($(a1)+ (3.0cm, 0.0cm)$)(a2){};
\vertex[right=3.0cm of a2] (a3) {\(q_j\)};
\vertex[above=3em of a1] (b1) {\(q_s\)};
\vertex[above=3em of a3] (b2) {\(q_s\)};
\vertex[below=1.5em of a3] (c1) {\(\ell_\alpha^{+}\)};
\vertex[below=3em of c1] (c3) {\(\ell_\beta^-\)};
\diagram* {
(a2) -- [fermion] (a3), 
(a1) -- [fermion] (a2),
(b1) -- [fermion] (b2),
(c1) -- [fermion] (a2),
(a2) -- [fermion] (c3),
};
\draw [decoration={brace}, decorate] (a1.south west) -- (b1.north west)
node [pos=0.5, left] {\(\mathcal P\)};
\draw [decoration={brace}, decorate] (b2.north east) -- (a3.south east)
node [pos=0.5, right] {\(\mathcal P^{\prime}/\mathcal V\)};
\end{feynman}
\end{tikzpicture}
\caption{\justifying Effective vertex for a LFV semi-leptonic decay of a pseudoscalar meson $\mathcal P$ to a pseudoscalar $\mathcal P^{\prime}$ or a vector meson $\mathcal V$. The (anti-)quark $q_i$ converts into another (anti-)quark $q_j$ and two different flavors of leptons $\ell_\alpha,\ell_\beta$. The other (anti-)quark $q_s$ represents the spectator (anti-)quark.}
\label{Fig:Generic PtoPp LFV}
\end{figure}


\begin{table}[h!] 
\centering
\begin{adjustbox}{max width=\textwidth}
\renewcommand{\arraystretch}{1.6}
	\begin{tabular}{!{\vrule width 1pt}c|c|c|c|c|c|c|c!{\vrule width 1pt}}
		\Xhline{1pt}
		\multirow{4}{*}{\shortstack{\phantom{b}\\Pseudoscalars\\$(\mathcal P)$}}&\multirow{2}{*}{\shortstack{Heavy\\ Quark\\($c,b$)}} & \textbf{$B^0$} & ~~~\textbf{$B^+$} ~~~& \textbf{$B_s^0$} & ~~~~\textbf{$D^0$}~~~~ & \textbf{$D^+$} & \textbf{$D_s^+$}  \\
		\cline{3-8}
		&& $d\bar{b}$ & $u\bar{b}$ & $s\bar{b}$ & $c\bar{u}$ & $c\bar{d}$ & $c\bar{s}$   \\
		\Xcline{2-8}{0.8pt}
		&\multirow{2}{*}{\shortstack{\phantom{b}\\Light\\ Quarks\\$(u,d,s)$}}  & \textbf{$K^+$} & \textbf{$K_{S(L)}^0$} & \textbf{$\pi^0$} & \textbf{$\pi^+$} & \textbf{$\eta$} & \textbf{$\eta'$} \\
		\cline{3-8}
		& & \rule{0pt}{1.9em}$u\bar{s}$ & $\displaystyle \frac{d\bar{s} \mp s\bar{d}}{\sqrt{2}}$ & $\displaystyle \frac{u\bar{u} - d\bar{d}}{\sqrt{2}}$ & $u\bar{d}$ & $\displaystyle \frac{u\bar{u} + d\bar{d} - 2s\bar{s}}{\sqrt{6}}$ & $\displaystyle \frac{u\bar{u} + d\bar{d} + s\bar{s}}{\sqrt{3}}$ \\[7pt]
		\Xhline{1pt}
		\multicolumn{2}{!{\vrule width 1pt}c|}{\multirow{2}{*}{\shortstack{\phantom{b}\\Vectors\\$\mathcal (\mathcal V)$}}}&\textbf{$\rho^0$} &$\Upsilon$ & \textbf{$\omega$} &  \textbf{$\phi$} &   \textbf{$J/\psi$} &\\
		\cline{3-8}
		 \multicolumn{2}{!{\vrule width 1pt}c|}{}&\rule{0pt}{1.9em}$\displaystyle \frac{u\bar{u} - d\bar{d}}{\sqrt{2}}$ & $b\bar{b}$ &  $\displaystyle \frac{u\bar{u} + d\bar{d}}{\sqrt{2}}$& $s\bar{s}$ & $c\bar{c}$  &  \\[7pt]
		\Xhline{1pt}
	\end{tabular}
\end{adjustbox}
\caption{\justifying All mesons and their quark compositions under consideration in this analysis. They take part in LFV processes following Fig.\,\ref{Fig:Generic PtoPp LFV} or Fig.\,\ref{Fig:Generic Ptolilj LFV}.}
\label{tab:mesons}
\end{table}


\subsubsection{Pseudoscalar meson to pseudoscalar meson decays ($\mathcal P\to \mathcal P'\ell_i^-\ell_j^+$)}

The exact expression of the branching ratio formula for semi-leptonic decays of pseudoscalar mesons can be quite involved (see \eg~Ref~\cite{Becirevic:2016zri} for $B$-meson decays). Therefore, for simplicity, we prefer to express the generic BR formula in terms of the effective coefficients ($a_i$) associated with the relevant WCs (Eq.(\ref{JMS WET basis tranform})) or their combinations, as:
\begin{align}
    \mathrm{BR}(\mathcal P \to \mathcal P' \ell_i^- \ell_j^+) &= a_{VV} \left| C_{VV} \right|^2 + a_{AV} \left| C_{AV} \right|^2 + a_{TT} \left( \left| C_T \right|^2 + \left| C_{T_5} \right|^2 \right) \nonumber\\
&\quad + a_{SS} \left| C_{SS} \right|^2 + a_{PS} \left| C_{PS} \right|^2 + a_{VS} \, \mathrm{Re} \left[ C_{VV} C_{SS}^{*} \right] \nonumber\\
&\quad + a_{AP} \, \mathrm{Re} \left[ C_{AV} C_{PS}^{*} \right] + a_{VT} \, \mathrm{Re} \left[ C_{VV} C_T^{*} \right] + a_{AT_5} \, \mathrm{Re} \left[ C_{AV} C_{T_5}^{*} \right].
\label{eq:Generic sudoscalar meson decays}
\end{align}


{\bf A comment on LFV Decays of $B,K$ and $D$ mesons:} 
(i) LFVBDs are one of the most extensively studied processes in the search for NP. These decays have both leptonic and semileptonic modes. The semileptonic processes involve a bottom $(b)$ quark to strange $(s)$ or down ($d$) quark transition along with two leptons $(\ell_{i,j},i\neq j)$, $b\to s(d)\ell_i\ell_j$. In literature, they are found to be most popularly studied in terms of operators $\cO_9, \cO_{10},\cO_S,\cO_P,\cO_T$ and their chiral counterparts~\cite{Aebischer:2015fzz, Buchalla:1995vs}. (ii) Similar to the LFVBDs, rare LFV $D$ meson decays can happen in several modes $\eg$, $D^0\to X^0e^{\pm}\mu^{\mp}$ where $X^0=\pi^0,K^0_S,\bar{K}^{*0},\rho^0,\phi,\omega,\eta$~\cite{BaBar:2020faa} or $D^+\to\pi^+e^{\pm}\mu^{\mp} \etc$. Such processes involve $c\to ue^\pm\mu^\mp$ transition through $|\Delta C|=1$ effective Lagrangian, which can be written with two-step matching at the $m_Z$ scale and the $m_b$ scale. Application of the EFT framework to these decays~\cite{Burdman:2001tf,deBoer:2015boa,Hazard:2017udp} also reveals that the differential decay rate in terms of the Wilson coefficients takes the general form of Eq.(\ref{eq:Generic sudoscalar meson decays}). It is to be noted here that, except $\pi^0$ and $K$ decay modes, other LFV decay modes couldn't be implemented in \textsf{Flavio} due to technical complexities. (iii) In the presence of strong interaction and neglecting the top quark mass, the $|\Delta S|=1$ interactions responsible for the rare leptonic and semi-leptonic $K$ decays under consideration are dominated by Gilman–Wise operators 
$\cO_{7X}~(X=V,A)$, with Wilson coefficients $C_{7X}$~\cite{Gilman:1979ud,Inami:1980fz,Dib:1988js}.
The decay rate for semileptonic processes like $K^+\to\pi^+\ell_1^+\ell_2^-$ and others takes the form of Eq.(\ref{eq:Generic sudoscalar meson decays}). Due to composition of $K_{L(S)}$ meson ($=\frac{1}{\sqrt 2}(\bar sd\pm \bar ds)$), one needs to make the replacement for $K_{L(S)}\to\pi^0$ decays as \cite{Plakias:2023esq}:
\begin{align}
    C_{XY}&\to\frac{(C_{XY})_{\alpha\beta12}\mp(C_{XY})_{\alpha\beta21}}{2} \hspace{1 em}\,;&C_{AB}&\to\frac{(C_{AB})_{\alpha\beta12}\pm(C_{AB})_{\alpha\beta21}}{2}
\end{align}
where $X,Y\in\{V,A\}$ and $A,B\in\{S,P\}$. The upper sign is for $K_L$ while the lower is for $K_S$.


\subsubsection{Pseudoscalar meson to vector meson decays ($\mathcal P\to \mathcal V\ell_i^-\ell_j^+$)}

Similarly to pseudoscalar-to-pseudoscalar transition, $\mathcal P\to \mathcal V$ transitions are also described by seven independent form factors which are usually determined by the light-cone QCD sum rule method~\cite{Ball:2004rg,Khodjamirian:2006st,Lu:2011jm} and can ultimately be expressed as,

\begin{align}
    \mathrm{BR}(\mathcal P \to \mathcal V \ell_i^- \ell_j^+) &= a_{VV} |C_{VV}|^2 + a_{VA} |C_{VA}|^2 + a_{AV} |C_{AV}|^2 + a_{AA} |C_{AA}|^2 \nonumber\\
&\quad + a_{PP} |C_{PP}|^2 + a_{SP} |C_{SP}|^2 + a_{AP} \, \text{Re}[C_{VA} C_{SP}^*] + a_{AS} \, \text{Re}[C_{AA} C_{PP}^*] \nonumber\\
&\quad + a_{TT} \left(|C_T|^2 + |C_{T_5}|^2 \right) + a_{ST} \, \text{Re}[C_{SP} C_T^*] + a_{PT} \, \text{Re}[C_{PP} C_T^*] \nonumber\\
&\quad + a_{ST_5} \, \text{Re}[C_{SP} C_{T_5}^*] + a_{PT_5} \, \text{Re}[C_{PP} C_{T_5}^*].
\label{eq: General sudoscalar to vector BR}
\end{align}


\subsection{LFV leptonic decays of Pseudoscalar Mesons ($\mathcal P\to \ell_i^-\ell_j^+$).}
\label{Subsec:Meson LFV decays}

The generic expression for a pseudoscalar meson $\mathcal P$ decaying to leptons ($\ell_i^-\ell_j^+$), corresponding to a generic Feynman diagram of Fig.\,\ref{Fig:Generic Ptolilj LFV}, is given as:
\begin{align}
    \mathrm{BR}(\mathcal P\to\ell_i^-\ell_j^+)&=\frac{\tau_\mathcal P}{128\pi\,m_\mathcal P^3v^4}\sqrt{\lambda(m_\mathcal P,m_i,m_j)}\nonumber\\
    &~~~~~\times\Bigg(\left(m_\mathcal P^2-(m_i+m_j)^2\right)\cdot\left|\sum_kf_\mathcal P^k\left(C^k_{VA}(m_i-m_j)+C^k_{SP}\frac{m_\mathcal P^2}{m_{q_1^k}+m_{q_2^k}}\right)\right|^2\nonumber\\
    &~~~~~~~~+\left(m_\mathcal P^2-(m_i-m_j)^2\right)\cdot\left|\sum_kf_\mathcal P^k\left(C^k_{AA}(m_i+m_j)+C^k_{PP}\frac{m_\mathcal P^2}{m_{q_1^k}+m_{q_2^k}}\right)\right|^2\Bigg)\,,
\label{eq:Generic LFV leptonic decays}    
\end{align}
where summation over $k$ runs over different quark-antiquark pairs of the meson,\footnote{The summation over $k$ is only required for mesons that are linear combinations of quark-antiquark pairs, such as $K_{L(S)}, \pi^0,\eta$, etc. For the others, $f_\mathcal P^2$ can be written in the prefactor.}
$v$ is the vacuum expectation value (vev), $\tau_\mathcal P$ and $m_\mathcal P$ are the lifetime and mass of $\mathcal P$ respectively. $m_{i,j}$ and $m_{q1,q2}$ are the masses of leptons and quarks respectively. The decay constant $f_\mathcal P$, defined in terms of hadronic matrix element, and K\"all\'en $\lambda$ function are given as:
\begin{align}
    \langle0|\bar q_2^k\gamma_\mu\gamma_5q^k_1|\mathcal P(p)\rangle&=ip_\mu f_\mathcal P^k , \label{eq:decay cont defn}\\
    \langle0|\bar q_2^k\gamma_5q^k_1|\mathcal P(p)\rangle&=if_\mathcal P^k \frac{m_\mathcal P^2}{m_{q_1^k}+m_{q_2^k}},\\
    \lambda(a,b,c)&=(a^2-(b-c)^2)(a^2-(b+c)^2)\,,
    \label{eq:lam defn}
\end{align} 
The hadronic matrix elements with scalar and vector densities vanish identically due to parity conservation. Also, due to the absence of the Lorentz structure, the tensor matrix element also vanishes. Thus, only the operators with pseudoscalar or axial-vector quark FCNC contribute, as evident from Eq.(\ref{eq:Generic LFV leptonic decays}).


\begin{figure}[h!]
\centering
\begin{tikzpicture}
\begin{feynman}
\vertex (a1) {\(q_j^k\)};
\vertex[right=2.5 cm of a1] (a2);
\vertex[above=2.5em of a1] (b1) {\(\overline q_i^k\)};
\vertex[above=2.5em of a2] (b2) ;
\vertex[above right =0.65em of a2,blob,shape=circle,fill=black,minimum height=1.0cm,minimum width=0.5cm] (c){};
\vertex[above right=2.5cm of c] (b3) {\(\ell_\alpha^-\)};
\vertex[below right=2.5cm of c] (a3) {\(\ell_\beta^+\)};
\diagram* {
 (a1) -- [fermion] (a2),
 (b2) -- [fermion] (b1),
 (a2) -- [half right, looseness=2.0] (b2),
 (c) -- [fermion](b3),
 (a3) -- [fermion] (c),
};
\draw [decoration={brace}, decorate] (a1.south west) -- (b1.north west)
node [pos=0.5, left] {\(\mathcal P/\mathcal V\)};
\end{feynman}
\end{tikzpicture}
\caption{\justifying Effective vertex of pseudoscalar/vector meson (or quarkonium) $(\mathcal P/\mathcal V)$ LFV decay to a pair of leptons via $(2q2\ell)$ operators.}
\label{Fig:Generic Ptolilj LFV}
\end{figure}


\subsubsection{LFV leptonic decays of mesons $(\pi^0,\eta,\eta^{\prime})$}
\label{subsec:LFV meson with light quarks}

The effective Lagrangian ($\cL_{\rm eff}$) for the decays of light-quark mesons $(\pi^0,\eta,\eta^{\prime})$ to two different flavors of leptons $(\ell_i,\ell_j)$, can be divided into a dimension-5 part, $\cL_{\rm dim-5}$; 
a dimension-6 part, $\cL_{\rm dim-6}$; and a dimension-7 gluonic part, $\cL_{\rm dim-7}$~\cite{Hazard:2016fnc,Hoferichter:2022mna,Calibbi:2022ddo} depending upon how one considers to include spin-dependent (SD) or independent contributions coming from the relevant operators. 
For our purpose, we shall consider only the $2q2\ell$ type of operators as the leading contributors, and the corresponding Lagrangian is given by Eq.(\ref{eq: master lagrangian}). The resulting branching ratio for $\mathcal P(\pi^0,\eta,\eta^{\prime})\to \ell_i^-\ell_j^+$ is given by Eq.(\ref{eq:Generic LFV leptonic decays}).\footnote{The scale relevant to $\pi^0$-decay ($\mu\sim100$\,MeV) is too low for {\sf Wilson} package to run the WCs. One can, in principle, calculate the approximate BR at the minimum possible scale for {\sf Wilson} ($\sim500$\,MeV). However, the constraints are relaxed due to its small lifetime ($\tau_{\pi^0}\sim10^{-16}$\,s). Thus, we do not consider these decays in this work.}


\subsubsection{LFV leptonic decays of $K$, $D$ and $B$ mesons}
\label{subsubsec:LFVKD}

In literature, EFT analyses of LFV Kaon decays are not plentiful, and mainly they are discussed in relation to LFVBDs \cite{Crivellin:2016vjc, Hazard:2017udp,Fayyazuddin:2018zww}. The decay rate for the process $K_L\to\ell_i^+\ell_j^-$ is given by Eq.(\ref{eq:Generic LFV leptonic decays}). Analyses of leptonic LFVDDs in the EFT scenario can be found in \cite{Ryd:2009uf,Hazard:2017udp}. The branching fraction for such decays can also be translated to the above generic form. The most general form of branching fraction that can be found in literature for such LFV decays is usually given in terms of $C_{9,10}$ \etc~\cite{Becirevic:2016zri}.


\subsection{Leptonic LFV decays of Vector Quarkonia ($\mathcal V\to \ell_i^-\ell_j^+$)}
\label{subsec:VLL_decays}

In this section, we shall discuss the LFV decays of a subset of vector mesons which have odd parity, the same as that of pseudoscalar mesons, but having total spin one. Vector mesons include rho $(\rho)$, phi $(\phi)$, omega $(\omega)$, $J/\psi$, $\Upsilon$ mesons $\etc.$ Out of these,  $J/\psi$ and $\Upsilon$ are collectively known as \textit{quarkonia} $(\mathcal V)$ as they consist of a heavy quark and its antiquark. The LFV decays of such states are known as LFV Quarkonium decays (LFVQDs). In the following, we shall limit our analysis to these decays, as their theoretical and experimental results are more robust.

The general expression of branching fraction for such processes can be written as \cite{Abada:2015zea,Hazard:2016fnc,Calibbi:2022ddo}

\begin{align}
    \mathrm{BR}(\mathcal V\to \ell_i^-\ell_j^+)&=\frac{\tau_\mathcal Vm_\mathcal V^3}{192\pi\,v^4}\left(1-\frac{m_{\ell_j}^2}{m_\mathcal V^2}\right)^2\nonumber\\
    &\quad\quad\times\Bigg[(f_\mathcal V)^2\left(\left|C_{VV}\right|^2+\left|C_{AV}\right|^2\right)\left(1+\frac{m_{\ell_j}^2}{2m_\mathcal V^2}\right)
    +8\left({f_\mathcal V^T}\right)^2\left(|\bar C_T|^2+|\bar C_{T5}|^2\right)\left(1+\frac{2m_{\ell_j}^2}{m_\mathcal V^2}\right)\nonumber\\
    &\quad\quad\quad\quad+12f_\mathcal Vf_\mathcal V^T\frac{m_{\ell_j}}{m_\mathcal V}\Re\left(\bar C_TC_{VV}^\ast-\bar C_{T5}C_{AV}^\ast\right)\Bigg],\quad\text{for }j>i.
    \label{eq:LFVQD expr}
\end{align}
For $j<i$, one can replace $m_{\ell_j}\to -m_{\ell_i}$. The mass of the lighter lepton is neglected. The decay constants $f_\mathcal V$ and $f_\mathcal V^T$ are defined as:
\begin{align}
    \langle 0|\bar q\gamma^\mu q|\mathcal V(p,\epsilon)\rangle=m_\mathcal V\epsilon^\mu_\mathcal Vf_\mathcal V,&&
    \langle0|\bar q\sigma^{\mu\nu }q|\mathcal V(p,\epsilon)\rangle=if_\mathcal V^T\left(\epsilon^\mu_\mathcal Vp^\nu-\epsilon^\nu_\mathcal Vp^\mu\right)\,.
\end{align}
 Other hadronic matrix elements vanish again due to parity conservation, resulting in contributions only coming from the operators with vector and tensor quark FCNC. The WCs $\bar C_{T(5)}$ are modified due to the leptonic dipole contributions as follows:
\begin{align}
    (\bar C_T)_{\alpha\beta ij}&=(C_T)_{\alpha\beta ij}-(eQ_i)\frac{{v}\, f_\mathcal V}{m_\mathcal Vf^T_\mathcal V}(C_D)_{\alpha\beta}\,\delta_{ij}\,,\nonumber\\
    (\bar C_{T5})_{\alpha\beta ij}&=(C_{T5})_{\alpha\beta ij}-(eQ_i)\frac{{v}\, f_\mathcal V}{m_\mathcal Vf^T_\mathcal V}(C_{D5})_{\alpha\beta}\,\delta_{ij}\,.
    \label{eq:dipole_contibution}
\end{align}


\subsection{LFV $\tau$ decays}
\label{Subsec:LFV tau decays}

The Feynman diagram of LFV $\tau$-decays through $2q2\ell$ effective vertex is given in Fig.\,\ref{Fig:tau LFV decays}. The quarks in the final states hadronise to form a meson. These LFV decays can also be further classified into two categories based on whether the final state meson is a pseudoscalar ($\tau\to\mathcal P\ell$) or a vector ($\tau\to\mathcal V\ell$) meson.

\begin{figure}[h!]
\centering
\begin{tikzpicture}
\begin{feynman}
\vertex (a1) {\(\tau\)};
\vertex[blob,shape=circle,fill=black,minimum height=1.0cm,minimum width=0.5cm] at ($(a1)+ (3.0cm, 0.0cm)$)(a2){};
\vertex[right=3.0cm of a2] (a3) {\(\ell_\alpha(e,\mu)\)};
\vertex[below=1.5em of a3] (c1) {\(\overline{q}_j\)};
\vertex[below=3em of c1] (c3) {\(q_i\)};
\diagram* {
(a2) -- [fermion] (a3), 
(a1) -- [fermion] (a2),
(c1) -- [fermion] (a2),
(a2) -- [fermion] (c3),
};
\end{feynman}
\end{tikzpicture}
\caption{\justifying Feynman diagram of LFV hadronic $\tau$ decays. The blob indicates the charged LFV vertex.}
\label{Fig:tau LFV decays}
\end{figure}


\subsubsection[Semileptonic $\tau$ decays  $\tau \to \mesonV \ell$]{Semileptonic $\tau$ decays  (${\tau \to \mesonV \ell}$)}
\label{App:semileptonicV}

The generic expression for the branching ratio of $\tau$ decays to vector meson and a lighter lepton $\ell\,(e/\mu)$ can be expressed as (neglecting $m_\ell$)~\cite{Plakias:2023esq}:
\begin{align}
    \text{BR}(\tau^\pm\to\ell^\pm\mathcal V)&=\tau_\tau\frac{ m_\tau^3}{256\pi  v^4}\left(1-\frac{m_\mathcal V^2}{m_\tau^2}\right)^2\nonumber\\
    &\quad\quad \times\Bigg[(f_\mathcal V)^2\left(1+\frac{2m_\mathcal V^2}{m_\tau^2}\right)\left(|C_{VV}|^2+|C_{AV}|^2\right)
    +32\left({f^T_\mathcal V}\right)^2\left(1+\frac{m_\mathcal V^2}{2m_\tau^2}\right)\left(|\bar C_T|^2+\left|\bar C_{T_5}\right|^2\right)\nonumber\\
    &\quad\quad\quad\quad+24f_\mathcal V^Tf_\mathcal V\frac{m_\mathcal V}{m_\tau}\Re\left(\bar C_TC_{VV}^\ast-\bar C_{T5}C_{AV}^\ast\right)\Bigg] \,,
\label{BR:tautoVecSemilep}    
\end{align}
where $m_{\mathcal V}$ and $m_{\tau}$ are the vector meson and $\tau$ mass respectively, $\tau_\tau$ is the lifetime of the $\tau$. Other notations carry a similar meaning as before. The only non-vanishing hadronic matrix elements are of the vector and tensor type, same as the leptonic decay of a vector meson (sec-\ref{subsec:VLL_decays}). As a result, only the vector and tensor WCs contribute, and $\bar C_{T(5)}$ are also modified due to the presence of leptonic dipole WCs, as given in Eq.(\ref{eq:dipole_contibution}).


\subsubsection[Semileptonic $\tau$ decays  $\tau \to \mesonP \ell$]{Semileptonic $\tau$ decays  (${\tau \to \mesonP \ell}$)}
\label{subsec:semileptonicP}

Similar to the vector mesons, we take the expressions for the $\tau$ decays to pseudoscalar mesons from Ref.~\cite{Aebischer:2018iyb}.  The branching ratio for $\tau\to \mesonP\ell$, with $\ell=e,\mu$, is given by (again, neglecting $m_\ell$)
\begin{align}
    \mathrm{BR}(\tau \to \ell_i \mesonP) &= \tau_\tau \frac{ m_\tau^3}{256\pi v^4} \left( 1 - \frac{m_\mesonP^2}{m_\tau^2} \right)^2\nonumber\\
    &\quad\times\left[
\left|\sum_k f_\mesonP^k\left(C_{VA}^{k} + \frac{m_\mesonP^2 C_{SP}^{k}}{m_\tau (m_{q_1^k} + m_{q_2^k})} \right)\right|^2 +
\left| \sum_k f_\mesonP^k\left(C_{AA}^{k} - \frac{m_\mesonP^2 C_{PP}^{k}}{m_\tau (m_{q_1^k} + m_{q_2^k})} \right)\right|^2
\right].
\end{align}
where $k$ runs over flavour eigenstate quark-antiquark pairs of the meson. Similar to the case of leptonic decays of pseudoscalar mesons (sec-\ref{Subsec:Meson LFV decays}), the only surviving hadronic matrix elements are the pseudoscalar and axial-vector, resulting in contributions only from operators having quark FCNC with these Lorentz structures. 


\subsection{Muon to electron conversion in nuclei (${\rm CR }(\mu \rightarrow e,N)$)}
\label{Subsec: CR-mue}

\begin{figure}[h!]
\centering
\begin{tikzpicture}
\begin{feynman}
\vertex (a1) {\(q_{\rm int}\)};
\vertex[blob,shape=circle,fill=black,minimum height=.5cm,minimum width=0.5cm] at ($(a1)+ (2.5cm, 0.0cm)$)(a2){};
\vertex[right=2.5cm of a2] (a3) {\(q_{\rm int}\)};
\vertex[above=3em of a1] (b1) {\(q_{\rm sp}^{(1)}\)};
\vertex[above=3em of a3] (b2) {\(q_{\rm sp}^{(1)}\)};
\vertex[above=1.5em of a1] (d1) {\(q_{\rm sp}^{(2)}\)};
\vertex[above=1.5em of a3] (d2) {\(q_{\rm sp}^{(2)}\)};
\vertex[below=3em of a1] (c1) {\(\mu^-\)};
\vertex[below=3em of a3] (c3) {\(e^-\)};
\diagram* {
(a2) -- [fermion] (a3), 
(a1) -- [fermion] (a2),
(b1) -- [fermion] (b2),
(d1) -- [fermion] (d2),
(c1) -- [fermion] (a2),
(a2) -- [fermion] (c3),
};
\draw [decoration={brace}, decorate] (a1.south west) -- (b1.north west)
node [pos=0.5, left] {\(p/n\)};
\draw [decoration={brace}, decorate] (b2.north east) -- (a3.south east)
node [pos=0.5, right] {\(p/n\)};
\end{feynman}
\end{tikzpicture}
\caption{\justifying Effective vertex of muon to electron conversion at a nucleus ($N$). $q\in\{u,d\}$ represents quarks with $q_{\rm int}$ and $q_{\rm sp}$, meaning interacting and spectator quarks respectively. Apart from the interaction with the valence quarks as shown, a muon can also interact with the sea quarks, resulting in scalar contributions from $s$-quarks as well~\cite{Kuno:1999jp}.}
\label{Fig:mue_conv}
\end{figure}
The LFV interaction Lagrangian for the $\mu$--$e$ transition (shown in Fig.~\ref{Fig:mue_conv}) in a nuclei is given by \cite{Kuno:1999jp}:

\begin{align}
    \text{CR}(\mu\to e, N)&=\frac{1}{\omega_{\rm capt}v^4}\sum_{X\in\{L,R\}}\left|{\frac{v}{2m_\mu}}C^X_{e\gamma}D_N+\sum_{f\in\{p,n\}}\left(g_{XS}^fS^f_N+g_{XV}^fV^f_N\right)\right|^2 \,,
    \label{eq:CRmue BR}
\end{align}
where $\omega_{\rm capt}$ is the muon capture rate by the nucleus $N$ and  $X$ runs over the chirality of the lepton FCNC. The dipole WCs can be obtained as $C^R_{e\gamma}=C_{e\gamma}$ and $(C^L_{e\gamma})_{\alpha\beta}=(C_{e\gamma})_{\beta\alpha}^\ast$. The fermion flavour indices $(\alpha\beta ij)$ are implicit. The effective couplings $g$'s are given in terms of LEFT WCs (Table-\ref{tab:JMSop}) by,
\bea
    g_{XS}^f &=& \sum_{q\in\{u,d,s\}}G_S^{q,f}(C_{eu(d)}^{S,XL}+C_{eu(d)}^{S,XR})\,, \label{eq:gfXS} \\
    g_{XV}^f &=& \sum_{q\in\{u,d\}}n_q^{f}(C_{eu(d)}^{V,XL}+C_{eu(d)}^{V,XR})\,,
    \label{eq:gfXV}
\eea
with $n_q^f$ being the number of valence $q$-quarks in $f$ (neutron/proton). While summing over $q$, the coefficient $C_{eu}(C_{ed})$ is used for $u$-type($d$-type) quarks. The coefficients $G_S^{q,f}$ are defined in terms of nucleon matrix elements as
\begin{align}
G_S^{q,f}=\frac{1}{2m_f}\langle f|\bar q q|f\rangle \hspace{1 em}\,; \quad f\in\{p,n\}\,,
\end{align} 
with $m_f$ is the mass of $f$. The matrix elements involving pseudo-scalar, axial-vector, and tensor quark currents vanish identically in this case. The numerical values of the relevant parameters, along with the values of the overlap integrals, $D_N$, $S_N$ and $V_N$~\cite{Kitano:2002mt,Kosmas:2001mv}, are provided in Table\,\ref{tab:SVG}.
\begin{table}[h!]
\setlength{\tabcolsep}{7pt}
\renewcommand{\arraystretch}{1.2}
    \centering
    \begin{tabular}{|c|c|c|c|c|c|c|}\hline
\multirow{2}{*}{$N$} & \multirow{2}{*}{\makecell[c]{$\omega_{\rm capt}$ \\ {[$10^{-18}$ GeV]}}}& \multirow{2}{*}{\makecell[c]{$D_N$\\$[m_\mu^{5/2}]$}}& \multicolumn{2}{c|}{$S\;[m_\mu^{5/2}]$} & \multicolumn{2}{c|}{$V\;[m_\mu^{5/2}]$} \\\cline{4-7}
 & & &$n$ & $p$ & $n$ & $p$ \\\hline\hline
Au & 8.603 & 0.189 & 0.0918 & 0.0614 & 0.146 & 0.0974 \\\hline
Ti & 1.705 & 0.0864 & 0.0435 & 0.0368 & 0.0468 & 0.0396 \\\hline
\end{tabular}~~~
\begin{tabular}{|c|c|c|}\hline
            & $p$ &$n$\\\hline
    $G_S^u$ & 5.1 & 4.3\\\hline
    $G_S^d$ & 4.3 & 5.1\\\hline
    $G_S^s$ & 2.5 & 2.5\\\hline
\end{tabular}
    \caption{\justifying Numerical values of parameters relevant to CR$(\mu\to e, N)$ used in \textsf{Flavio}. The values of factors $D,S$ and $V$ are in units of $m_\mu^{5/2}$.}
    \label{tab:SVG}
\end{table}


\section{Calculation methods}
\label{sec:Calculation method}

As already mentioned, we use the \textsf{Flavio}~\cite{Aebischer:2018bkb} package for computing the observables, which itself relies on the {\sf Wilson} package for the running and matching of the SMEFT and LEFT WCs. Most of the observables considered in this work are already implemented in the \textsf{Flavio} package. However, several observables relevant to our analysis are not included in the default implementation. To address this, we extend the \textsf{Flavio} framework by incorporating these additional observables manually, using the generic expressions mentioned in sec-\ref{sec:observables}. The parameters and other input data used for this purpose are provided in Tables-\ref{tab:effco} and \ref{tab:decay constants}.


\begin{table}[H]
    \centering
\renewcommand{\arraystretch}{1.2}
\begin{tabular}{|l|c|c|c|c|c|c|c|c|c|}
    \hline
    Decay Process& $a_{VV}$ & $a_{AV}$ & $a_{SS}$ & $a_{PS}$ & $a_{VS}$ & $a_{AP}$ & $a_{TT}$ & $a_{VT}$ & $a_{AT5}$ \\
    \hline
    $D^+ \to \pi^+e^-\mu^+$     &0.0166 &0.0166 &0.0337 & 0.0337 & $-0.0066$ & 0.0066 & 0.019 & 0.0081 & 0.0081 \\
    $D^0 \to \pi^0e^-\mu^+$     & 0.00326 & 0.00326 & 0.0066 & 0.0066 & $-0.0013$ & 0.0013 & 0.0038 & 0.0016 & 0.0016 \\
    $D_s^+ \to K^+e^-\mu^+$     & 0.00622 & 0.00622 & 0.00893 & 0.00893 & $-0.002364$ & 0.002364 & 0.00377 & 0.0025 & 0.0025 \\
    $K_L \to \pi^0e^-\mu^+$     & 0.690 & 0.693 & 12.18 &12.23 & $-3.19$ & 3.24 & -- & -- & -- \\
    $K^+ \to \pi^+e^-\mu^+$     & 0.157 & 0.1578 &2.71 & 2.72& $-0.723$ & 0.735 & -- & -- & -- \\
    \hline
\end{tabular}
    \caption{\justifying Values of coefficients mentioned in Eq.(\ref{eq:Generic sudoscalar meson decays}) for various meson decay modes~\cite{Plakias:2023esq}. Entries marked with ``--'' indicate non-relevant data since tensor type WCs for down-type transitions are not contributing at dim-6. The coefficient for opposite lepton sign can be obtained by replacing $a_{VS}\to-a_{VS}$. }
    \label{tab:effco}
\end{table}
\begin{table}[h!]
    \renewcommand{\arraystretch}{1.8}
	\begin{tabular}{|c|c|c|c|c|c|c|}
		\hline
		\bf Meson & $\pi^0$ & $K_L$ & $K_S$ & $D^0$ & $\eta$ & $\eta'$ \\\hline
		\bf Lifetime [GeV$^{-1}$] 
		& $1.28\times 10^{8}$ 
		& $7.77\times 10^{16}$ 
		& $1.36\times 10^{14}$ 
		& $6.23\times 10^{11}$ 
		& $7.63\times 10^{5}$ 
		& $5.02\times 10^{3}$ \\\hline
	\end{tabular}
	\\[10pt]
	\begin{tabular}{|c|c|c|c|cc|cc|c|c|}
		\hline
		\bf Decay constant &
		\bf $f_{K_L}^{\bar sd}=f_{K_L}^{\bar ds}$ &
		\bf $f_{K_S}^{\bar sd}=-f_{K_S}^{\bar ds}$ &
		\bf $f_{\pi^0}^{\bar uu}=-f_{\pi^0}^{\bar dd}$ &
		\bf $f_\eta^{\bar uu}=f_\eta^{\bar dd}$ &
		\bf ~~~$f_\eta^{\bar ss}$~~ &
		\bf $f_{\eta'}^{\bar uu}=f_{\eta'}^{\bar dd}$ &
		\bf ~~~$f_{\eta'}^{\bar ss}$~~ &
		\bf ~~~$f_{D^0}^{\bar uc}$~~ \\\hline
		\bf Value [MeV]&
		$109.9$ & 
		$109.9$ & 
		$130.2$ & 
		$100.3$ & 
		$-152.3$ & 
		$72.9$ & 
		$195.3$ & 
		$212.0$ \\\hline
	\end{tabular}
    \caption{\justifying Lifetimes and decay constants used to implement the observables not included in \textsf{Flavio}.}
    \label{tab:decay constants}
\end{table}

To determine the bounds on the WCs in our analysis, we use a numerical minimiser. For each WC, the minimiser searches for the value for which the absolute difference between the prediction and the experimental limits is minimised ($\left|\BR_{\rm pred}-\BR_{\rm exp}\right|\to0$). This effectively identifies the largest value of the WC consistent with experimental data. Similarly, for the UV scale $\Lambda$, the minimiser finds the smallest possible value of $\Lambda$ for which the predicted observable still satisfies the experimental constraint.

\section{Results and discussion}
\label{sec:results}

In order to obtain constraints on the LFV $2q2\ell$ operators, we first identify the SMEFT operators $(\cO_{\alpha\beta ij})$, which are responsible for decays under consideration in this analysis, with exact flavour indices. Since we are considering charged lepton flavour-violating processes, it is necessary to have operators with lepton FCNC ($\alpha\ne\beta$).
However, the quark current in the operator can either be FCNC ($i\ne j$, \eg~for LFVBDs) or flavour conserving ($i=j$, \eg~for LFVQDs) as shown in Tables-\ref{Tab:Operator_different-quark_indices} and \ref{Tab:Operator_same-quark_indices} respectively. Therefore, we plan to classify these two different kinds of coefficients separately and quote bounds on them coming from different experiments. Such classification also provides insight into specific operators that contribute dominantly to particular processes. For the convenience of mentioning different SMEFT vector WC classes repeatedly, we categorise them based on the following features:
\begin{itemize}
\item Based on the constituent quark current chirality:
\begin{align}
    &\text{WCs with left chiral quark (doublet) current:}&\mathcal C_L&\in\{\mathcal C_{\ell q}^{(1)},\mathcal C_{\ell q}^{(3)},\mathcal C_{qe}\},\nonumber\\
    &\text{WCs with right chiral quark (singlet) current:}&\mathcal C_R&\in\{\mathcal C_{\ell u},\mathcal C_{\ell d},\mathcal C_{eu},\mathcal C_{ed}\};
    \label{eq:WCSetCh}
\end{align}

\item Based on the type of constituent quarks:
\begin{align}
    &\text{WCs with up-type quark neutral current:}&\mathcal C_u&\in\{\mathcal C_{\ell q}^{(1)},\mathcal C_{\ell q}^{(3)},\mathcal C_{qe},\mathcal C_{\ell u},\mathcal C_{eu}\},\nonumber\\
    &\text{WCs with down-type quark neutral current:}&\mathcal C_d&\in\{\mathcal C_{\ell q}^{(1)},\mathcal C_{\ell q}^{(3)},\mathcal C_{qe},\mathcal C_{\ell d},\mathcal C_{ed}\},
    \label{eq:WCSetQ}
\end{align}
here, since the WCs involving left-chiral doublet currents ($\mathcal C_{\ell q}^{(1,3)},\mathcal C_{qe}$) include both $u$ and $d$, they are included in both sets.
\end{itemize}


\subsection{1-D Analysis}
\label{subsec:1d analysis}

In the analysis of one-operator-at-a-time, we can extract two types of information which are complementary to each other. First, we can get constraints for different operators from different LFV processes. Second, assuming a perturbative SMEFT coefficient $(|\mathcal C(\Lambda)|\leq 1)$ we can probe the highest NP scale for each LFV observable. In the following, we discuss our results separately for $\mu$ decay and $\tau$ decay channels. 


\subsubsection{$\mu$--$e$ sector}

The availability of multiple LFV decay channels in the $\mu$--$e$ sector provides a diverse set of complementary bounds on the Wilson coefficients considered in this analysis. This is evident from the fact that WCs can have a relatively weaker constraint $\sim 10^{2}$ (\eg~see $\eta$ LFV decay for $(\lQ1)_{1212}$), or it can have a very strong constraint $\sim 10^{-6}$ for some others (\eg~see CR$(\mu\to e)$ in Gold nuclei for $(\lQ1)_{1211}$). Naturally, in the presence of a single operator at the high-energy scale, the constraints on the WCs would be proportional to the precision of the experimental limit. 

\begin{table}[h!]
\centering
\renewcommand{\arraystretch}{1.7}
\begin{adjustbox}{width=1\textwidth}
\small
\begin{tabular}{|c|rl|c|c|c|c|}
\hline
\multirow{2}{*}{Observables} & \multicolumn{2}{c|}{\multirow{2}{*}{Quark-level Process}} &  \multicolumn{2}{c|}{Vector} &  Scalar & Tensor  \\ 
\cline{4-7}
& \multicolumn{2}{c|}{} & $\cO_{\ell q},\cO_{\ell d},\cO_{ed},\cO_{\ell u},\cO_{eu}$  & $\cO_{qe}$ & $\cO_{\ell equ}^{(1)}$, $\cO_{\ell edq}$ & $\cO_{\ell equ}^{(3)}$ \\
\hline\hline
BR$(K_L^0 \to e^{\pm} \mu^{\mp}$)   &  $\frac{d\bar{s}+s\bar{d}}{\sqrt{2}}\to $ &  $e^{\pm}\mu^{\mp}$ & $[\cO]_{1212},[\cO]_{1221}$ & $[\cO]_{1212},[\cO]_{1221}$ &  $[\cO]_{1212},[\cO]_{1221},[\cO]_{2112},[\cO]_{2121}$ & -   \\
BR$(K_L^0 \to \pi^0 e^\pm \mu^\mp)$ &  $\frac{d\bar{s}+s\bar{d}}{\sqrt{2}}\to$ &  $ {d\bar{d}} e^\pm\mu^\mp$ & $[\cO]_{1212},[\cO]_{1221}$ & $[\cO]_{1212},[\cO]_{1221}$ & $[\cO]_{1212},[\cO]_{1221},[\cO]_{2112},[\cO]_{2121}$ & - \\ 
BR$(K^+\to \pi^+ \mu^- e^+)$  & $u\bar{s}\to$ &  $ u\bar{d}e^+\mu^-$ & $[\cO]_{1212}$ & $[\cO]_{1212}$ &  $[\cO]_{1212},[\cO]_{2121}$ & -  \\
BR$(K^+\to \pi^+ \mu^+ e^-)$  & $u\bar{s}\to $ &  $u\bar{d}e^-\mu^+$ & $[\cO]_{1221}$ & $[\cO]_{1221}$  & $[\cO]_{1221},[\cO]_{2112}$ & -  \\
\hline
BR$(D^0 \to e^{\pm} \mu^{\mp}$) & $c\bar{u}\to $ &  $e^{\pm}\mu^{\mp}$ & $[\cO]_{1212},[\cO]_{1221}$ & $[\cO]_{1212},[\cO]_{1221}$ & $[\cO]_{1212},[\cO]_{1221},[\cO]_{2112},[\cO]_{2121}$  & $[\cO]_{1212},[\cO]_{1221},[\cO]_{2112},[\cO]_{2121}$ \\
BR$\left(D^+(D_s^+) \to \pi^+(K^+) e^+ \mu^- \right)$ &  $c\bar{d}(\bar s)\to$ &  $ u\bar{d}(\bar s) e^+ \mu^-$ & $[\cO]_{1212}$ & $[\cO]_{1212}$ & $[\cO]_{1212},[\cO]_{2121}$  & $[\cO]_{1212},[\cO]_{2121}$  \\ 
BR$(D^+(D_s^+) \to \pi^+(K^+) e^- \mu^+ )$ &  $c\bar{d}(\bar s)\to $ &  $u\bar{d}(\bar s)e^- \mu^+$ & $[\cO]_{1221}$ & $[\cO]_{1221}$ & $[\cO]_{1221},[\cO]_{2112}$  & $[\cO]_{1221},[\cO]_{2112}$  \\
\hline
BR$(B^{0(+)}\to K^{*0(+)}e^{+}\mu^{-})$ &  $d(u)\bar{b}\to $ &  $d(u)\bar{s}e^{+}\mu^{-}$ & $[\cO]_{1223}$ & $[\cO]_{2312}$  & $[\cO]_{1223}$,$[\cO]_{2132}$ & -  \\
BR$(B^{0(+)}\to K^{*0(+)}e^{-}\mu^{+})$ &  $d(u)\bar{b}\to $ &  $d(u)\bar{s}e^{-}\mu^{+}$ & $[\cO]_{1232}$ & $[\cO]_{2321}$  & $[\cO]_{1232}$,$[\cO]_{2123}$ & -  \\
BR$(B^{0(+)}\to \pi^{0(+)}e^{\pm}\mu^{\mp})$ & $d(u)\bar{b}\to $ &  $d(u)\bar d\mu^{\pm}e^{\mp}$ & $[\cO]_{1213},[\cO]_{1231}$ & $[\cO]_{1312},[\cO]_{1321}$ &  $[\cO]_{1213},[\cO]_{1231},[\cO]_{2113},[\cO]_{2131}$ & -  \\
BR$(B^0\to \mu^{\mp}e^{\pm})$ & $d\bar{b}\to$ &  $\mu^{\pm}e^{\mp}$ & $[\cO]_{1213},[\cO]_{1231}$ & $[\cO]_{1312},[\cO]_{1321}$  & $[\cO]_{1213},[\cO]_{1231},[\cO]_{2113},[\cO]_{2131}$ & -  \\
BR$(B_s^0\to \mu^{\mp}e^{\pm})$ & $s\bar{b}\to$ &  $\mu^{\pm}e^{\mp}$ & $[\cO]_{1223},[\cO]_{1232}$ & $[\cO]_{2312},[\cO]_{2321}$  & $[\cO]_{1223},[\cO]_{1232},[\cO]_{2123},[\cO]_{2132}$ & -  \\
BR$(B_s^0\to \phi\mu^{\pm}e^{\mp})$ & $s\bar{b}\to $ &  $s\bar{s}\mu^{\pm}e^{\mp}$ & $[\cO]_{1223},[\cO]_{1232}$ & $[\cO]_{2312},[\cO]_{2321}$  & $[\cO]_{1223},[\cO]_{1232},[\cO]_{2123},[\cO]_{2132}$ & -  \\
\hline
\end{tabular}
\end{adjustbox}
\caption{\justifying Operators with different lepton and quark indices under discussion. The second column represents the quark and lepton composition of the process. All vector operators under consideration have the same order of lepton and quark indices except $\cO_{qe}$. Columns 5 and 6 show scalar and tensor operators, respectively.}
\label{Tab:Operator_different-quark_indices}
\end{table}
\begin{table}[h!]
\renewcommand{\arraystretch}{1.7}
\begin{adjustbox}{width=1\textwidth}
		\begin{tabular}{|c|rl|c|c|c|c|c|}
			\hline
			\multirow{2}{*}{Observables} & \multicolumn{2}{c|}{\multirow{2}{*}{Quark-level Process}} &  \multicolumn{2}{c|}{Vector} &  Scalar & Tensor  \\ 
			\cline{4-7}
			&\multicolumn{2}{c|}{} & $\cO_{\ell q},\cO_{\ell d},\cO_{ed},\cO_{\ell u},\cO_{eu}$  & $\cO_{qe}$ & $\cO_{\ell equ}^{(1)}$, $\cO_{\ell edq}$ & $\cO_{\ell equ}^{(3)}$  \\
			\hline\hline
			CR$(\mu\to e, N)$ & $\mu~(u/d/s)\to$&$ e~(u/d/s)$ & $[\cO]_{1211}$ & $[\cO]_{1112}$ & $[\cO]_{1211},[\cO]_{2111}$ & $[\cO]_{1211},[\cO]_{2111}$ \\ 
			\hline
			BR$(J/\Psi \to e^{\pm}\mu^{\mp})$ & $c\bar{c}\to$&$ e^{\pm}\mu^{\mp}$ & $[\cO]_{1222}$ & $[\cO]_{2212}$ &  - & $[\cO]_{1222},[\cO]_{2122}$\\
			\hline
			BR$(\Upsilon \to e^{\mp}\mu^{\pm})$ &  $b\bar{b}\to$&$ e^{\pm}\mu^{\mp}$ & $[\cO]_{1233}$ & $[\cO]_{3312}$  &- & -\\
			\hline
			BR$(\pi^0\to e^{\pm}\mu^{\mp})$ & $\frac{u\bar{u}-d\bar{d}}{\sqrt{2}}\to $&$e^{\pm}\mu^{\mp}$ & $[\cO]_{1211}$ & $[\cO]_{2111}$ & $[\cO]_{1211},[\cO]_{2111}$  & $[\cO]_{1211},[\cO]_{2111}$  \\ \hline
			BR$(\eta^0\to e\mu)$ &  $\frac{u\bar{u}+d\bar{d}-2s\bar{s}}{\sqrt{2}}\to $&$e^{\pm}\mu^{\mp}$ & $[\cO]_{1211}, [\cO]_{1222}$ & $[\cO]_{1112}, [\cO]_{2212}$ & $[\cO]_{1211}, [\cO]_{2111}, [\cO]_{1222}, [\cO]_{2122}$ & $[\cO]_{1211}, [\cO]_{2111}, [\cO]_{1222}, [\cO]_{2122}$\\
			BR$(\eta^{0\prime}\to e\mu)$ &  $\frac{u\bar{u}+d\bar{d}+s\bar{s}}{\sqrt{2}}\to$&$ e^{\pm}\mu^{\mp}$ & $[\cO]_{1211}, [\cO]_{1222}$ & $[\cO]_{1112}, [\cO]_{2212}$ & $[\cO]_{1211}, [\cO]_{2111}, [\cO]_{1222}, [\cO]_{2122}$ & $[\cO]_{1211}, [\cO]_{2111}, [\cO]_{1222}, [\cO]_{2122}$\\ \hline
		\end{tabular}
\end{adjustbox}
\caption{\justifying Operators with different lepton indices but same quark indices. Similar to Table-\ref{Tab:Operator_different-quark_indices}, the last three columns represent vector, scalar and tensor operators respectively.}
\label{Tab:Operator_same-quark_indices}
\end{table}

\begin{figure}[h!]
    \centering
  \includegraphics[width=\linewidth]{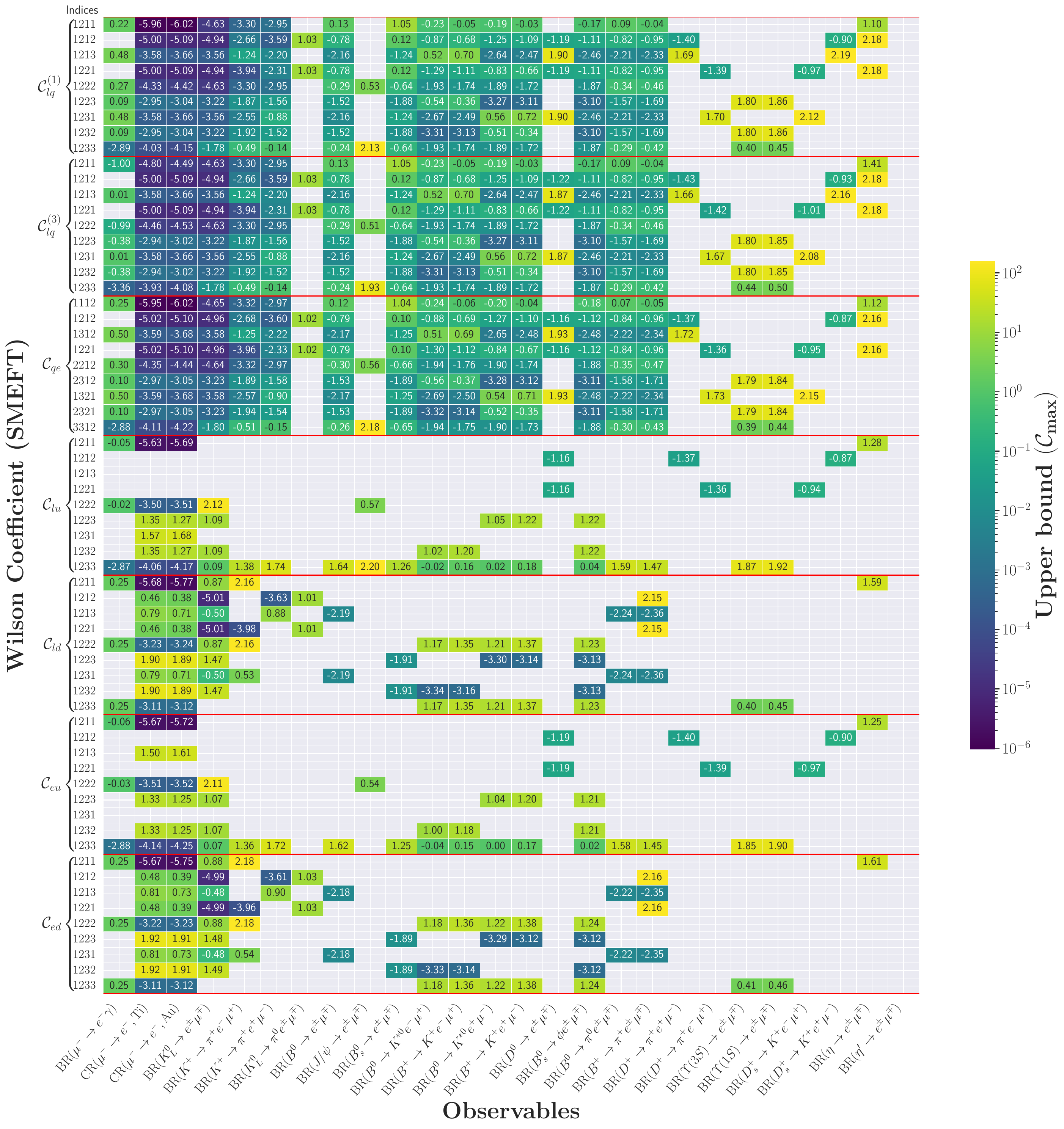}
\caption{\justifying Upper bounds on the $2q2\ell$ Wilson coefficient ($\mathcal C_{\rm max}$) involving first- and second-generation charged leptons ($e^\pm$, $\mu^\pm$), considering one at a time at scale $\Lambda=1$\,TeV. Each box shows the constraint from a specific observable, colour-coded accordingly. The numerical value of the limit is given as $\mathcal C_{\rm max} = 10^{x}$, where $x$ is the value displayed in the box. The blank spaces indicate that the observable does not constrain the corresponding WC.}
\label{Fig:emuplot}
\end{figure}

Along its vertical axis, Fig.\,\ref{Fig:emuplot} represents the limits on WCs of all vector LFV $2q2\ell$ operators with indices given in both Tables-\ref{Tab:Operator_different-quark_indices} and \ref{Tab:Operator_same-quark_indices}. In many cases, instead of a combined value, the experimental data for LFV decay channels with $\ell_1^-\ell_2^+$ and $\ell_1^+\ell_2^-$ are reported separately. Therefore, we include both the combinations in our analysis and present the corresponding results separately. 
If we look from bottom-up, for a particular observable, the figure gives us a complete picture of the relative strengths of all vector operators with different chiralities. The horizontal axis shows the constraints on a particular WC coming from different processes.


All the observables under consideration are arranged in ascending order of their experimental limits (\ie, in descending order of stringency), starting from $\CRmue(\sim 10^{-13})$, followed by $K_L^0\to e^{\pm}\mu^{\mp}(\sim 10^{-12})$ and so on. As explained in the caption of Fig.\,\ref{Fig:emuplot}, the numbers shown represent the exponent 
$x$ in the relation $\mathcal C_{\rm max}=10^x$, providing the maximum allowed value of the WC that satisfies the current experimental limit. The most stringent constraint on a given WC corresponds to the smallest (most negative) value of $x$.  The boxes are also colour coded, with lighter shades starting from yellow, representing weaker constraints, whereas the darker boxes refer to a stronger constraint, as shown in the legend of the figure. A missing entry corresponding to a particular WC-observable pair indicates that the operator is not relevant for that process and thus are not constrained by the corresponding observable. In the following, we shall explain the limits in the figure systematically.

One can readily see that almost all the WCs of vector operators consisting of left-handed quark doublet current 
($\mathcal C_L$'s) are constrained by most of the observables involving interaction of $d$-type quarks, despite the difference in the quark-flavour indices relevant to the process and that of the operator (see Tables-\ref{Tab:Operator_different-quark_indices} and \ref{Tab:Operator_same-quark_indices} for details). Here we note that according to our convention, the running mass matrix of $u_L$-type quarks is diagonal at $\Lambda$ (\textit{Warsaw up} basis in \textsf{Wilson}), due to which WCs of operators with $d_L$-type quarks involve the CKM rotation as,
\begin{align}
v^2\,\mathcal L_{\rm LEFT}\supset (C^{V,XL}_{ed})_{\alpha\beta ij}
    (O^{V,XL}_{ed})_{\alpha\beta ij}
    &\equiv(C^{V,XL}_{ed})_{\alpha\beta ij}\Big(\bar e_\alpha\gamma^\mu P_X e_\beta\Big)\left(\bar d_i\gamma_\mu P_L d_j\right)\nonumber\\
    &=(C^{V,XL}_{ed})_{\alpha\beta ij}\Big(\bar e_\alpha\gamma^\mu P_X e_\beta\Big)\left((\bar{\tilde  d}_kV^\ast_{ki})\gamma_\mu P_L (V_{jl}\tilde d_l)\right)\nonumber\\
    &=\left[V^\ast_{ki} (C^{V,XL}_{ed})_{\alpha\beta ij}V_{jl}\right] \Big(\bar e_\alpha\gamma ^\mu P_X e_\beta\Big)\left(\bar{\tilde d}_k\gamma_\mu  P_L\tilde d_l\right)\nonumber\\
    &= (\tilde C^{V,XL}_{ed})_{\alpha\beta kl}\Big(\bar e_\alpha\gamma^\mu P_X e_\beta\Big)\left(\bar{\tilde d}_k\gamma_\mu  P_L\tilde d_l\right)\nonumber\\
    \Rightarrow\quad (\tilde C^{V,XL}_{ed})_{\alpha\beta kl}&=V^\ast_{ki}(C^{V,XL}_{ed})_{\alpha\beta ij}V_{jl}\,,
    \label{eq:CKMmixingV}
\end{align}
where $X\in\{L,R\}$ is the chirality of lepton vector FCNC. Summations over repeated indices are implicit following Einstein's convention. $\tilde C$ and $C$ represent the WCs of the LEFT operator involving quarks in mass eigenstates ($\tilde O$) and flavour eigenstates ($O$), respectively, and $V$ represents the CKM matrix. Similarly,
\begin{align}
    (\tilde C_{ed}^{Z,XL})_{\alpha\beta ij}&=( C_{ed}^{Z,XL})_{\alpha\beta ik}V_{kj}\,,\nonumber\\
    \quad(\tilde C_{ed}^{Z,XR})_{\alpha\beta ij}&=V_{ik}^\ast( C_{ed}^{Z,XR})_{\alpha\beta kj}\,,
    \label{eq:CKMmixingS}
\end{align}
are for scalar ($Z=S$) and tensor ($Z=T$) WCs.
Thus for operators involving left-handed $d$-type quark(s), a mass-eigenstate WC receives contributions from all the flavour-eigenstate WCs with different flavour indices, weighted by the corresponding CKM matrix element(s). Thus, depending on these elements, the observable becomes sensitive to WCs with all the flavour indices. Thus, one finds a constraint for the WC.

On the other hand, observables with $u$ and $c$ quarks, namely decays of $J/\psi$ and $D$-mesons requiring $u$-type LEFT operators, do not involve CKM. 
These observables provide constraints only on the relevant WCs mentioned in Tables-\ref{Tab:Operator_different-quark_indices} and \ref{Tab:Operator_same-quark_indices}. 
Similarly, SMEFT WCs of the operators with right-handed quark current ($\mathcal C_R$'s) also do not undergo the CKM rotation and are thus typically not constrained by the observables with a different quark current.

Another feature of the figure is that the processes considered with a combined limit on $e^\pm\mu^\mp$ channels (e.g., $B^0 \to \pi^0 e^\pm \mu^\mp$) provide limits on WCs with both flavour index combinations $\alpha\beta ij$ and $\alpha\beta ji$ while the ones with separate limits constrain each index combination independently. This is because the operator with indices $(\beta\alpha ij)$, where lepton FCNC flow is opposite, is related to the one with indices ($\alpha\beta ji)$ by a hermitian conjugation, 
\begin{align}
    \left(\mathcal{C}_{\alpha\beta ji}(\bar e_\alpha\Gamma e_\beta)(\bar q_j\Gamma q_i)\right)^\dagger&=\mathcal{C}_{\alpha\beta ji}^\ast(\bar e_\beta\Gamma e_\alpha)(\bar q_i\Gamma q_j)\,.
\end{align}
Both of these operators appear in the Lagrangian to keep it real.
Thus, while one of the indices combinations ($\alpha\beta ij)$ and ($\alpha\beta ji$) is relevant to the process with $e_\alpha^+ e_\beta^-$, the other contributes to $e_\alpha^- e_\beta^+$. Hence, a combined experimental limit constrains both WCs simultaneously. Although decays of the $K_L^0$ meson are considered with combined limits, they would exhibit similar behaviour due to its quark composition ($\sim d\bar s + s\bar d$) even if the limits were not combined.

The classification of operators with indices according to the fermion composition of the decay process, as presented in Tables-\ref{Tab:Operator_different-quark_indices} and \ref{Tab:Operator_same-quark_indices}, is particularly useful for interpreting the numerical results shown in Fig.\,\ref{Fig:emuplot}. For example, it is evident from Table-\ref{Tab:Operator_same-quark_indices} that for the vector WCs $(\mathcal C_{u,d})_{1211}$ (see Table-\ref{tab:Dim-6 2q2l operators1} for details)\footnote{In the case of $\mathcal{C}_{qe}$, the quark flavor indices precede the lepton ones, in contrast to the convention used for other classes. To maintain clarity, we do not explicitly write the indices. Nonetheless, for this class, one should interpret the indices replacement as $\alpha\beta ij\to ij\alpha\beta$.}, stringent constraints might come from the limit on $\CRmue$ and indeed, it provides a strong bound $\sim 10^{-6}$. The observable is equally sensitive to both left- and right-chiral WCs (see Eqs.(\ref{eq:CRmue BR}-\ref{eq:gfXV})), and receives contributions from both vector WCs involving $u$ and $d$ quarks with equal strengths (see Table-\ref{tab:SVG}). 
This explains why all of them receive comparable constraints from this process. The WC $(\mathcal C_{\ell q}^{(3)})_{1211}$, however, is relatively less constrained due to the opposite contributions of the WC to LEFT WCs, $(C_{eu}^{V,LL})_{1211}$ and $(C_{ed}^{V,LL})_{1211}$, during the matching (for details, see Appendix \ref{appndx:LEFT SMEFT matching}). 
As a result, the proton and neutron contributions to the total amplitude interfere destructively. 

Turning our attention to the quark-flavour-diagonal (QFD) WCs $(\mathcal C_{u,d})_{1222}$ and $(\mathcal C_{u,d})_{1233}$, again, we see that they also receive the strongest constraints, $\sim 10^{-4}$ and $\sim 10^{-3}$ respectively, from the $\CRmue$ experiment.
Although WCs with $u$-type quarks, $(\mathcal C_u)_{1222}$, are directly relevant to the decay $J/\psi\to e^\pm\mu^\mp$, the branching ratio of this LFV process gets suppressed due to its small lifetime ($\sim10^{-20}$\,sec), and thus provides a very weak constraint. 
Also, as it involves $u$-type quark current, it does not constrain WCs with other indices as explained earlier.
For $\Upsilon$ LFV decay, in addition to its small lifetime  ($\sim10^{-20}$\,sec), weak constraint on the relevant WC ($(\mathcal C_d)_{1233}$) is also caused by a weak experimental limit. On the other hand, the sensitivity of $\CRmue$ on these WCs arises from two main effects: the RGEs of the SMEFT WCs and the SMEFT-to-LEFT matching.
Unlike off-diagonal WCs, QFD WCs mix significantly with each other due to RGEs. As a result, even if a specific QFD WC is assumed to be absent initially at the high scale $\Lambda$, it can be radiatively generated through operator mixing while running down to the electroweak scale (for a detailed discussion, see Refs.~\cite{Alonso:2013hga,Alonso:2014csa}). An additional enhancement occurs during the matching, when the $Z$-boson is integrated out (see Appendix \ref{appndx:LEFT SMEFT matching}). The sensitivity of CR$(\mu\to e)$ as well as BR$(\mu\to e\gamma)$ (see Appendix-\ref{App:muegamma}) towards the operators involving top-quark $(\mathcal C_u)_{1233}$ arises due to the induction of leptonic dipole operator $(C_{e\gamma})_{12}$ through $(\mathcal C_{\ell edq})_{1233,2133}$ as explained below. 

Similarly, from Table-\ref{Tab:Operator_different-quark_indices} we find that the SMEFT WCs $(\mathcal C_d)_{1212,1221}$ and $(\mathcal C_u)_{1212,1221}$ are relevant to the LFVKDs and LFVDDs respectively, and therefore they should receive the most stringent constraints, specifically from these decays. We find this to be true for WCs with right-handed $d$-type quark current operators ($(\mathcal C_R)_{1212,1221}$). However, for the left-handed WCs $((\mathcal C_L)_{1212,1221})$, a strong experimental limit helps CR$(\mu\to e)$ to provide constraints little stronger than $\KLemu$ through the CKM mixing as explained earlier. LFVDDs, on the other hand, are typically less constraining due to their comparatively weaker experimental limits.

For the coefficients ${(\mathcal C_d)}_{1213}$ and ${(\mathcal C_d)}_{1231}$, Table-\ref{Tab:Operator_different-quark_indices} indicates that limits should come from $B^0\to e^{\pm}\mu^{\mp}$ and $B^+(B^0)\to \pi^+(\pi^0)e^{\pm}\mu^{\mp}$ processes. While this is the case for WCs from classes involving $d_R$-type quarks ($\ed,\ld$) with these indices, the results of Fig.\,\ref{Fig:emuplot} show that the strongest limit of $10^{-3}$ on WCs involving left-handed quarks ($(\mathcal{C}_L)_{1213,1231}$) again derives from $\CRmue$, due to reasons explained earlier. 
Also, as evident from Table-\ref{Tab:Operator_different-quark_indices}, the coefficient $(\mathcal C_d)_{1223}$  and $(\mathcal C_d)_{1232}$ are mostly responsible for processes involving $s$-$b$ quark FCNC and thus several LFVBD processes are expected to put strong constraints on these coefficients. 
Therefore, it explains why the most stringent constraints of $\sim10^{-3}$ for these WCs come from the processes $B^0\to K^{*0}e^-\mu^+$ and $B^0\to K^{*0}e^+\mu^-$ respectively.

\begin{figure}[h!]
    \centering
 \includegraphics[width=\linewidth]{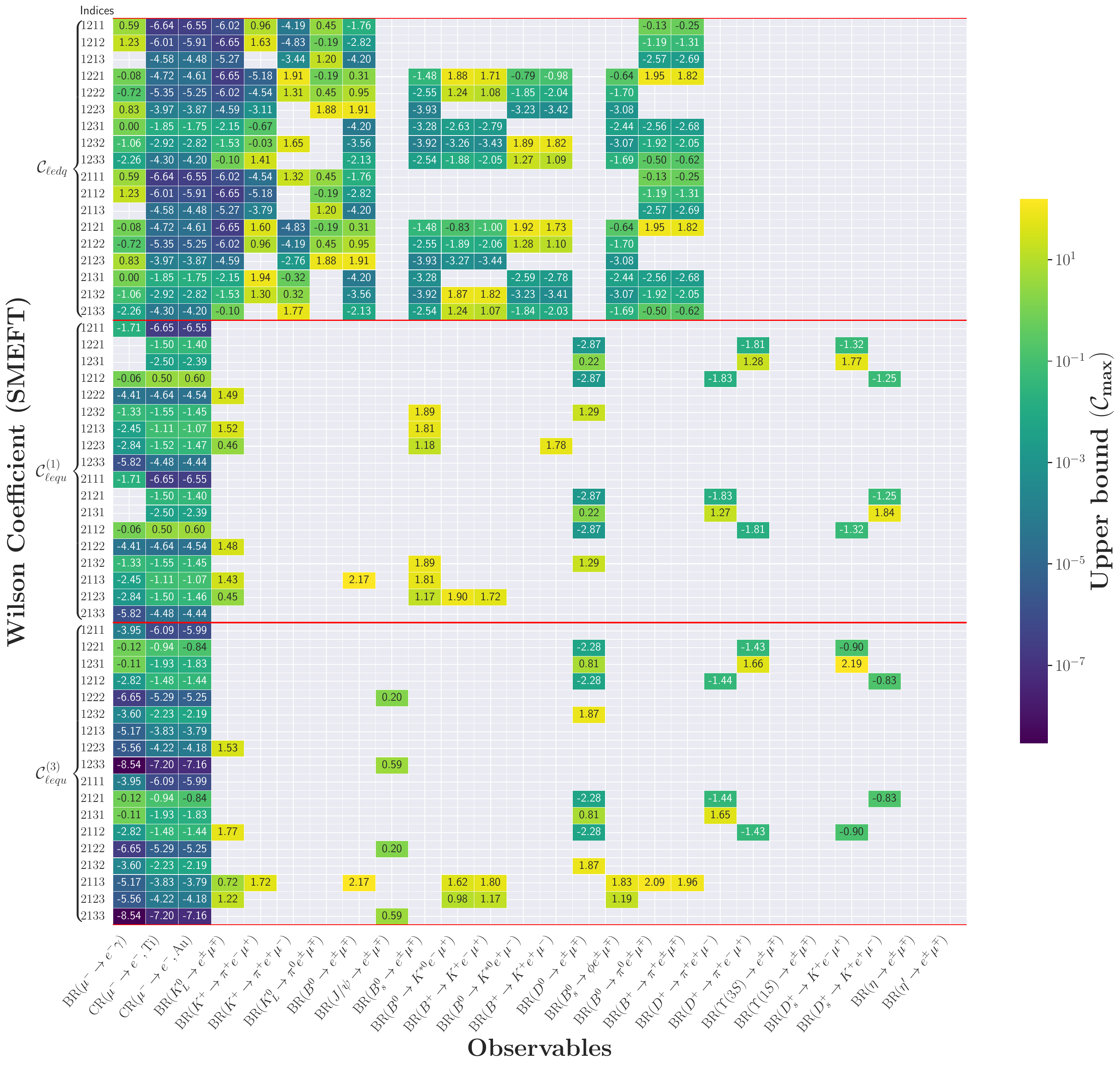}
\caption{Same as Fig.\,\ref{Fig:emuplot} but with scalar and tensor $2q2\ell$ operators.}
\label{Fig:emu scalar plot}
\end{figure}

Similar to Fig.\,\ref{Fig:emuplot}, Fig.\,\ref{Fig:emu scalar plot} represents the upper limits on WCs of scalar and tensor operators. 
Again Tables-\ref{Tab:Operator_different-quark_indices} and \ref{Tab:Operator_same-quark_indices} guide us in understanding the results of this figure. 
As before, $\CRmue$ proved to be the most sensitive process for the SMEFT WC $(\mathcal C_j)_{1211}$  of both scalar and tensor operator classes ($\mathcal C_j\in\{\mathcal C_{\ell edq},\mathcal C_{\ell equ}^{(1)},C_{\ell equ}^{(3)}\}$), with a limit $\sim 10^{-6} \GeV$. 
Since most of the observables under consideration involve $d$-type quark current, the scalar operator of class $\ledq$ receives significant constraints from several of those processes involving such current. 
Additionally, the CKM mixing (see Eq.(\ref{eq:CKMmixingS})) also makes the observable sensitive to WCs with different indices.
Thus we find that the process $\KLemu$ provides the strongest constraints for several WCs of scalar class $(\mathcal C_{\ell edq})$, out of which, the relevent WCs $(\mathcal C_{\ell edq})_{1212,1221,2112,2121}$ are most constrained. Likewise $(\mathcal C_{\ell edq})_{1213,1231,2113,2131}$, and $(\mathcal C_{\ell edq})_{1223,1232,2123,2132}$ with $b$-$d$ and $b$-$s$ quark scalar FCNC, respectively, receive the strongest constraints from several LFVBD processes.
To be specific, the former set of WCs is constrained by $\BZmue$ $(\sim 10^{-4})$ whereas, the latter set is constrained by $B_s\to e^\pm\mu^\mp$ process. The QFD WCs $(\ledq)_{1233,2133}$ $(\sim 10^{-2})$, are again constrained through CKM mixing. The contribution of $\mathcal{C}_{\ell equ}^{(1),(3)}$ is usually quite negligible in these LFV processes because they involve $u$-type quark current and thus they receive very weak constraints, as is evident from the results of this table.

An interesting observation from Fig.\,\ref{Fig:emu scalar plot} is the strong constraint $(\sim 10^{-7})$  on $(\lequ3)_{1233,2133}$ coming from BR$(\mu\to e\gamma)$ and  $\CRmue$. This is due to the contributions from the LEFT lepton-photon dipole operator ${C}_{e\gamma}$. The WC ${C}_{e\gamma}$ is related linearly to the leptonic dipole operators $\mathcal C_{eW}$ and $\mathcal C_{eB}$ (see Eq.(\ref{eq:cegammaMatching})), which are strongly dependent on these QFD tensor operators during RGE running in the following way,
\begin{align}
(\dot{\mathcal C}_{eW})_{\alpha\beta} &\sim -2g_2 (\mathcal C_{\ell equ}^{(3)})_{\alpha\beta ij} \left[Y_u\right]_{ji}+\dots \,,& (\dot{\mathcal C}_{eB})_{\alpha\beta} &\sim 4g_1 (\mathcal C_{\ell equ}^{(3)})_{\alpha\beta ij} \left[Y_u\right]_{ji}+\dots \,,
\label{eq:CeW RGE}
\end{align}
where, 
\begin{align*}
    \dot{\mathcal C_i}\equiv16\pi^2\mu\frac{{\rm d}\mathcal C_i}{{\rm d}\mu}\,,
\end{align*}
$g_i$'s are the gauge couplings and $Y_u$ represents the Yukawa matrix of $u$-type quark. Similarly, we also observe a limit $(\sim 10^{-4})$ for the same flavour indices but for the scalar operator of class $\cO_{\ell equ}^{(1)}$. The origin of this constraint can be traced to the Gauge RGEs~\cite{Alonso:2013hga} of these operators instead, where they are related to each other as well as with the dipole operator $\mathcal C_{eW}$, as can be seen from the equation below
\begin{align}
(\dot{\mathcal C}_{\ell equ}^{(3)})_{\alpha\beta ij} = \frac{3}{8}g_2^2({\mathcal C}_{\ell equ}^{(1)})_{\alpha\beta ij} + \left(N_c-\frac{1}{N_c}\right)g_3^2 (\mathcal C_{\ell equ}^{(3)})_{\alpha\beta ij}-\frac{3}{2}g_2(\mathcal C_{eW})_{\alpha\beta}\big[Y_u^{\dagger}\big]_{ij}+\cdots \,,
\end{align}
where $N_c=3$ is the number of colours. Moreover, CR$(\mu\to e)$ provides constraints on scalar WCs $(\mathcal C_{\ell edq},\mathcal C_{\ell equ}^{(1)})$ diagonal in charm and bottom quark flavours, due to contribution of gluonic operators\footnote{Since the gluonic form factor of the Au/Ti nucleus is non-negligible, the dim-7 gluonic operators, $\ell_i\ell_jG^a_{\mu\nu}G_{a}^{\mu\nu}$, also contribute to CR$(\mu\to e)$ through heavy quark loops~\cite{Petrov:2013vka,Cirigliano:2009bz}. Although the matching and contributions of dim-7 operators are not included in {\sf Wilson-2.5.2} and {\sf Flavio-2.6.2}, we implement them using references~\cite{Petrov:2013vka, Plakias:2023esq,Cirigliano:2009bz}} that are induced through heavy quark $(c,b,t)$ loops~\cite{Shifman:1978bx}. In the case of WCs with top-quark, the contributions from LEFT leptonic dipole operators dominate. The off-diagonal WCs involving $b$-quark are also constrained, again due to CKM mixing.

\begin{figure}[h!]
    \centering
  \includegraphics[width=1\linewidth]{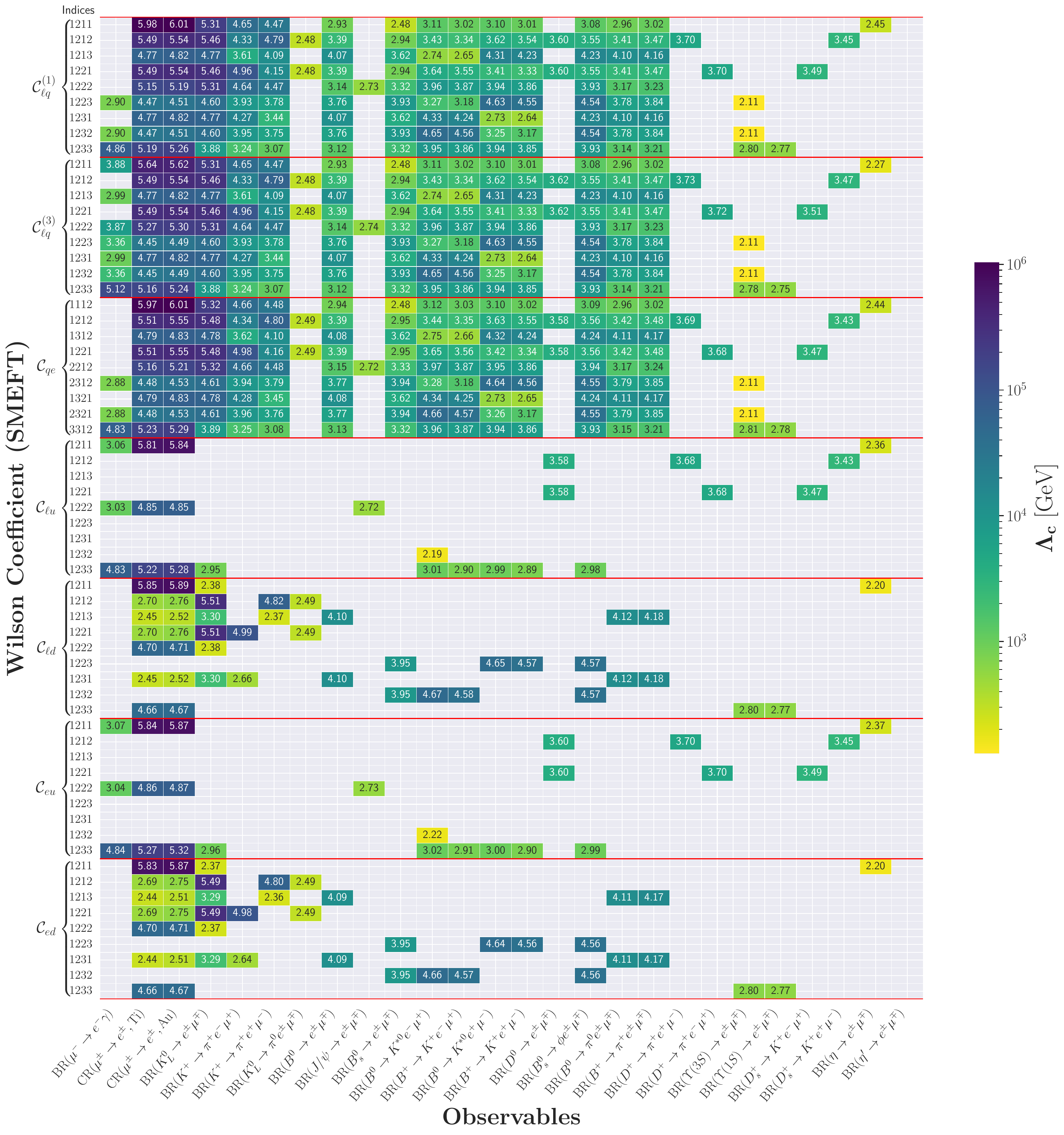}
\caption{\justifying Upper bounds on the UV scale for each $2q2\ell$ Operator ($\Lambda_{\rm c}$) involving first- and second-generation charged leptons ($e^\pm$, $\mu^\pm$), considering $\mathcal C=1$ . Each box shows the constraint from a specific observable, colour-coded accordingly. The numerical value of the limit is given as $\Lambda_{\rm c} = 10^{x}$, where $x$ is the value displayed in the box, thus representing a stringent constraint for larger numbers (darker box).}
    \label{Fig:emuLamplot}
\end{figure}

Fig.\,\ref{Fig:emuLamplot} shows the maximum values of the new physics scale $\Lambda_{\rm c}$ accessible by each of these operators in the $\mu$--$e$ sector for single operator analysis. Different colour codes represent different energy values, from $10^2 \GeV$ to $10^6 \GeV$, as shown in the legend, and the numbers in boxes give the log values.
To produce these numbers, we kept the value of the particular perturbative WC fixed at unity and varied $\Lambda$ to satisfy the experimental limit of each observable (minimum entry-level value of this figure was kept fixed at $125 \GeV$). Since an increasing value of $\Lambda$ is effectively equivalent to a decreasing WC, we also find a similar trend, as in the limits of WCs, existing among the numbers of these two figures. 
Clearly, $\CRmue$ is the most sensitive process for most of the diagonal elements, $(\mathcal{C}_{u,d})_{1211,1222,1233}$, ranging from $10^4-10^6 \GeV$ irrespective of the chirality, as explained earlier. 
This is followed by the process $\KLemu$, which is almost equally sensitive to $\mathcal C_L$'s. 
Therefore, it can be concluded that for a single-operator dominance, WCs of left-handed quark current operators with all the quark-flavour indices, except $(\mathcal C_L)_{1211,1233}$, can be probed competitively by $\CRmue$ and $\KLemu$ processes. For the right-handed quark current operators, situations are a little more distributed. For the QFD WCs, $\CRmue$ can probe the highest energy scale, whereas to probe $(\mathcal C_L)_{1212}$, the relevant experiment is $\KLemu$.  

\begin{figure}[h!]
    \centering
  \includegraphics[width=\linewidth]{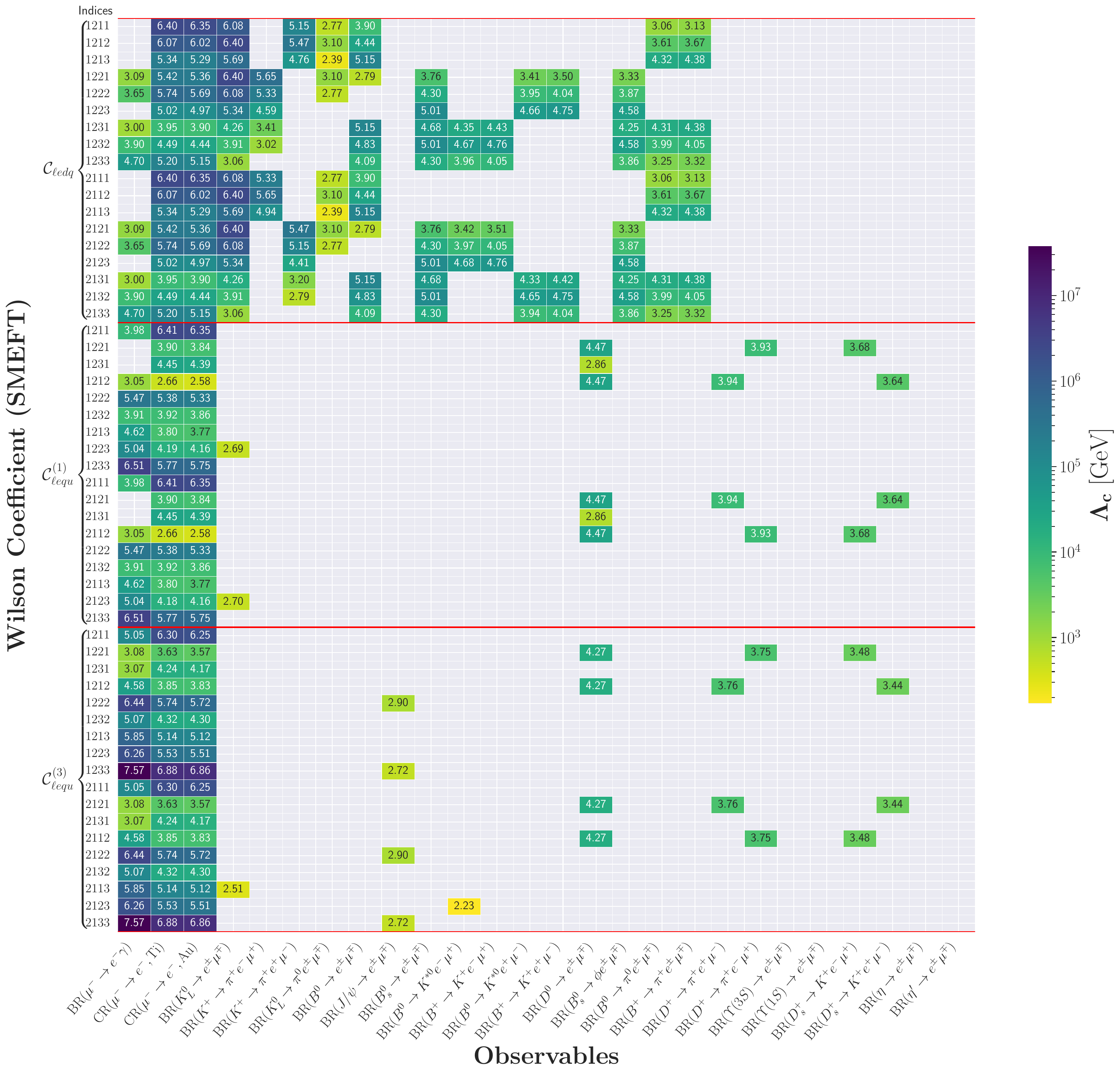}
\caption{Same as Fig.\,\ref{Fig:emuLamplot} but with scalar and tensor operators.}
\label{Fig:emu scalar Lamplot}
\end{figure}

On the other hand, several LFVBD processes probe the WCs ($(\mathcal C_d)_{1213,1223}$) at scale $\sim 10^3-10^5 \GeV$.
Also, LFVQDs are found to be least sensitive except $(\mathcal C_u)_{1222}$ for $J/\Psi$ and $(\mathcal C_d)_{1233}$ for $\Upsilon$. While LFVDD processes can probe WCs $(\mathcal{C}_u)_{1212}$ of operators with $u_R$-type quark-current at $\sim 10^3 \GeV$, their chiral counterparts are dominated by $\CRmue$. Therefore we can conclude that if any physics that might generate these operators with couplings $(\mathcal C/\Lambda^2)$ within the range $10^{-12}-10^{-8}\,\GeV^{-2}$, it is expected that $\CRmue$, $\KLemu$ or several LFVBDs can be observed in future experiments (see Table-\ref{Tab:LFVlimits}). 

The maximum energy values that can be probed by the scalar and tensor operators are shown in Fig.\,\ref{Fig:emu scalar Lamplot}. Looking at the range of $\Lambda_{\rm c}$ we can again conclude that for any new physics that might generate the scalar operator $\ledq$ with couplings $(\mathcal C/\Lambda^2)$ within the range $10^{-12}-10^{-4}$\,GeV$^{-2}$, it is expected that $(\mu\to e\gamma)$, $\CRmue$, LFVKDs or several LFVBDs can be observed in future experiments. On the other hand, for the scalar WC class $\leQu1$ or the tensor WC class $\lequ3$, the same range of coupling is relevant for $\CRmue$ and LFVDDs.

\begin{figure}[h!]
    \centering
    \includegraphics[width=\linewidth]{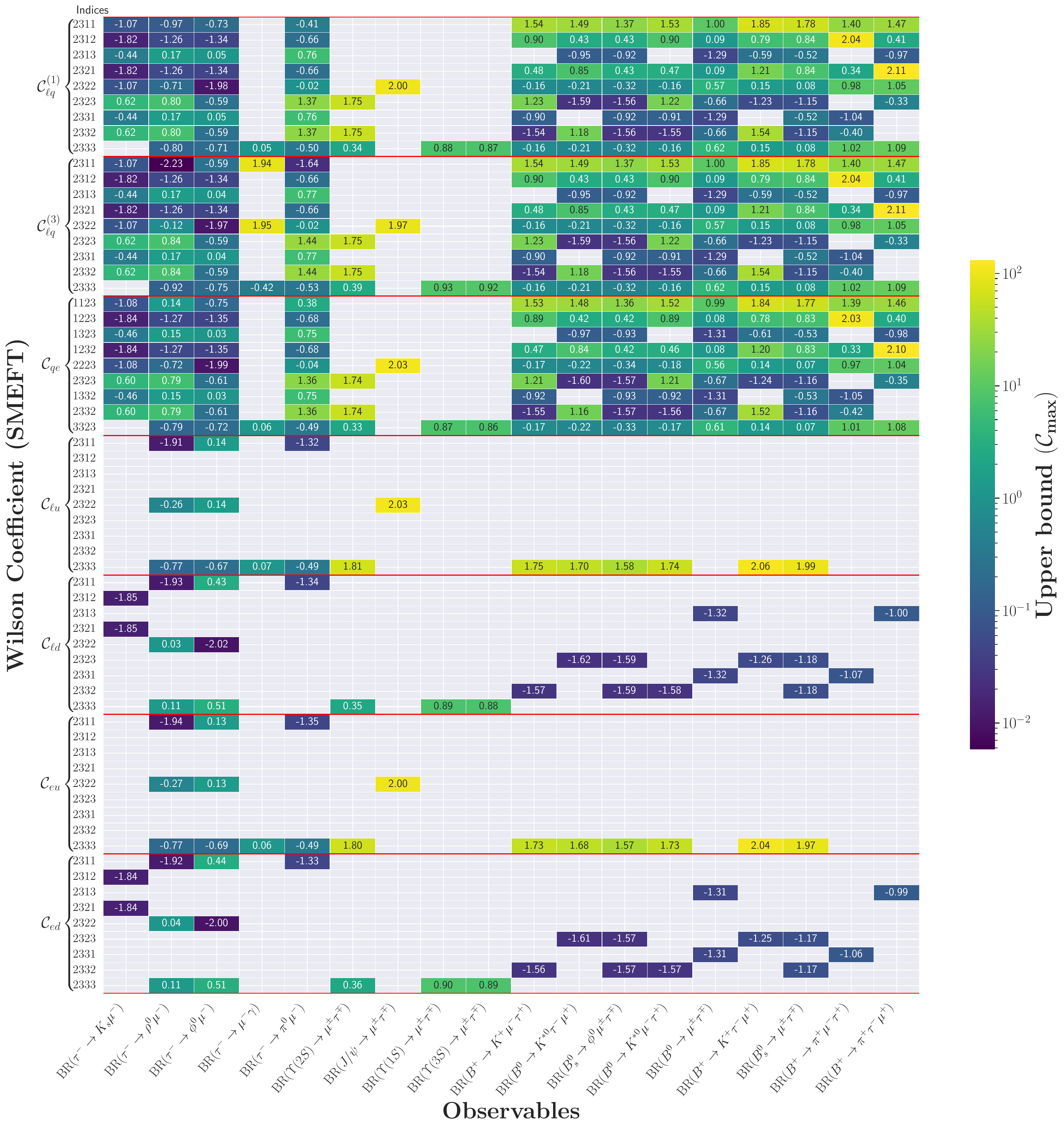}
    \caption{\justifying Same as Fig.\,\ref{Fig:emuplot} but with observables and operators with second- and third-generation charged leptons $(\mu^\pm,\tau^\pm)$.}
    \label{Fig:mutauplot}
\end{figure}

\subsubsection{$\tau$--$\ell(e,\mu)$ sector}

\begin{figure}[h!]
    \centering
  \includegraphics[width=\linewidth]{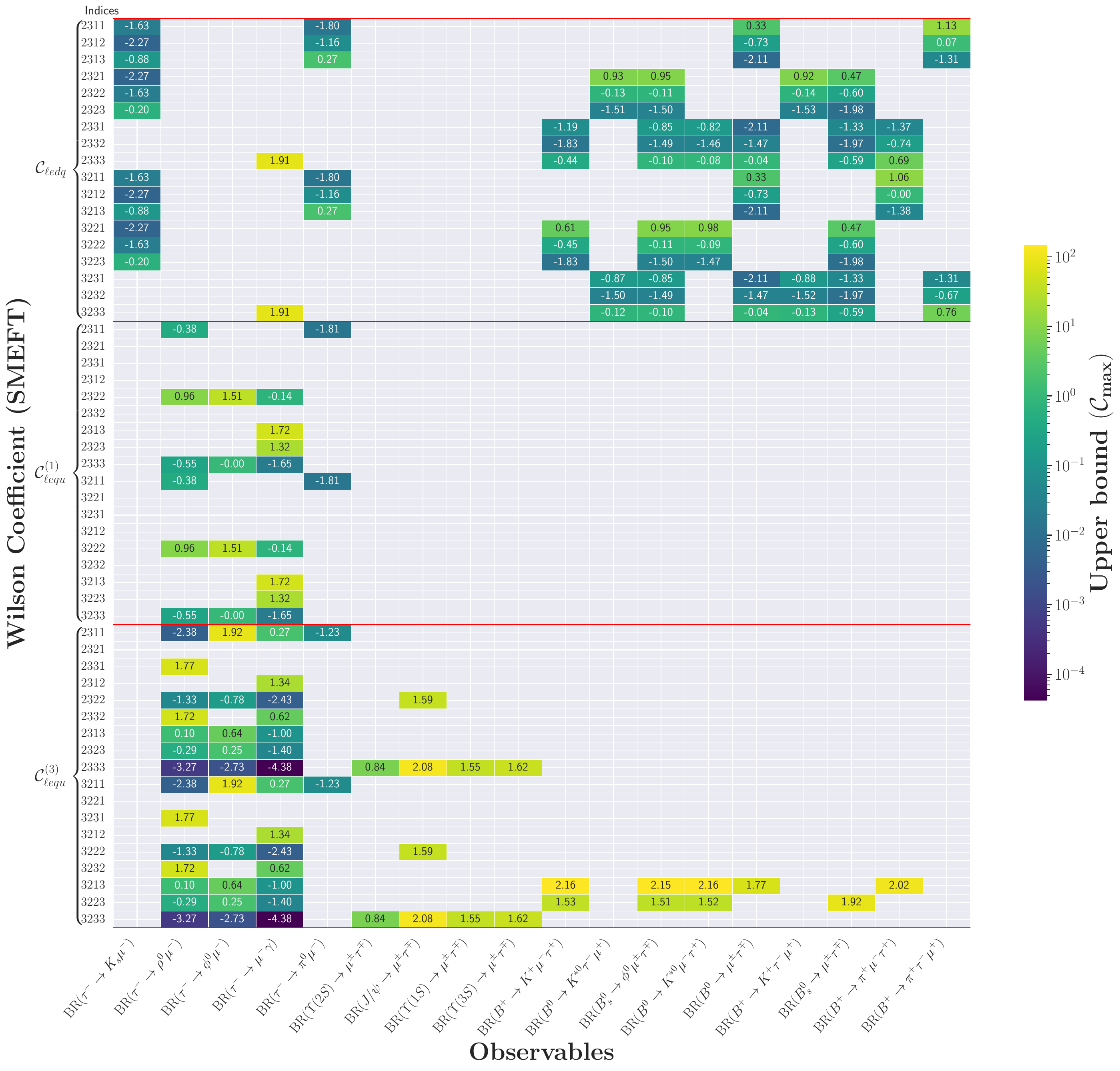}
\caption{Same as Fig.\,\ref{Fig:mutauplot} but with scalar and tensor $2q2\ell$ operators.}
\label{Fig:mutau_scalar_plot}
\end{figure}

\begin{figure}[h!]
    \centering
    \includegraphics[width=\linewidth]{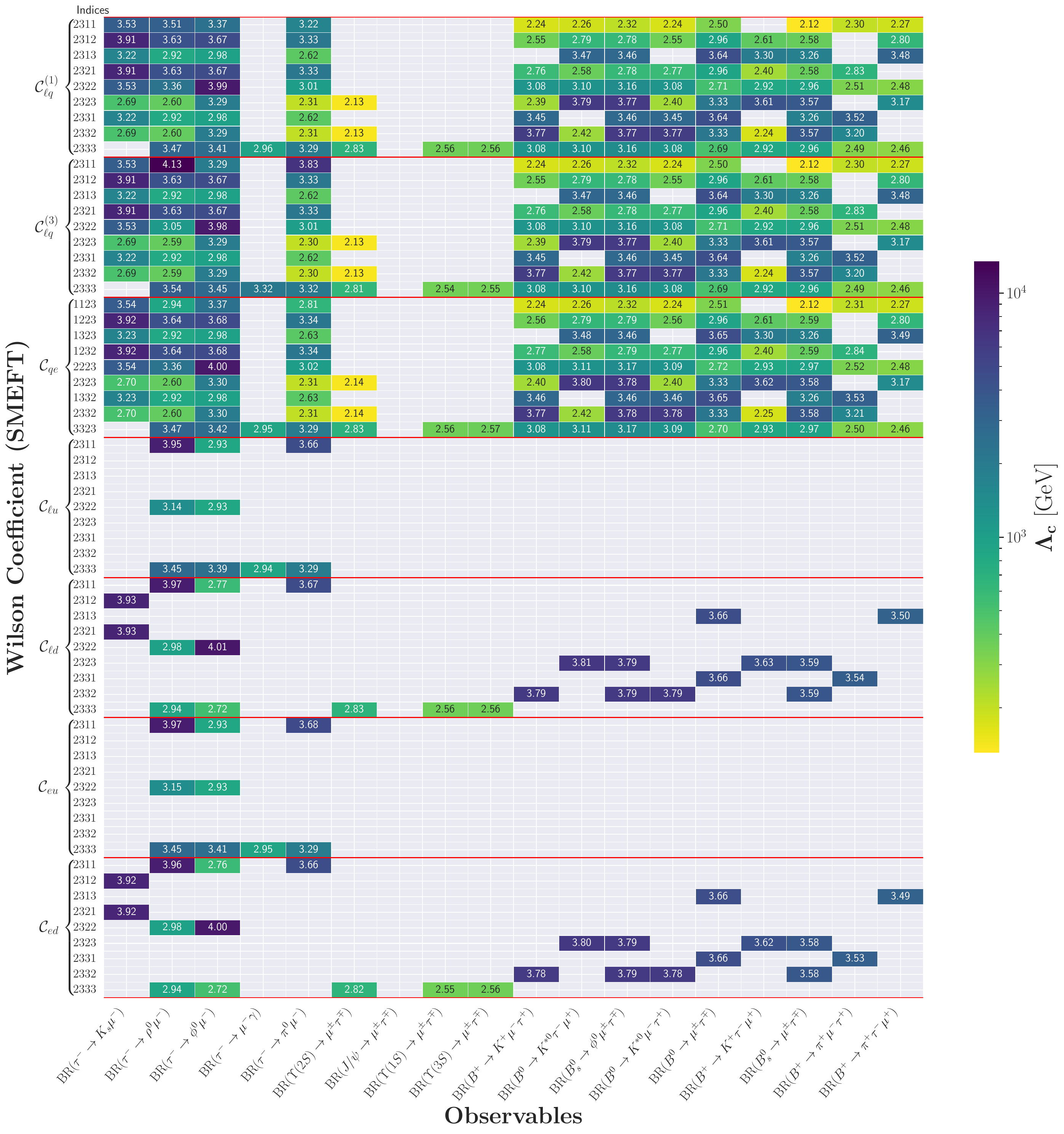}
    \caption{\justifying Same as Fig.\,\ref{Fig:emuLamplot} but with observables and operators with second- and third-generation charged leptons $(\mu^\pm,\tau^\pm)$.}
    \label{Fig:mutauLamplot}
\end{figure}

\begin{figure}[H]
    \centering
  \includegraphics[width=\linewidth]{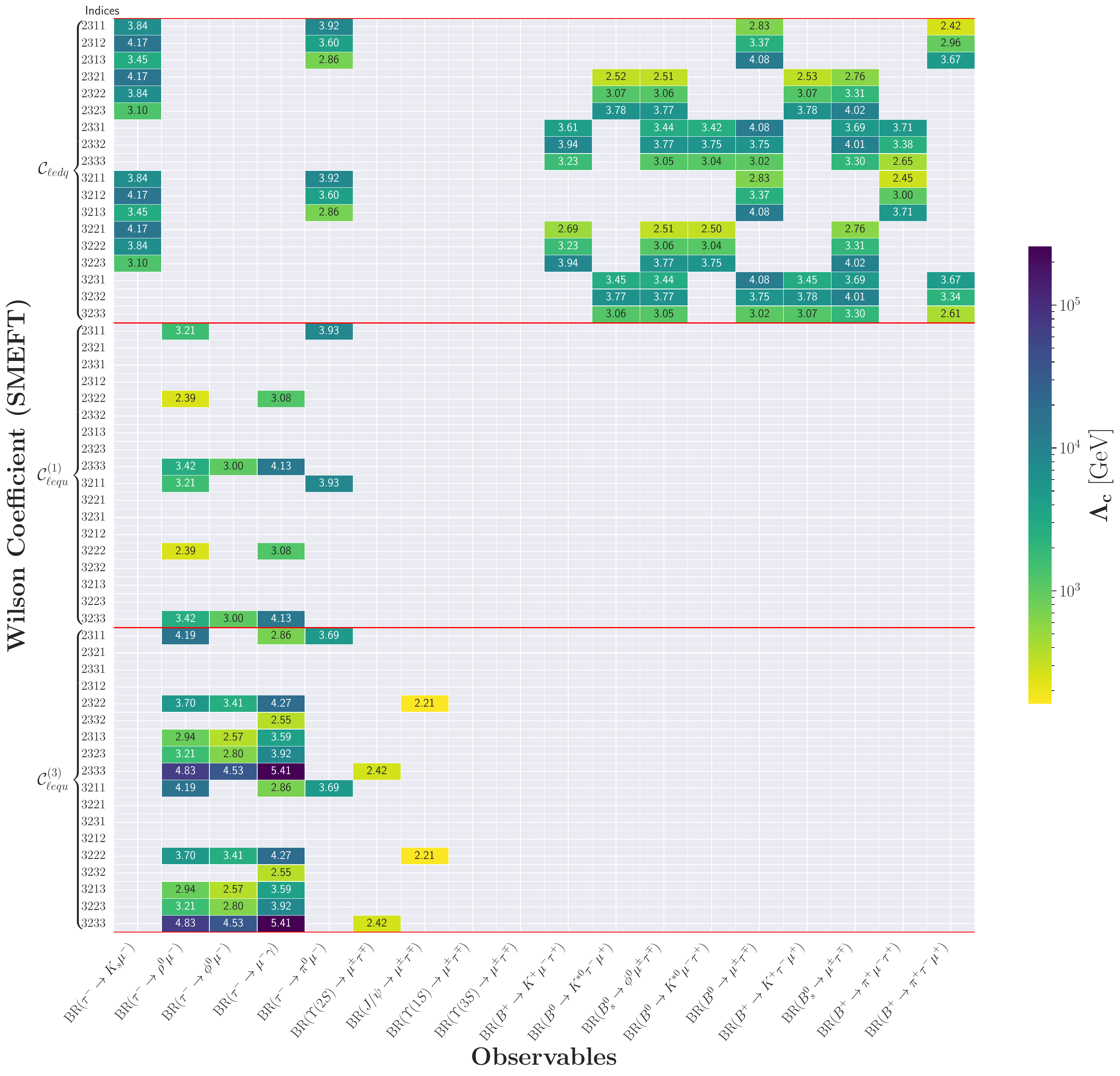}
\caption{Same as Fig.\,\ref{Fig:mutauLamplot} but with scalar and tensor operators.}
\label{Fig:mutau_scalar_Lam_plot}
\end{figure}

Similar to $\mu$--$e$ sector, $\tau$--$\mu$ sector results are shown in Figs.\,\ref{Fig:mutauplot}-\ref{Fig:mutau_scalar_Lam_plot}. 
A couple of significant differences from $\mu$--$e$ sector are the following. First, the WCs in this sector are significantly less constrained.
Compared to the 8 orders of magnitude spread of WCs in the $\mu$--$e$ sector, WCs relevant for the $\tau$--$\ell$ sector vary only 4 orders of magnitude, from $10^{-2}$ to $10^{2}$. 
The reason for this is that the experimental values of LFV decays in $\tau$--$\ell$ sector are a few orders of magnitude less precise compared to those of $\mu$--$e$ sector. 
The second difference is the presence of more empty boxes for WCs with right-handed $u$-type quark current (in both $\tau$--$e$ and $\tau$--$\mu$ sector). 
On the contrary, constraints on WCs with right-handed $d$-type quark current ($\ld, \ed$) are relatively significant because all LFV processes considered here involve interaction of $d$-type quark current only, except $\tau^-\to\mu^-(\rho^0/\pi^0)$, which can occur through interactions of $u$-type current as well.
In the following, we shall discuss the results of decays in $\tau$--$\mu$ sector only.
We observe a similar trend of results for $\tau$--$ e$ sector and those are included in the Appendix- \ref{Appndx:tau-e results} (Figs.\,\ref{Fig:etauplot}-\ref{Fig:etau_scalar_Lam_plot}) for completeness.

\begin{table}[h!]
\renewcommand{\arraystretch}{1.6}
\begin{adjustbox}{width=1\textwidth}
\centering
\begin{tabular}{|c|rl|c|c|c|c|c|c|}
\hline
\multirow{2}{*}{Observables} & \multicolumn{2}{c|}{\multirow{2}{*}{Quark-level Process}} &  \multicolumn{2}{c|}{Vector} &  Scalar & Tensor  \\ 
\cline{4-7}
& \multicolumn{2}{c|}{}& $\cO_{\ell q},\cO_{\ell d},\cO_{ed},\cO_{\ell u},\cO_{eu}$  & $\cO_{qe}$ & $\cO_{\ell equ}^{(1)}$, $\cO_{\ell edq}$ & $\cO_{\ell equ}^{(3)}$ \\
\hline\hline
BR$(\tau^+\to K_s^0 \mu^+)$
    & $\tau^+\to$  & $ \frac{1}{\sqrt2}\left(d\bar{s}-\bar ds\right)\mu^+$ & $[\cO]_{2312},[\cO]_{2312}$ & $[\cO]_{1223}$ & $[\cO]_{2312},[\cO]_{2321},[\cO]_{3212},[\cO]_{3221}$ & -\\ 
 BR$(\tau^+\to \pi^0 \mu^+)$ & $\tau^{+}\to$  & $ \frac{1}{\sqrt{2}}\left(u\bar{u}-d\bar{d}\right)\mu^{+}$ & $[\cO]_{2311}$ & $[\cO]_{1123}$ & $[\cO]_{2311},[\cO]_{3211}$ & $[\cO]_{2311},[\cO]_{3211}$ \\ 
\hline

BR$(\tau^+\to\rho^0\mu^+)$ & $\tau^{+}\to$  & $ \frac{1}{\sqrt{2}}\left(u\bar{u}-d\bar{d}\right)\, \mu^{+}$ & $[\cO]_{2311}$ & $[\cO]_{1123}$ & - & $[\cO]_{2311},[\cO]_{3211}$\\ 
BR$(\tau^+\to\phi^0\mu^+)$    & $\tau^{+}\to $  & $s\bar{s} \mu^{+}$ & $[\cO]_{2322}$ & $[\cO]_{2223}$ & - & -\\ 
BR$(B^0\to\tau^\pm\mu^\mp)$ & $d\bar b\to $  & $\tau^\pm\mu^\mp$ & $[\cO]_{2313},[\cO]_{2331}$ & $[\cO]_{1323},[\cO]_{1332}$ & $[\cO]_{2313},[\cO]_{2331}, [\cO]_{3213},[\cO]_{3231}$ & -\\
BR$(B^{0(+)}\to K^{*0(+)}\tau^{+}\mu^{-})$ &  $d(u)\bar{b}\to $  & $d(u)\bar{s}\tau^{+}\mu^{-}$ & $[\cO]_{2332}$ & $[\cO]_{2332}$ &  $[\cO]_{2332},[\cO]_{3223}$ & - \\
BR$(B^{0(+)}\to K^{*0(+)}\tau^{-}\mu^{+})$ &  $d(u)\bar{b}\to $  & $d(u)\bar{s}\tau^{-}\mu^{+}$ & $[\cO]_{2323}$ & $[\cO]_{2323}$ &  $[\cO]_{2323},[\cO]_{3232}$ & - \\
BR$(B^+\to\pi^+\tau^+\mu^-)$ & $u\bar b\to$  & $ u\bar d\tau^+\mu^-$ & $[\cO]_{2331}$ & $[\cO]_{1332}$ & $[\cO]_{2331}, [\cO]_{3213}$ & -
\\
BR$(B^+\to\pi^+\tau^-\mu^+)$ & $u\bar b\to$  & $ u\bar d\tau^-\mu^+$ & $[\cO]_{2313}$ & $[\cO]_{1323}$ & $[\cO]_{2313}, [\cO]_{3231}$ & -
\\
BR$(B_s^0\to\tau^\pm\mu^\mp)$ & $s\bar b\to$  & $ \tau^\pm\mu^\mp$ & $[\cO]_{2323},[\cO]_{2332}$ & $[\cO]_{2323},[\cO]_{2332}$ & $[\cO]_{2323},[\cO]_{2332}, [\cO]_{3223},[\cO]_{3232}$ & -\\
BR$(B_s^0\to\phi^0\tau^\pm\mu^\mp)$ & $s\bar b\to$  & $ s\bar s\tau^\pm\mu^\mp$ & $[\cO]_{2323},[\cO]_{2332}$ & $[\cO]_{2323},[\cO]_{2332}$ & $[\cO]_{2323},[\cO]_{2332}, [\cO]_{3223},[\cO]_{3232}$ & -
\\\hline
BR$(J/\psi\to\tau^\pm\mu^\mp)$ & $c\bar c\to$  & $\tau^\pm\mu^\mp$ & $[\cO]_{2322}$ & $[\cO]_{2223}$ & - & $[\cO]_{2322},[\cO]_{3222}$ \\
BR$(\Upsilon\to\tau^\pm\mu^\mp)$ & $b\bar b\to$  & $\tau^\pm\mu^\mp$ & $[\cO]_{2333}$ & $[\cO]_{3323}$ & - & - \\
\hline
\end{tabular}
\end{adjustbox}
\caption{\justifying Operators with indices for different LFV decay modes which include $\tau$ and $\mu$ leptons. The operators relevant for the processes where $e$ is in the final states instead of $\mu$ can be obtained by replacing the indices as $\{{ij23},{ij32}\}\to\{{ij13},{ij31}\}$ for $\cO_{qe}$ and $\{{23ij},{32ij}\}\to\{{13ij},{31ij}\}$ for other operators.}
\label{Tab:Operator_indices_tau}
\end{table}
Following Table-\ref{Tab:Operator_indices_tau} we find that the WCs $(\mathcal{C}_{u,d})_{2311}$ are relevant for a couple of $\tau$-decays, namely, $\tau^-\to\mu^-(\rho^0/\pi^0)$ and thus the strongest constraint is expected from them. Indeed it is found from Fig.\,\ref{Fig:mutauplot} that the process $\tau^{-}\to\mu^{-}\rho^0$ provides the strongest constraint, although weak ($\sim 10^{-2}$), for these WCs except the ones from $\lQ1$ and $\qe$. For these two WCs, a similar order of constraints is obtained from $\tau^-\to\mu^- K_s$ decay. This is because $\tau^-\to\mu^-(\pi^0/\rho^0)$ decays involve contributions from both $u$ and $d$ operators with opposite signs, which interfere destructively due to similar contributions from these operators. The QFD WCs $(\mathcal{C}_d)_{2322}$ and $(\mathcal{C}_u)_{2322}$ are primarily responsible for $\tau^-\to\mu^-\phi$ or $J/\psi\to\tau^{\pm}\mu^{\mp}$ decays respectively. The former one provides the strongest constraint ($\sim 10^{-2}$) as the latter one is highly suppressed due to a small lifetime, allowing larger WC values. For the operators with right-handed quark current and QFD WCs, $(\mathcal C_R)_{23ii}$, the constraints from the processes $\tau\to\mu^-(\rho^0,\pi^0,\phi^0)$ come due to RGE running and intermixing. The process $\Upsilon\to\mu^\pm\tau^\mp$, similar to the case of $J/\psi$, is less constraining due to a small lifetime of $\Upsilon$ quarkonia.

Similarly, the WCs $(\mathcal{C}_d)_{2312,2321}$ are relevant to the decay $\tau^-\to K_s\mu^-$, as mentioned in Table-\ref{Tab:Operator_indices_tau} and thus receive the strongest constraint from this decay. The WCs $(\mathcal{C}_d)_{2313,2331}$ and $(\mathcal{C}_d)_{2323,2332}$ are responsible for LFVBDs since they involve $b$-$d$ and $b$-$s$ quark transitions respectively as shown in Ref.~\cite{Ali:2023kua}. We find our results to be consistent with this work as constraints of similar magnitude $(\sim 10^{-2})$ on these coefficients come from different LFVBDs. Being responsible for $u$-type quark transition only, WCs of classes $\lu$ and $\eu$ do not contribute to these LFVBDs.

Fig.\,\ref{Fig:mutau_scalar_plot} presents the limits on scalar and tensor SMEFT WCs, similar to Fig.\,\ref{Fig:mutauplot}. Except $\tau^-\to\mu^-\mathcal V,\;\mathcal V\in\{\rho^0,\phi^0\}$ and LFVQDs ($J/\psi,\Upsilon$), other observables are sensitive to the WCs $(\mathcal C_{\ell edq})_{23ij}$, either directly or through CKM mixing. This is due to the vanishing scalar and pseudoscalar hadronic matrix elements for such processes as explained in sec-\ref{Subsec:LFV tau decays}.

We find that only QFD WCs of the scalar operator $\mathcal C_{\ell equ}^{(1)}$ receive constraints from the LFV processes under consideration, and those constraints again come only from $\tau^-\to\mu^-(\gamma,\rho^0,\phi^0,\pi^0)$ processes. 
Among these processes, $\tau^-\to \mu^-\gamma$ occurs through dipole operators that are induced through RGE, similar to the case of $\mu^-\to e^-\gamma$. 
The processes $\tau^-\to\mu^-(\rho^0,\pi^0)$ involve both $u\bar u$ and $d\bar d$ interactions (see Table~\ref{Tab:Operator_indices_tau}) and their sensitivity towards these WCs arises through the induced RGE effect of $\mathcal C_{\ell equ}^{(3)}$, which is directly relevant for the processes. On the other hand, the sensitivity of the process $\tau\to\mu^-\phi^0$ towards these QFD WCs arises through previously mentioned dipole effects.

Although WCs of class $\mathcal{C}_{\ell equ}^{(3)}$ are only directly relevant for $\tau^-\to\mu^-\rho^0$, several other WCs are constrained by different processes due to Yukawa RGEs of heavier quarks as explained before. For the QFD WCs, especially with indices $2333,3233$, processes $\tau^-\to\mu^-\gamma$ and $\tau^-\to\mu^-\mathcal V$ are quite sensitive due to the generation of leptonic dipole operators, similar to the case of CR$(\mu\to e)$ and $\mu^-\to e^-\gamma$.

Figs.\,\ref{Fig:mutauLamplot} and \ref{Fig:mutau_scalar_Lam_plot}, shows the energy scales accessible by each experiment for each of the $\mu$--$\tau$ $2q2\ell$ SMEFT WCs. Here again, as increasing $\Lambda$ is equivalent to decreasing WC, the behaviour and explanations are exactly as those for plots showing limits on WCs.

\section{2-D Analysis}
\label{sec:2d analysis}

Although the 1-D analyses mentioned above are quite simple to perform, these analyses neglect the possible interferences in the amplitudes if more than one operator has non-negligible/comparable WCs at SMEFT scale $\Lambda$. Such interferences can be captured by a more realistic, 2-D analysis, where two of the operators have non-vanishing WCs, with the rest set to zero at $\Lambda$. Such an analysis can provide valuable insights into the interplay between different operators arising from RGE evolution, and can also reveal possible correlations by identifying the most constrained combinations. Therefore, in this section, we present the results from the 2-D analysis. 
Among the many possible pairs, we present a representative subset that exhibits particularly interesting features. We employ the hard-cut (direct exclusion) method: theoretical predictions that exceed the experimental upper limits are ruled out. This avoids the complexity of constructing a global likelihood and, more importantly, provides a transparent view of the interference patterns between operators in the 2-D analysis.

\begin{figure}[h!]
    \centering
     \includegraphics[width=0.49\linewidth]{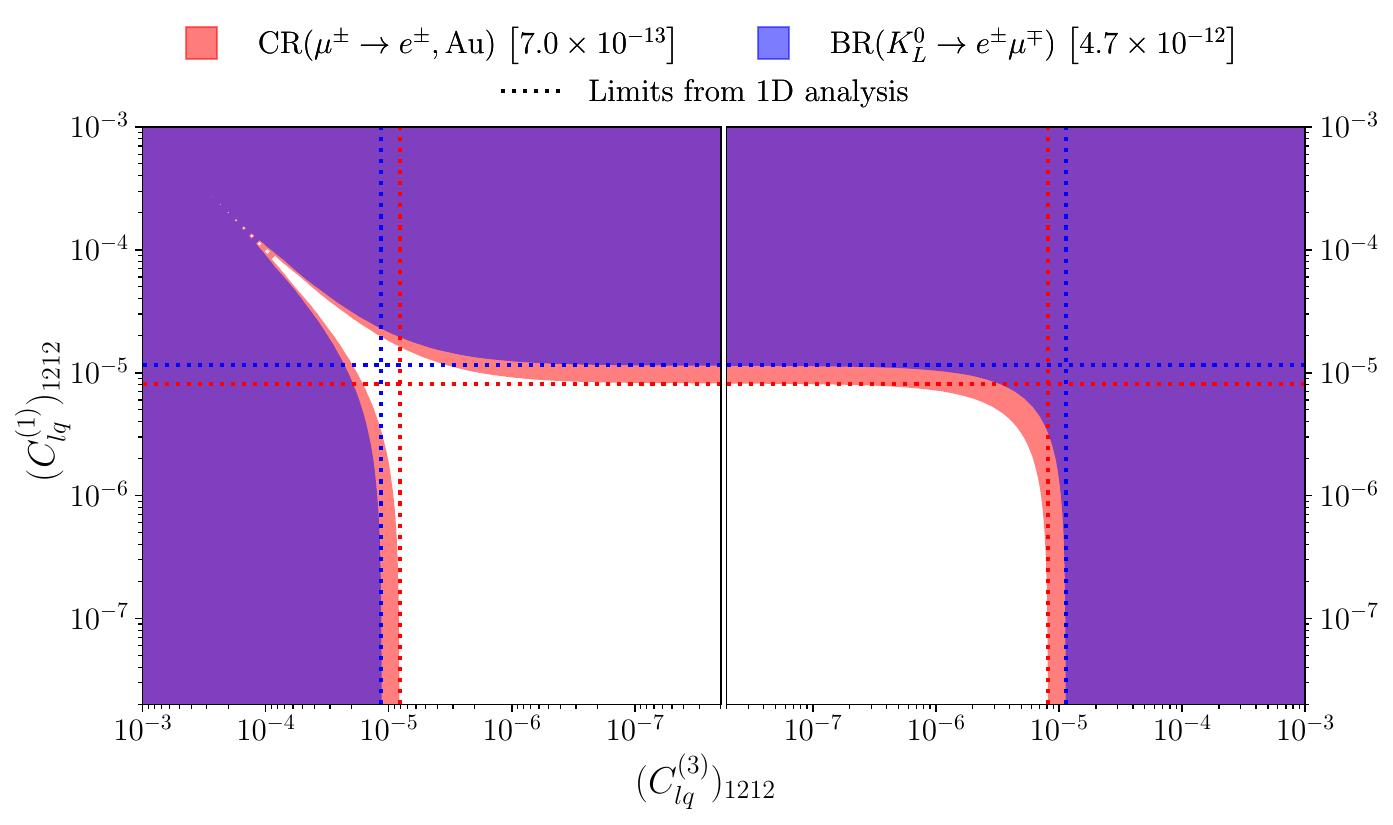}
    \includegraphics[width=0.49\linewidth]{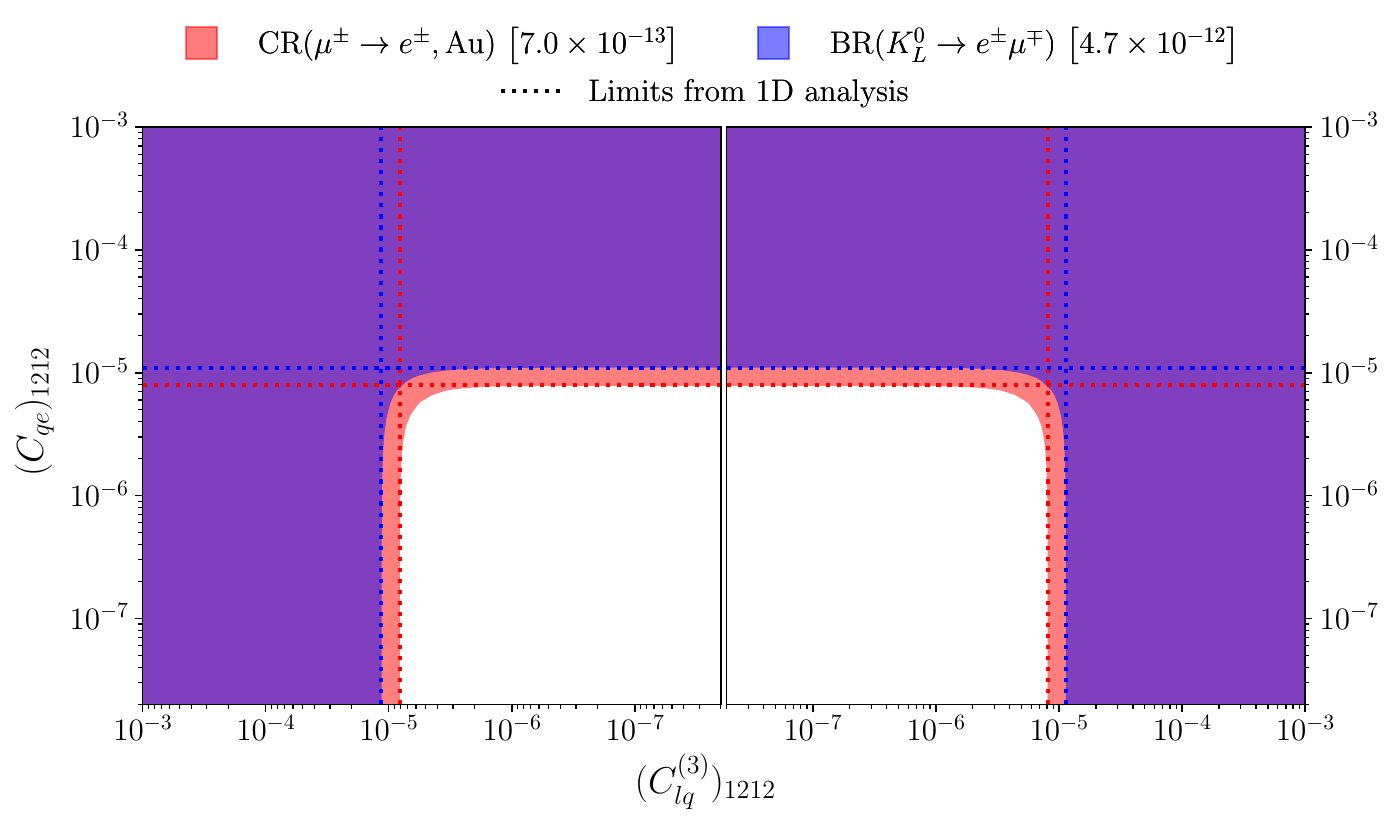}
    \includegraphics[width=0.49\linewidth]{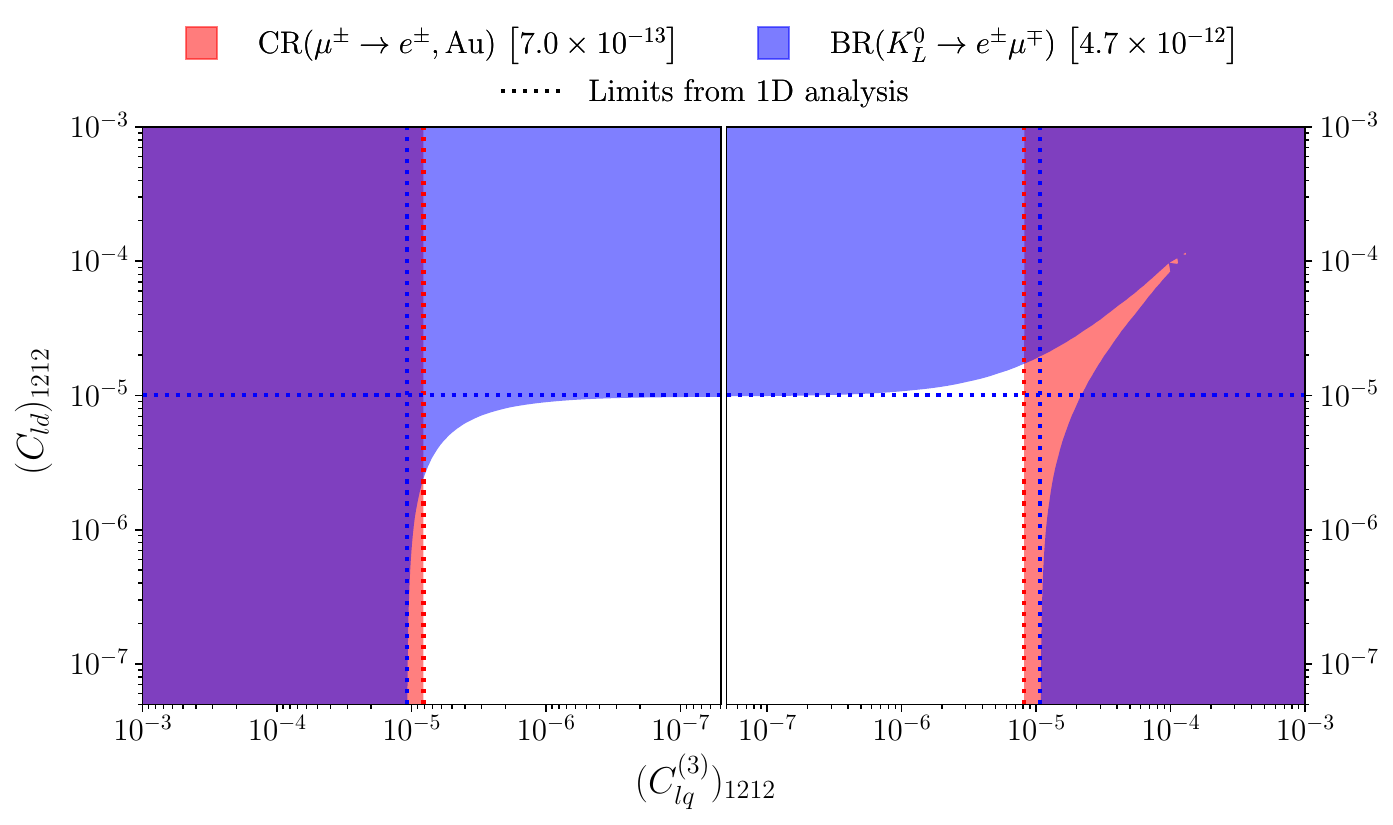}
    \includegraphics[width=0.49\linewidth]{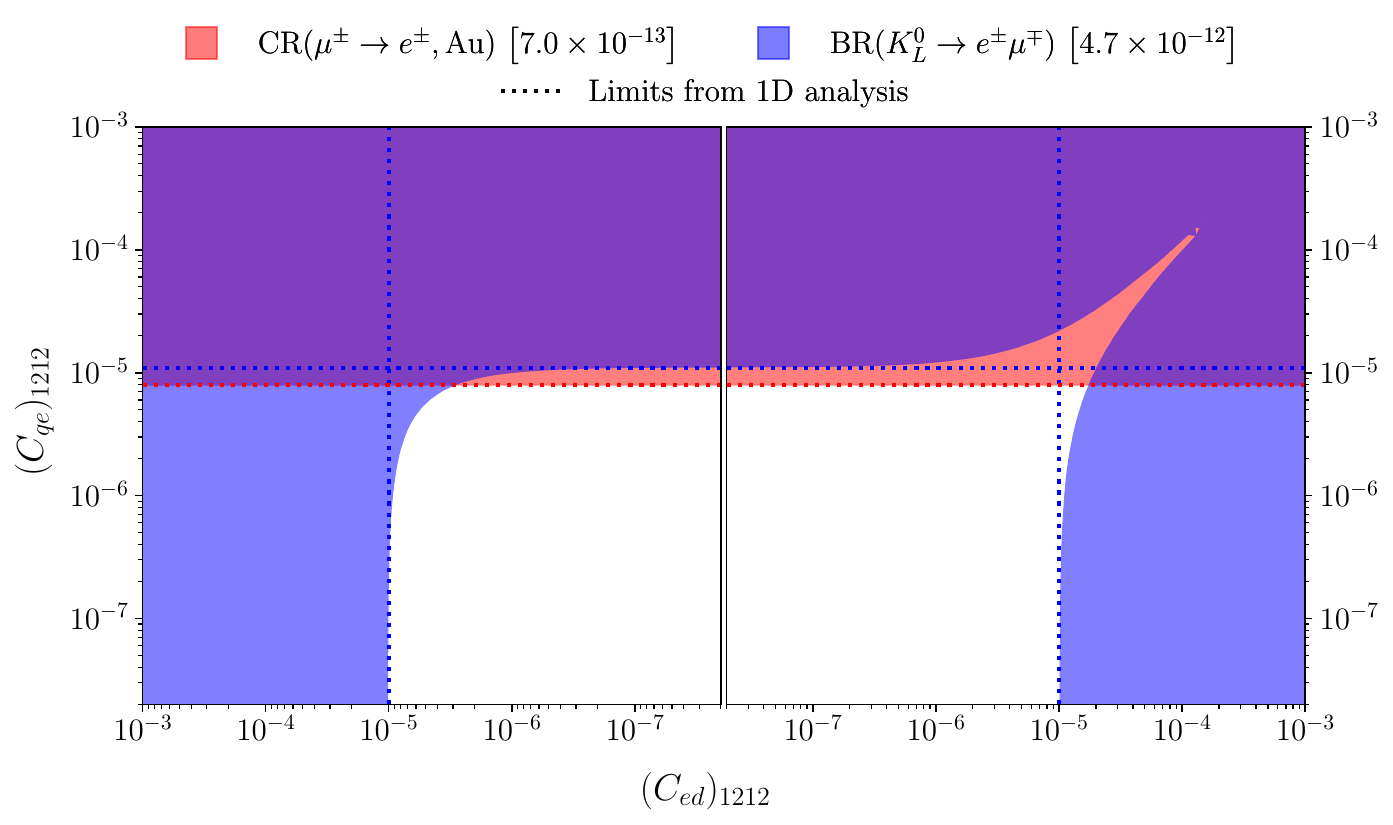}
    \caption{\justifying  Limits on 2D WC parameter space from two of the most constraining observables. White color shows the allowed region for both WCs.
    }
    \label{fig:clq3-ced-1212}
\end{figure}

In Fig.\,\ref{fig:clq3-ced-1212} we show the contours of branching ratios of two most constraining LFV processes, $\CRmue$ and $\KLemu$, on the 2-D WC space of different SMEFT operators with the same flavour indices but with different chiralities of quark and lepton FCNC on a logarithmic scale. We choose to show only the positive values of the WC along the vertical axis against both positive and negative values of the WC along the horizontal axis as the relative sign between the WCs affects the observable. The dotted lines represent the limits on the two-dimensional WC space obtained from 1-D analysis to satisfy current experimental data. The regions in red and blue are excluded from $\CRmue$ and $\KLemu$, respectively.

One can readily see the extended region in the corners for some pairs of WCs, where the observable under consideration fails to constrain the WC space, known as the {\it flat directions}. Primarily, the origin of these flat directions can be understood by looking at the leading order running and matching conditions between SMEFT and LEFT operators. For example, in case of $\CRmue$ (top-left panel), the matching of $d_L$-type LEFT WC involves contributions from both $\lQ1$ and $\lq3$ as, 
\begin{align}
    \CRmue&\sim |(C^{V,LL}_{ed})_{1211}|^2\sim |V_{12}(C^{V,LL}_{ed})_{1212}|^2\sim |V_{12}(\lQ1+\lq3)_{1212}|^2\,,
\end{align}
where $V$ is the CKM matrix (see Eq.(\ref{eq:CKMmixingV})). Thus, in the region where these SMEFT WC are typically equal and opposite, the $d_L$-type LEFT WC vanishes, resulting in a suppressed $\CRmue$. Thus, in the region where these SMEFT WC are typically equal and opposite, the $d_L$-type LEFT WC vanishes, resulting in a suppressed $\CRmue$. For $\KLemu$, in addition to this (top-left), the \textit{flat direction} arises also through the cancellation in the amplitudes of diagrams involving currents with opposite chirality (bottom panels) as,
\begin{align}
    \BR[\KLemu]&\sim \left|\left(C^{ed}_{VA}\right)_{1212}\right|^2+\left|\left(C^{ed}_{AA}\right)_{1212}\right|^2\nonumber\\
    &=\left|\left(-C_{ed}^{V,LL}+C_{ed}^{V,LR}-C_{ed}^{V,RL}+C_{ed}^{V,RR}\right)_{1212}\right|^2+\left|\left(C_{ed}^{V,LL}-C_{ed}^{V,LR}-C_{ed}^{V,RL}+C_{ed}^{V,RR}\right)_{1212}\right|^2\nonumber\\
    &\sim\left|\left(-\left(\lQ1+\lq3\right)+\ld-\qe+\ed\right)_{1212}\right|^2+\left|\left(\left(\lQ1+\lq3\right)-\ld-\qe+\ed\right)_{1212}\right|^2\,,
\end{align}
where the relative sign difference between the WCs is to be noted. Thus, $\KLemu$ (and other pseudoscalar meson decays) is expected to feature a \textit{flat direction} in a space consisting of a WC-pair with the same chirality of the lepton FCNC as can be seen in the bottom panels of Fig.\,\ref{fig:clq3-ced-1212}

Similarly, we can find other pairs of WCs with flat directions, which is clearly due to the interplay of RGE matching and running, but it must be noted that the results obtained from these 2-D analyses are consistent with the findings of our 1-D analysis.

\section{Conclusion}
\label{sec:conclusion}

In this paper, we have considered a comprehensive list of low-energy LFV observables which are directly dependent on LFV $2q2\ell$ operators. Based on the fermion constituents of these processes we have categorized the WCs of these operators by their specific indices in Tables~\ref{Tab:Operator_different-quark_indices}-\ref{Tab:Operator_indices_tau} and found constraints on each of them coming from all the observables. In order to do so, we have implemented several LFV decay modes in the open-source package \textsf{Flavio} for the first time, the details of which are provided in sec-\ref{sec:Calculation method}. The main findings of this analysis, which was done by assuming the presence of only a single operator at $\Lambda$, can be summarized as follows.

\begin{itemize}
    \item In the $\mu$--$e$ sector, almost all the WCs of vector operators consisting of left-handed quark doublet current ($\mathcal C_{\ell q}^{(1)},\mathcal C_{\ell q}^{(3)}$ and $\mathcal{C}_{qe}$) are constrained by most of the observables involving interaction of $d$-type quarks. This is because of the convention used in \textit{Warsaw up} basis in \textsf{Wilson} where the running mass matrix of $u_L$-type quarks is considered as diagonal at $\Lambda$. As a result WCs of operators with $d_L$-type operators require CKM rotation and subsequent enhancement due to the CKM matrix elements becomes significant for $d_L$ and $s_L$ quarks. On the other hand, observables with $u$ and $c$ quarks (quarkonium and LFVDDs) or WCs of the operators with right-handed quark current ($\mathcal C_{\ell d},\ed,\mathcal C_{\ell u}$ and $\eu$) do not require such a large CKM rotation are thus not constrained or weakly constrained by the observables with different quark FCNC.
    \item The coefficient $(\mathcal C_{L,R})_{1211}$ for all the vector LFV $2q2\ell$ operators, except $(\lq3)_{1211}$, receives stringent constraint $(\sim 10^{-6})$ from $\CRmue$ experiment due to its high precision value. $(\mathcal C_{\ell q}^{(3)})_{1211}$ is relatively less constrained due to the equal and opposite contribution of SMEFT operators to LEFT operators, $(C_{eu}^{V,LL})_{1211}$ and $(C_{ed}^{V,LL})_{1211}$, resulting in a destructive interference between the proton and neutron contributions to the total amplitude.
    \item Similarly, other two diagonal WCs $(\mathcal C_{L})_{1222}$ and $(\mathcal C_{L})_{1233}$ operators with left-handed quark current, also receive strongest constraints, $\sim 10^{-4}$ and $\sim 10^{-3}$ respectively, from the $\CRmue$ experiment.
    \item Although $J/\psi$ LFV decay has a strong experimental bound $(\sim 10^{-9})$, it generates very weak bound for the coefficient $(\mathcal C_u)_{1222}$ (relevant for its LFV decay) because of its small lifetime ($\tau\lesssim10^{-20}\,\rm s$).
    \item  LFVKDs turn out to be the provider of the most stringent constraint ($\sim 10^{-5}$) for the WCs of right-handed $d$-type quark current operators, namely $(\ld)_{1212,1221}$ and $(\ed)_{1212,1221}$.
    \item The coefficients $(\mathcal C_{L,R})_{1223}$ and $(\mathcal C_{L,R})_{1232}$ are constrained by LFVBD processes $B^0\to K^{*0}e^-\mu^+$ and $B^0\to K^{*0}e^+\mu^-$ respectively with limit  $\sim10^{-3}$.  
    \item The process $\KLemu$ provides strongest constraints for several scalar WCs $(\ledq)_{1212,1213,1222}$. Likewise  $(\ledq)_{1231,1232,1233}$ receive strongest constraints from several LFVBD processes.
    \item  For the tensor WC $(\lequ3)_{1233}$ a strong constraint $(\sim 10^{-7})$ comes from $\CRmue$ through leptonic dipole operators. Whereas the coefficient $(\lequ3)_{1221,1221,2112,2121}$ can be probed in LFVDDs. 
    \item $\CRmue$ is the most sensitive process for most of the diagonal elements of vector operators, $((\mathcal C_{L,R})_{1211,1222,1233})$, ranging from $10^4-10^6 \GeV$ irrespective of the chirality. Whereas several LFVBD processes are most sensitive in probing at $\sim 10^3-10^5 \GeV$ for left-handed and $d$-type right-handed quark WCs $((\mathcal C_{d})_{1213,1231,1223,1232})$. Also, except $(\mathcal C_u)_{1222}$ for $J/\Psi$ or $(\mathcal C_d)_{1233}$ for $\Upsilon$, LFVQDs are found to be least sensitive to other operators and coefficients. For LFVDD processes, they can probe the WCs $(\mathcal C_{u_R})_{1212,1221}$ having right-handed $u$-type quark-current operators $\sim 10^3 \GeV$, but their chiral counterparts are dominated by $\CRmue$.
    \item Clearly, compared to $\mu$--$e$ sector, $\tau$--$\ell$ sector is more promising for vector LFV $2q2\ell$ operators as the relevant WCs are significantly less constrained $(\mathcal{C}_{\rm max}\sim10^{-4})$ and thus can be probed within the energy limit of $10^4 \GeV$.
    \item  When the NP effects are not assumed to be produced by a single SMEFT operator at $\Lambda$, cancellations are possible among different contributing elements to the LFV decay rates, resulting in the so-called \textit{flat directions}.
\end{itemize}

\section*{Acknowledgment}
The computations in this project were partially supported by SAMKHYA, the high-performance computing (HPC) facility provided by the Institute of Physics, Bhubaneswar (IOPB). The authors thank Subhadip Bisal for valuable discussions.

\appendix


\section{Matching between LEFT and SMEFT basis}
\label{appndx:LEFT SMEFT matching}

At the matching scale, the tree-level matching relations between WCs of the $2q2\ell$ operators in LEFT ($C$) and those in SMEFT ($\mathcal{C}$)  are given by~\cite{Jenkins:2013wua}:
\begin{align}
C_{\substack{eu(d)\\\alpha\beta ij}}^{V,LL}  &=  \frac{v^2}{\Lambda^2}\left[\mathcal{C}^{(1)}_{ \substack {\ell q \\ \alpha\beta ij } }  \mp \mathcal{C}^{(3)}_{ \substack {\ell q \\ \alpha\beta ij } }\right]  -\frac{\bar{g}_Z^2v^2}{ m_Z^2} \left[ Z_{e_L}\right]_{\alpha\beta}\left[ Z_{u(d)_L}\right]_{ij}\,, \quad &
C_{\substack{eu(d)\\\alpha\beta ij}}^{V,RR}  &= \frac{v^2}{\Lambda^2}\left[\mathcal{C}_{\substack{ eu(d) \\ \alpha\beta ij}}\right] -\frac{\bar{g}_Z^2v^2}{ M_Z^2}   \left[Z_{e_R} \right]_{\alpha\beta} \left[Z_{u(d)_R} \right]_{ij} \,,
\\
C_{\substack{eu(d)\\\alpha\beta ij}}^{V,LR}  &= \frac{v^2}{\Lambda^2}\left[\mathcal{C}_{\substack{ \ell u(d)\\ \alpha\beta ij}}\right] 
-\frac{\bar{g}_Z^2v^2}{M_Z^2}   \left[Z_{e_L} \right]_{\alpha\beta} \left[Z_{u(d)_R} \right]_{ij}\,, \quad &
C_{\substack{u(d)e\\ij\alpha\beta}}^{V,LR}  &= \frac{v^2}{\Lambda^2}\left[\mathcal{C}_{\substack{ qe \\ ij\alpha\beta}}\right]
 -\frac{\bar{g}_Z^2v^2}{M_Z^2}   \left[Z_{u(d)_L} \right]_{ij} \left[Z_{e_R} \right]_{\alpha\beta}\,,\\
C_{\substack{eu\\\alpha\beta ij}}^{S,RL}  &= 0 \,, \quad & {C}_{\substack{ed\\\alpha\beta ij}}^{S,RL}  &= \frac{v^2}{\Lambda^2}\mathcal{C}_{\substack{ \ell edq \\ \alpha\beta ij}}\,,\\
C_{\substack{eu\\\alpha\beta ij}}^{S,RR}  &= -\frac{v^2}{\Lambda^2}\mathcal{C}^{(1)}_{\substack{ \ell equ \\ \alpha\beta ij}} \,, \quad & C_{\substack{ed\\\alpha\beta ij}}^{S,RR}  &= 0 \,,\\
C_{\substack{eu\\\alpha\beta ij}}^{S,LL}  &= -\frac{v^2}{\Lambda^2}\mathcal{C}^{(1)\ast}_{\substack{ \ell equ \\ \beta\alpha ji}} \,, \quad & C_{\substack{ed\\\alpha\beta ij}}^{S,LL}  &= 0 \,,\\
C_{\substack{eu\\\alpha\beta ij}}^{T,RR}  &= -\frac{v^2}{\Lambda^2}\mathcal{C}^{(3)}_{\substack{ \ell equ \\ \alpha\beta ij}} \,, \quad & C_{\substack{ed\\\alpha\beta ij}}^{T,RR}  &= 0 \,,
\label{eq:LEFT SMEFT matching}
\end{align}    
where, in the first equation, ($-$) sign between $\mathcal{C}_{lq}^{(1)}$ and $\mathcal{C}_{lq}^{(3)}$ is for $C_{eu}$ and ($+$) is for $C_{ed}$. $Z_{ij(\alpha\beta)}$, with $i,j(\alpha,\beta)$ being quark(lepton) flavour indices, represents the $Z$-boson coupling of the neutral currents in the SMEFT operators. These couplings are given as:
\bea
\left[Z_{e_L}\right]_{\alpha\beta} &=& \left[\delta_{\alpha\beta}\left(-\frac{1}{2}+s^2\right)-\frac{1}{2}\frac{v^2}{\Lambda^2}\left( \mathcal{C}_{H\ell}^{(1)}+\mathcal{C}_{H\ell}^{(3)}\right)_{\alpha\beta}\right]\\
\left[Z_{e_R}\right]_{\alpha\beta} &=& \left[\delta_{\alpha\beta}\left(+s^2 \right) - \frac12 \frac{v^2}{\Lambda^2}  \mathcal{C}_{\substack {He \\  \alpha\beta}}  \right]\\
\left[Z_{u_L}\right]_{ij} &=& \left[\delta_{ij}\left(\frac{1}{2}-\frac{2}{3}s^2\right)-\frac{1}{2}\frac{v^2}{\Lambda^2}\left( \mathcal{C}_{Hq}^{(1)}-\mathcal{C}_{Hq}^{(3)}\right)_{ij}\right]
\\
\left[Z_{d_L}\right]_{ij} &=& \left[\delta_{ij}\left(-\frac{1}{2}+\frac{1}{3}s^2\right)-\frac{1}{2}\frac{v^2}{\Lambda^2}\left( \mathcal{C}_{Hq}^{(1)}+\mathcal{C}_{Hq}^{(3)}\right)_{ij}\right],\\
\left[Z_{u_R}\right]_{ij} &=& \left[\delta_{ij}\left(-\frac{2}{3}s^2\right)-\frac{1}{2}\frac{v^2}{\Lambda^2}\left(\mathcal{C}_{Hu}\right)_{ij}\right]\\
\left[Z_{d_R}\right]_{ij} &=& \left[\delta_{ij}\left(\frac{1}{3}s^2\right)-\frac{1}{2}\frac{v^2}{\Lambda^2}\left(\mathcal{C}_{Hd}\right)_{ij}\right]
\eea
Here $s=\sin\theta_W$, $\theta_W$ being the electroweak mixing angle, $v$ is the  Higgs vacuum expectation value (vev), and $\mathcal{C}_{Hf}^{(1)/(3)}$ are the associated WC of dimension-6 Higgs-fermion operators. We neglect the modifications in the symmetry-breaking parameters ($v,\theta_W$ \etc) arising from the modified electroweak symmetry breaking~\cite{Jenkins:2013wua,Jenkins:2017jig}, as the dimension-6 Higgs self-interaction operators in SMEFT are considered to be absent. The matching of leptonic dipole WCs is given as~\cite{Jenkins:2013wua},
\begin{align}
    ({C}_{e\gamma})_{\alpha\beta}\sim\frac{1}{\sqrt 2} \frac{v^2}{\Lambda^2} \left[c\left(\mathcal C_{eB}\right)_{\alpha\beta}- s\left(\mathcal C_{eW}\right)_{\alpha\beta}\right]\label{eq:cegammaMatching}
\end{align}


\section{Radiative Lepton Generation Transition ($\ell_\alpha\to \ell_\beta\gamma$)}
\label{App:muegamma}
\begin{figure}[h!]
\centering
\begin{tikzpicture}
\begin{feynman}
\vertex (a1) {$\ell_i^-$};
\vertex[blob,shape=circle,fill=black,minimum height=.8cm,minimum width=0.5cm] at ($(a1)+(2.0cm,0.0cm)$) (a2) {};
\vertex[right=2.0cm of a2] (a3) {$\ell_j^-$};
\vertex[above right=1.2cm and 1cm of a2] (p1) {$\gamma$};
\diagram*{
(a1) -- [fermion] (a2) -- [fermion] (a3),
(a2) -- [photon, black] (p1),
};
\end{feynman}
\end{tikzpicture}
\caption{Feynman diagram of the lepton flavor violating decay $\ell_i^- \to \ell_j^-\gamma$. }
\label{Fig:muegamma}
\end{figure}

Radiative lepton generation transitions, such as $\mu\to e \gamma$, are one of the most constrained LFV processes. The processes are primarily driven by the leptonic dipole operators, which can be induced from SMEFT $2q2\ell$ operators through heavy quark loops. The branching ratio of these processes is given as $(\alpha>\beta)$,
\begin{align}
    {\rm BR}(\ell_\alpha\to \ell_\beta\gamma)&=\frac{\tau_{\ell_\alpha}m_{\ell_\alpha}^3}{4\pi v^2}\left(\left|C_{e\gamma}^{\alpha\beta}\right|^2+\left|C_{e\gamma}^{\beta\alpha\,\ast}\right|^2\right)
\end{align}
where, $\tau_\ell$ and $m_\ell$ are lifetime and mass of lepton $\ell$, respectively.

\section{$\tau$--$e$ Results}
\label{Appndx:tau-e results}

\begin{figure}[h!]
    \centering
    \includegraphics[width=\linewidth]{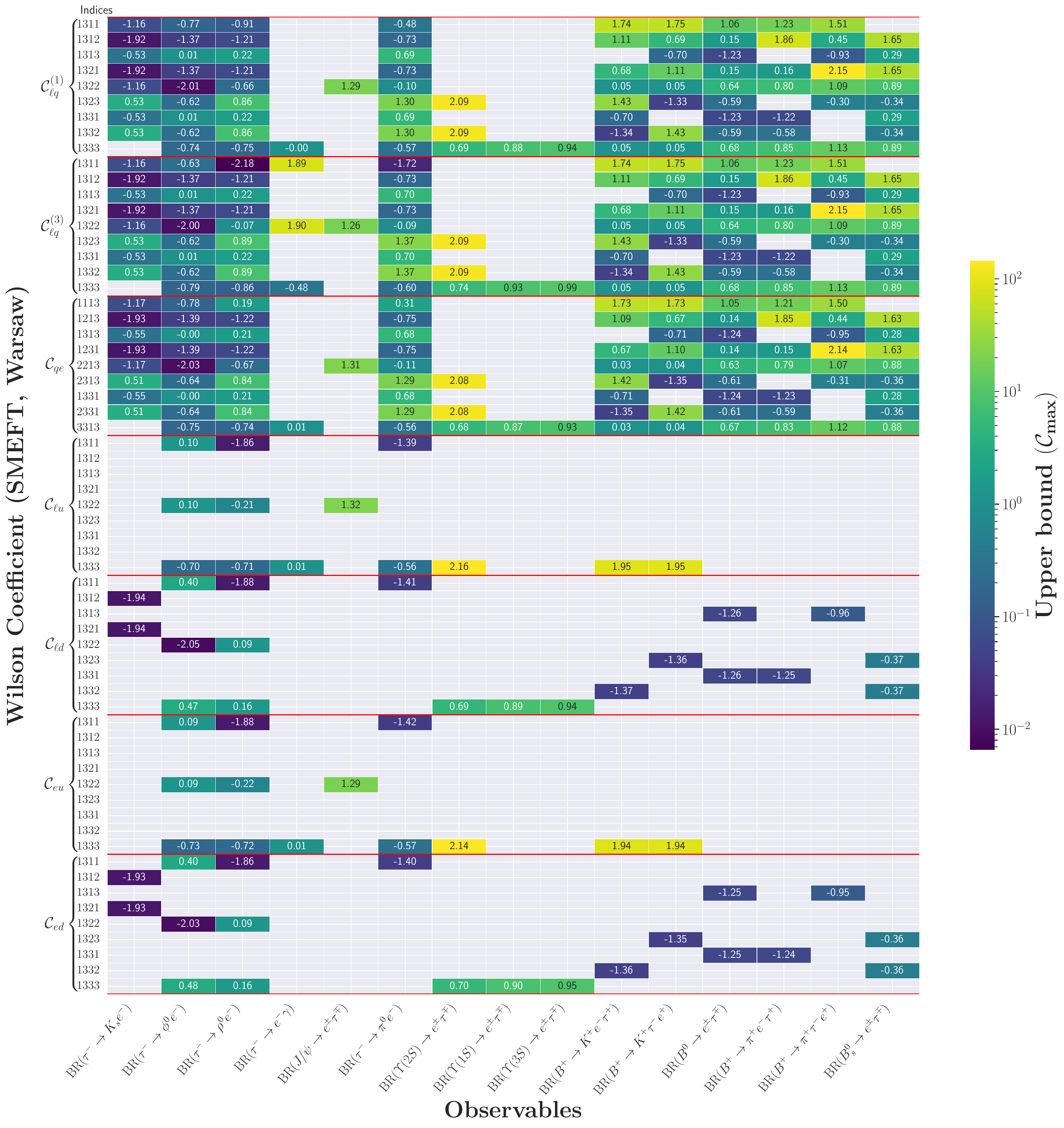}
    \caption{Same as Fig.\,\ref{Fig:emuplot} but with observables and operators with first- and third-generation charged leptons $(e^\pm,\tau^\pm)$.}
    \label{Fig:etauplot}
\end{figure}

\newpage
\begin{figure}[h!]
    \centering
    \includegraphics[width=\linewidth]{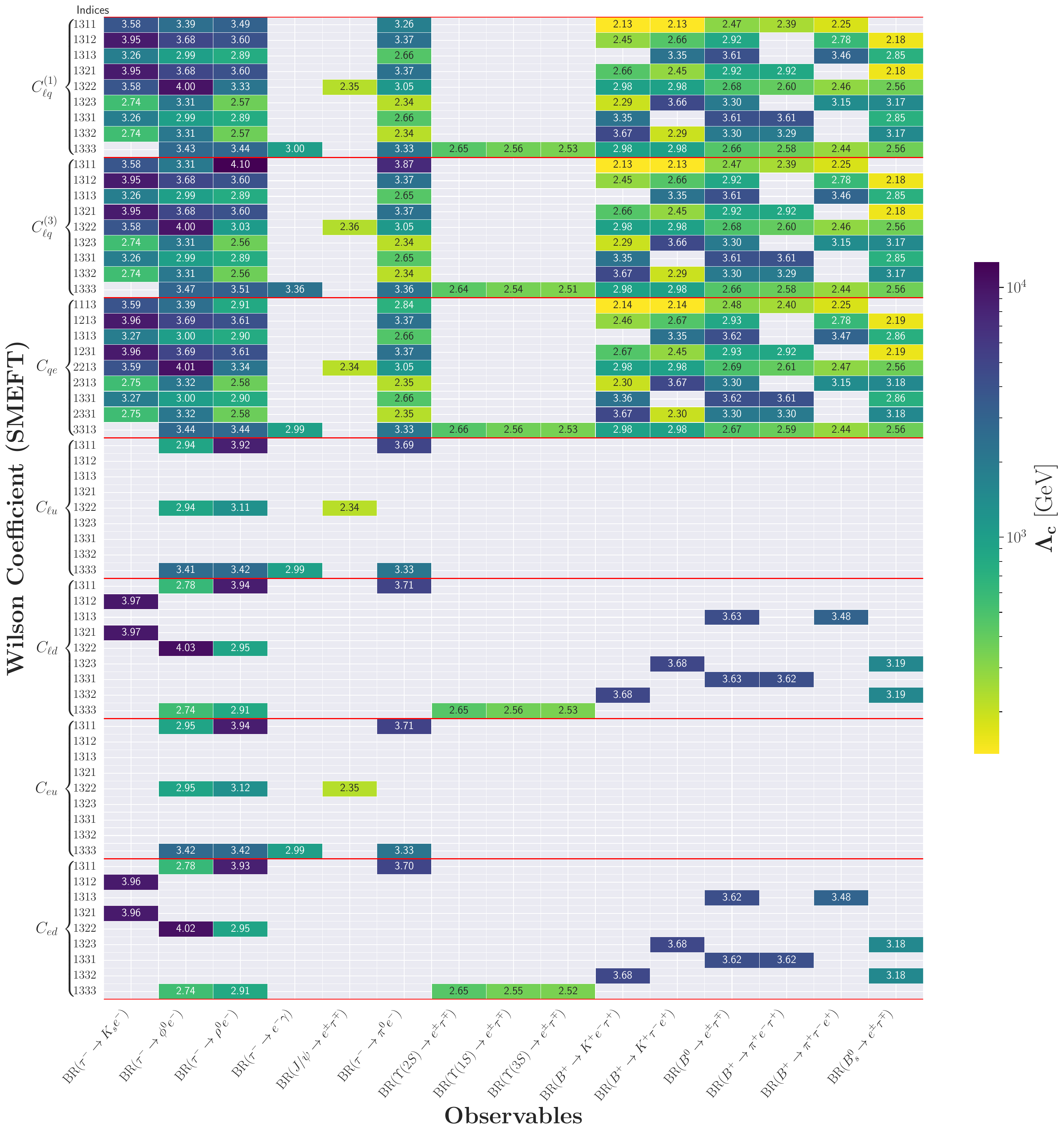}
    \caption{Same as Fig.\,\ref{Fig:emuLamplot} but with observables and operators with first- and third-generation charged leptons $(e^\pm,\tau^\pm)$.}
    \label{Fig:etauLamplot}
\end{figure}

\begin{figure}[H]
    \centering
\includegraphics[width=\linewidth]{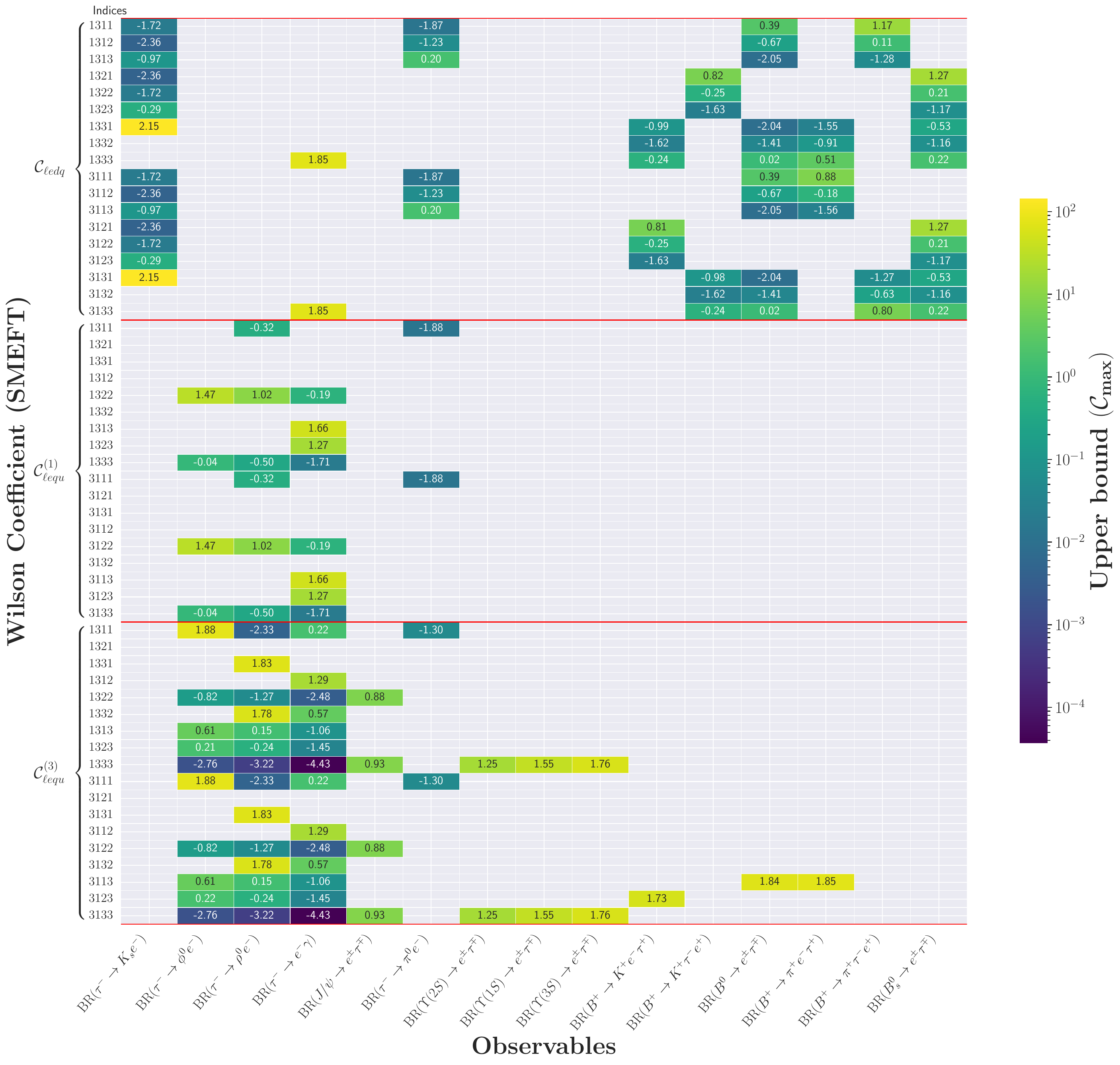}
\caption{Same as Fig.\,\ref{Fig:etauplot} but with scalar and tensor $2q2\ell$ operators.}
\label{Fig:etau_scalar_plot}
\end{figure}

\begin{figure}[H]
    \centering
  \includegraphics[width=\linewidth]{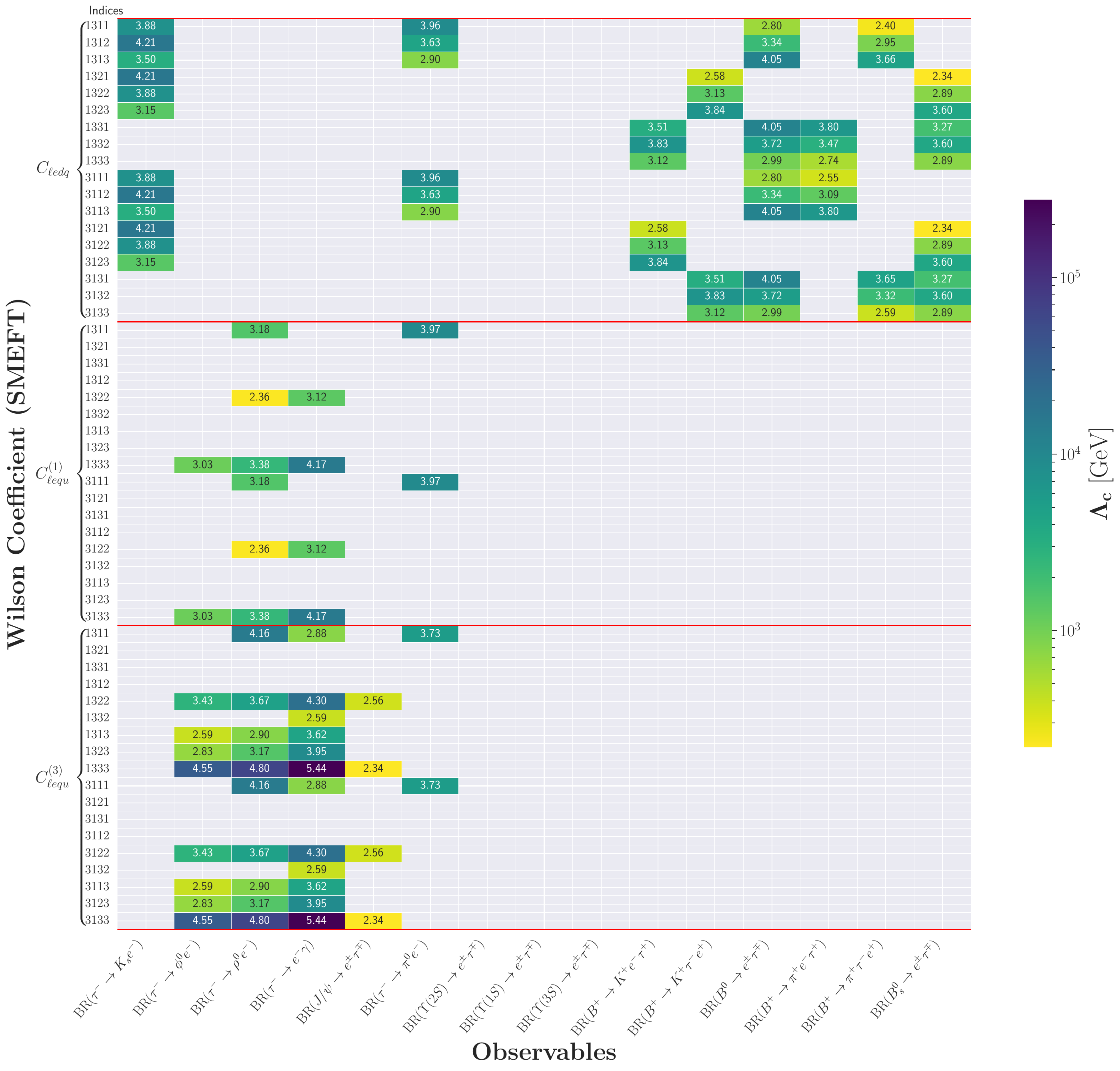}
\caption{Same as Fig.\,\ref{Fig:etauLamplot} but with scalar and tensor operators.}
\label{Fig:etau_scalar_Lam_plot}
\end{figure}


\bibliographystyle{JHEPCust.bst}
\bibliography{refs_lfvkd.bib}

@article{BaBar:2021loj,
    author = "Lees, J. P. and others",
    collaboration = "BaBar",
    title = "{Search for Lepton Flavor Violation in~$\Upsilon (3S)\rightarrow e^{\pm}\mu^{\mp}$}",
    eprint = "2109.03364",
    archivePrefix = "arXiv",
    primaryClass = "hep-ex",
    reportNumber = "BABAR-PUB-21/003, SLAC-PUB-17617",
    doi = "10.1103/PhysRevLett.128.091804",
    journal = "Phys. Rev. Lett.",
    volume = "128",
    number = "9",
    pages = "091804",
    year = "2022"
}

@article{Belle:2023iln,
    author = "Dhamija, R. and others",
    collaboration = "Belle",
    title = "{Search for charged-lepton flavor violation in $\Upsilon(2S) \to \ell^\mp\tau^\pm$ ($\ell=e,\mu$) decays at Belle}",
    eprint = "2309.02739",
    archivePrefix = "arXiv",
    primaryClass = "hep-ex",
    reportNumber = "Belle Preprint 2023-14, KEK Preprint 2023-19",
    doi = "10.1007/JHEP02(2024)187",
    journal = "JHEP",
    volume = "02",
    pages = "187",
    year = "2024"
}

@article{MEGII:2023ltw,
    author = "Afanaciev, K. and others",
    collaboration = "MEG II",
    title = "{A search for $\mu^+ \rightarrow \textrm{e}^+ \gamma $ with the first dataset of the MEG-II experiment}",
    eprint = "2310.12614",
    archivePrefix = "arXiv",
    primaryClass = "hep-ex",
    doi = "10.1140/epjc/s10052-024-12416-2",
    journal = "Eur. Phys. J. C",
    volume = "84",
    number = "3",
    pages = "216",
    year = "2024",
    note = "[Erratum: Eur.Phys.J.C 84, 1042 (2024)]"
}

@article{Petrov:2013vka,
    author = "Petrov, Alexey A. and Zhuridov, Dmitry V.",
    title = "{Lepton flavor-violating transitions in effective field theory and gluonic operators}",
    eprint = "1308.6561",
    archivePrefix = "arXiv",
    primaryClass = "hep-ph",
    reportNumber = "WSU-HEP-1307",
    doi = "10.1103/PhysRevD.89.033005",
    journal = "Phys. Rev. D",
    volume = "89",
    number = "3",
    pages = "033005",
    year = "2014"
}

@article{Cirigliano:2009bz,
    author = "Cirigliano, Vincenzo and Kitano, Ryuichiro and Okada, Yasuhiro and Tuzon, Paula",
    title = "{On the model discriminating power of mu ---{\ensuremath{>}} e conversion in nuclei}",
    eprint = "0904.0957",
    archivePrefix = "arXiv",
    primaryClass = "hep-ph",
    reportNumber = "IFIC-09-14, FTUV-09-0402, KEK-TH-1308, LA-UR-09-01473, TU-845",
    doi = "10.1103/PhysRevD.80.013002",
    journal = "Phys. Rev. D",
    volume = "80",
    pages = "013002",
    year = "2009"
}

@article{Shifman:1978bx,
    author = "Shifman, Mikhail A. and Vainshtein, A. I. and Zakharov, Valentin I.",
    title = "{QCD and Resonance Physics. Theoretical Foundations}",
    reportNumber = "ITEP-73-1978, ITEP-80-1978",
    doi = "10.1016/0550-3213(79)90022-1",
    journal = "Nucl. Phys. B",
    volume = "147",
    pages = "385--447",
    year = "1979"
}

@article{Aebischer:2025qhh,
    author = "Aebischer, Jason and Buras, Andrzej J. and Kumar, Jacky",
    title = "{SMEFT ATLAS: The Landscape Beyond the Standard Model}",
    eprint = "2507.05926",
    archivePrefix = "arXiv",
    primaryClass = "hep-ph",
    reportNumber = "AJB-25-1, CERN-TH-2025-129, LA-UR-24-24665",
    month = "7",
    year = "2025"
}

@article{Gherardi:2020det,
    author = "Gherardi, Valerio and Marzocca, David and Venturini, Elena",
    title = "{Matching scalar leptoquarks to the SMEFT at one loop}",
    eprint = "2003.12525",
    archivePrefix = "arXiv",
    primaryClass = "hep-ph",
    doi = "10.1007/JHEP07(2020)225",
    journal = "JHEP",
    volume = "07",
    pages = "225",
    year = "2020",
    note = "[Erratum: JHEP 01, 006 (2021)]"
}

@article{Dekens:2018bci,
    author = "Dekens, W. and de Vries, J. and Jung, M. and Vos, K. K.",
    title = "{The phenomenology of electric dipole moments in models of scalar leptoquarks}",
    eprint = "1809.09114",
    archivePrefix = "arXiv",
    primaryClass = "hep-ph",
    reportNumber = "SI-HEP-2018-29, QFET-2018-18, ACFI-T18-14",
    doi = "10.1007/JHEP01(2019)069",
    journal = "JHEP",
    volume = "01",
    pages = "069",
    year = "2019"
}

@article{Bobeth:2017ecx,
    author = "Bobeth, Christoph and Buras, Andrzej J.",
    title = "{Leptoquarks meet $\varepsilon'/\varepsilon$ and rare Kaon processes}",
    eprint = "1712.01295",
    archivePrefix = "arXiv",
    primaryClass = "hep-ph",
    reportNumber = "TUM-HEP-1114-17, TUM-HEP-1114/17",
    doi = "10.1007/JHEP02(2018)101",
    journal = "JHEP",
    volume = "02",
    pages = "101",
    year = "2018"
}

@book{Buras:2020xsm,
    author = "Buras, Andrzej",
    title = "{Gauge Theory of Weak Decays}",
    doi = "10.1017/9781139524100",
    isbn = "978-1-139-52410-0, 978-1-107-03403-7",
    publisher = "Cambridge University Press",
    month = "6",
    year = "2020"
}

@article{Kosmas:2001mv,
    author = "Kosmas, T. S. and Kovalenko, Sergey and Schmidt, Ivan",
    title = "{Nuclear muon- e- conversion in strange quark sea}",
    eprint = "hep-ph/0102101",
    archivePrefix = "arXiv",
    doi = "10.1016/S0370-2693(01)00657-8",
    journal = "Phys. Lett. B",
    volume = "511",
    pages = "203",
    year = "2001"
}

@article{BaBar:2010vxb,
    author = "Lees, J. P. and others",
    collaboration = "BaBar",
    title = "{Search for Charged Lepton Flavor Violation in Narrow Upsilon Decays}",
    eprint = "1001.1883",
    archivePrefix = "arXiv",
    primaryClass = "hep-ex",
    reportNumber = "SLAC-PUB-13898, BABAR-PUB-09-032",
    doi = "10.1103/PhysRevLett.104.151802",
    journal = "Phys. Rev. Lett.",
    volume = "104",
    pages = "151802",
    year = "2010"
}

@article{Belle:2022cce,
    author = "Patra, S. and others",
    collaboration = "Belle",
    title = "{Search for charged lepton flavor violating decays of $\Upsilon (1S)$}",
    eprint = "2201.09620",
    archivePrefix = "arXiv",
    primaryClass = "hep-ex",
    reportNumber = "Belle Preprint 2022-02; KEK Preprint 2021-59",
    doi = "10.1007/JHEP05(2022)095",
    journal = "JHEP",
    volume = "05",
    pages = "095",
    year = "2022"
}

@article{Appel:2000wg,
    author = "Appel, R. and others",
    title = "{An Improved limit on the rate of decay K+ ---\ensuremath{>} pi+ muon+ e-}",
    eprint = "hep-ex/0005016",
    archivePrefix = "arXiv",
    doi = "10.1103/PhysRevLett.85.2450",
    journal = "Phys. Rev. Lett.",
    volume = "85",
    pages = "2450--2453",
    year = "2000"
}

@article{Sher:2005sp,
    author = "Sher, Aleksey and others",
    title = "{An Improved upper limit on the decay K+ ---\ensuremath{>} pi+ mu+ e-}",
    eprint = "hep-ex/0502020",
    archivePrefix = "arXiv",
    doi = "10.1103/PhysRevD.72.012005",
    journal = "Phys. Rev. D",
    volume = "72",
    pages = "012005",
    year = "2005"
}

@article{Belle:2023ziz,
    author = "Tsuzuki, N. and others",
    collaboration = "Belle",
    title = "{Search for lepton-flavor-violating \ensuremath{\tau} decays into a lepton and a vector meson using the full Belle data sample}",
    eprint = "2301.03768",
    archivePrefix = "arXiv",
    primaryClass = "hep-ex",
    reportNumber = "Belle Preprint 2022-33; KEK Preprint 2022-46",
    doi = "10.1007/JHEP06(2023)118",
    journal = "JHEP",
    volume = "06",
    pages = "118",
    year = "2023"
}

@article{BaBar:2012azg,
    author = "Lees, J. P. and others",
    collaboration = "BaBar",
    title = "{A search for the decay modes $B^{+-} \to h^{+-} \tau^{+-}l$}",
    eprint = "1204.2852",
    archivePrefix = "arXiv",
    primaryClass = "hep-ex",
    reportNumber = "SLAC-PUB-14951, BABAR-PUB-11-023",
    doi = "10.1103/PhysRevD.86.012004",
    journal = "Phys. Rev. D",
    volume = "86",
    pages = "012004",
    year = "2012"
}

@article{White:1995jc,
    author = "White, D. B. and others",
    title = "{Search for the decays eta ---\ensuremath{>} mu e and eta ---\ensuremath{>} e+ e-}",
    reportNumber = "PRINT-95-237 (UCLA)",
    doi = "10.1103/PhysRevD.53.6658",
    journal = "Phys. Rev. D",
    volume = "53",
    pages = "6658--6661",
    year = "1996"
}

@article{BES:2004jiw,
    author = "Ablikim, M. and others",
    collaboration = "BES",
    title = "{Search for the lepton flavor violation processes J / psi ---\ensuremath{>} mu tau and e tau}",
    eprint = "hep-ex/0406018",
    archivePrefix = "arXiv",
    doi = "10.1016/j.physletb.2004.08.005",
    journal = "Phys. Lett. B",
    volume = "598",
    pages = "172--177",
    year = "2004"
}

@article{BESIII:2021slj,
    author = "Ablikim, Medina and others",
    collaboration = "BESIII",
    title = "{Search for the charged lepton flavor violating decay $J/\psi\to e\tau$}",
    eprint = "2103.11540",
    archivePrefix = "arXiv",
    primaryClass = "hep-ex",
    doi = "10.1103/PhysRevD.103.112007",
    journal = "Phys. Rev. D",
    volume = "103",
    number = "11",
    pages = "112007",
    year = "2021"
}

@article{Wintz:1998rp,
    author = "Wintz, P.",
    editor = "Klapdor-Kleingrothaus, H. V. and Krivosheina, I. V.",
    title = "{Results of the SINDRUM-II experiment}",
    journal = "Conf. Proc. C",
    volume = "980420",
    pages = "534--546",
    year = "1998"
}

@article{Belle-II:2022cgf,
    author = "Aggarwal, Latika and others",
    collaboration = "Belle-II",
    title = "{Snowmass White Paper: Belle II physics reach and plans for the next decade and beyond}",
    eprint = "2207.06307",
    archivePrefix = "arXiv",
    primaryClass = "hep-ex",
    month = "7",
    year = "2022"
}

@article{Ohshima:2007zz,
    author = "Ohshima, T.",
    editor = "Cei, Fabrizio and Ferrante, Isidoro and Lusiani, Alberto",
    collaboration = "Belle",
    title = "{Study of LFV in tau decay at Belle}",
    doi = "10.1016/j.nuclphysbps.2007.02.116",
    journal = "Nucl. Phys. B Proc. Suppl.",
    volume = "169",
    pages = "174--185",
    year = "2007"
}

@article{Mu2e:2014fns,
    author = "Bartoszek, L. and others",
    collaboration = "Mu2e",
    title = "{Mu2e Technical Design Report}",
    eprint = "1501.05241",
    archivePrefix = "arXiv",
    primaryClass = "physics.ins-det",
    reportNumber = "FERMILAB-TM-2594, FERMILAB-DESIGN-2014-01",
    doi = "10.2172/1172555",
    month = "10",
    year = "2014"
}

@article{SINDRUMII:2006dvw,
    author = "Bertl, Wilhelm H. and others",
    collaboration = "SINDRUM II",
    title = "{A Search for muon to electron conversion in muonic gold}",
    doi = "10.1140/epjc/s2006-02582-x",
    journal = "Eur. Phys. J. C",
    volume = "47",
    pages = "337--346",
    year = "2006"
}

@article{COMET:2018auw,
    author = "Abramishvili, R. and others",
    collaboration = "COMET",
    title = "{COMET Phase-I Technical Design Report}",
    eprint = "1812.09018",
    archivePrefix = "arXiv",
    primaryClass = "physics.ins-det",
    doi = "10.1093/ptep/ptz125",
    journal = "PTEP",
    volume = "2020",
    number = "3",
    pages = "033C01",
    year = "2020"
}

@article{BaBar:2009hkt,
    author = "Aubert, Bernard and others",
    collaboration = "BaBar",
    title = "{Searches for Lepton Flavor Violation in the Decays tau+- ---\ensuremath{>} e+- gamma and tau+- ---\ensuremath{>} mu+- gamma}",
    eprint = "0908.2381",
    archivePrefix = "arXiv",
    primaryClass = "hep-ex",
    reportNumber = "SLAC-PUB-13753, BABAR-PUB-09-026",
    doi = "10.1103/PhysRevLett.104.021802",
    journal = "Phys. Rev. Lett.",
    volume = "104",
    pages = "021802",
    year = "2010"
}

@article{Belle:2021ysv,
    author = "Abdesselam, A. and others",
    collaboration = "Belle",
    title = "{Search for lepton-flavor-violating tau-lepton decays to $\ell\gamma$ at Belle}",
    eprint = "2103.12994",
    archivePrefix = "arXiv",
    primaryClass = "hep-ex",
    doi = "10.1007/JHEP10(2021)019",
    journal = "JHEP",
    volume = "10",
    pages = "19",
    year = "2021"
}

@article{Belle-II:2018jsg,
    author = "Altmannshofer, W. and others",
    editor = "Kou, E. and Urquijo, P.",
    collaboration = "Belle-II",
    title = "{The Belle II Physics Book}",
    eprint = "1808.10567",
    archivePrefix = "arXiv",
    primaryClass = "hep-ex",
    reportNumber = "KEK Preprint 2018-27, BELLE2-PUB-PH-2018-001, FERMILAB-PUB-18-398-T, JLAB-THY-18-2780, INT-PUB-18-047, UWThPh 2018-26",
    doi = "10.1093/ptep/ptz106",
    journal = "PTEP",
    volume = "2019",
    number = "12",
    pages = "123C01",
    year = "2019",
    note = "[Erratum: PTEP 2020, 029201 (2020)]"
}

@article{Bechtle:2022tck,
    author = {Bechtle, Philip and Chall, Cristin and King, Martin and Kraemer, Michael and Maettig, Peter and St\"oltzner, Michael},
    title = "{Bottoms Up: Standard Model Effective Field Theory from a Model Perspective}",
    eprint = "2201.08819",
    archivePrefix = "arXiv",
    primaryClass = "physics.hist-ph",
    month = "1",
    year = "2022"
}

@article{Ibarra:2004pe,
    author = "Ibarra, Alejandro and Masso, Eduard and Redondo, Javier",
    title = "{Systematic approach to gauge-invariant relations between lepton flavor violating processes}",
    eprint = "hep-ph/0410386",
    archivePrefix = "arXiv",
    reportNumber = "CERN-PH-TH-2004-196, UAB-FT-574",
    doi = "10.1016/j.nuclphysb.2005.03.017",
    journal = "Nucl. Phys. B",
    volume = "715",
    pages = "523--535",
    year = "2005"
}

@article{Zarnecki:1999je,
    author = "Zarnecki, Aleksander Filip",
    title = "{Global analysis of $e e q q$ contact interactions and future prospects for high-energy physics}",
    eprint = "hep-ph/9904334",
    archivePrefix = "arXiv",
    reportNumber = "DESY-99-074, IFD-02-99",
    doi = "10.1007/s100520050653",
    journal = "Eur. Phys. J. C",
    volume = "11",
    pages = "539--557",
    year = "1999"
}

@article{Crivellin:2013hpa,
    author = "Crivellin, Andreas and Najjari, Saereh and Rosiek, Janusz",
    title = "{Lepton Flavor Violation in the Standard Model with general Dimension-Six Operators}",
    eprint = "1312.0634",
    archivePrefix = "arXiv",
    primaryClass = "hep-ph",
    doi = "10.1007/JHEP04(2014)167",
    journal = "JHEP",
    volume = "04",
    pages = "167",
    year = "2014"
}

@article{Pruna:2014asa,
    author = "Pruna, Giovanni Marco and Signer, Adrian",
    title = "{The $\mu\to e\gamma$ decay in a systematic effective field theory approach with dimension 6 operators}",
    eprint = "1408.3565",
    archivePrefix = "arXiv",
    primaryClass = "hep-ph",
    reportNumber = "PSI-PR-14-07, ZU-TH-25-14",
    doi = "10.1007/JHEP10(2014)014",
    journal = "JHEP",
    volume = "10",
    pages = "014",
    year = "2014"
}

@article{Crivellin:2017rmk,
    author = "Crivellin, Andreas and Davidson, Sacha and Pruna, Giovanni Marco and Signer, Adrian",
    title = "{Renormalisation-group improved analysis of $\mu\to e$ processes in a systematic effective-field-theory approach}",
    eprint = "1702.03020",
    archivePrefix = "arXiv",
    primaryClass = "hep-ph",
    reportNumber = "PSI-PR-17-01, ZU-TH-01-17",
    doi = "10.1007/JHEP05(2017)117",
    journal = "JHEP",
    volume = "05",
    pages = "117",
    year = "2017"
}

@article{Grinstein:1988me,
    author = "Grinstein, Benjamin and Savage, Martin J. and Wise, Mark B.",
    title = "{B ---\ensuremath{>} X(s) e+ e- in the Six Quark Model}",
    reportNumber = "LBL-25014, CALT-68-1489",
    doi = "10.1016/0550-3213(89)90078-3",
    journal = "Nucl. Phys. B",
    volume = "319",
    pages = "271--290",
    year = "1989"
}

@article{Buchalla:1995vs,
    author = "Buchalla, Gerhard and Buras, Andrzej J. and Lautenbacher, Markus E.",
    title = "{Weak decays beyond leading logarithms}",
    eprint = "hep-ph/9512380",
    archivePrefix = "arXiv",
    reportNumber = "SLAC-PUB-7009, SLAC-PUB-95-7009, MPI-PH-95-104, TUM-T31-100-95, FERMILAB-PUB-95-305-T",
    doi = "10.1103/RevModPhys.68.1125",
    journal = "Rev. Mod. Phys.",
    volume = "68",
    pages = "1125--1144",
    year = "1996"
}

@article{Alonso:2014csa,
    author = "Alonso, Rodrigo and Grinstein, Benjamin and Martin Camalich, Jorge",
    title = "{$SU(2)\times U(1)$ gauge invariance and the shape of new physics in rare $B$ decays}",
    eprint = "1407.7044",
    archivePrefix = "arXiv",
    primaryClass = "hep-ph",
    doi = "10.1103/PhysRevLett.113.241802",
    journal = "Phys. Rev. Lett.",
    volume = "113",
    pages = "241802",
    year = "2014"
}

@article{Crivellin:2015era,
    author = "Crivellin, Andreas and Hofer, Lars and Matias, Joaquim and Nierste, Ulrich and Pokorski, Stefan and Rosiek, Janusz",
    title = "{Lepton-flavour violating $B$ decays in generic $Z'$ models}",
    eprint = "1504.07928",
    archivePrefix = "arXiv",
    primaryClass = "hep-ph",
    reportNumber = "CERN-PH-TH-2015-091, TTP15-018",
    doi = "10.1103/PhysRevD.92.054013",
    journal = "Phys. Rev. D",
    volume = "92",
    number = "5",
    pages = "054013",
    year = "2015"
}

@article{Aebischer:2015fzz,
    author = "Aebischer, Jason and Crivellin, Andreas and Fael, Matteo and Greub, Christoph",
    title = "{Matching of gauge invariant dimension-six operators for $b\to s$ and $b\to c$ transitions}",
    eprint = "1512.02830",
    archivePrefix = "arXiv",
    primaryClass = "hep-ph",
    reportNumber = "CERN-PH-TH-2015-278",
    doi = "10.1007/JHEP05(2016)037",
    journal = "JHEP",
    volume = "05",
    pages = "037",
    year = "2016"
}

@article{Becirevic:2016zri,
    author = "Be\v{c}irevi\'c, Damir and Sumensari, Olcyr and Zukanovich Funchal, Renata",
    title = "{Lepton flavor violation in exclusive $b\rightarrow s$ decays}",
    eprint = "1602.00881",
    archivePrefix = "arXiv",
    primaryClass = "hep-ph",
    reportNumber = "LPT-15-57",
    doi = "10.1140/epjc/s10052-016-3985-0",
    journal = "Eur. Phys. J. C",
    volume = "76",
    number = "3",
    pages = "134",
    year = "2016"
}

@article{Descotes-Genon:2023pen,
    author = "Descotes-Genon, S\'ebastien and Faroughy, Darius A. and Plakias, Ioannis and Sumensari, Olcyr",
    title = "{Probing lepton flavor violation in meson decays with LHC data}",
    eprint = "2303.07521",
    archivePrefix = "arXiv",
    primaryClass = "hep-ph",
    doi = "10.1140/epjc/s10052-023-11860-w",
    journal = "Eur. Phys. J. C",
    volume = "83",
    number = "8",
    pages = "753",
    year = "2023"
}

@article{Calibbi:2021pyh,
    author = "Calibbi, Lorenzo and Marcano, Xabier and Roy, Joydeep",
    title = "{Z lepton flavour violation as a probe for new physics at future $e^+e^-$ colliders}",
    eprint = "2107.10273",
    archivePrefix = "arXiv",
    primaryClass = "hep-ph",
    reportNumber = "TUM-HEP 1352/21",
    doi = "10.1140/epjc/s10052-021-09777-3",
    journal = "Eur. Phys. J. C",
    volume = "81",
    number = "12",
    pages = "1054",
    year = "2021"
}

@article{Cullen:2020zof,
    author = "Cullen, Jonathan M. and Pecjak, Benjamin D.",
    title = "{Higgs decay to fermion pairs at NLO in SMEFT}",
    eprint = "2007.15238",
    archivePrefix = "arXiv",
    primaryClass = "hep-ph",
    reportNumber = "IPPP/20/31",
    doi = "10.1007/JHEP11(2020)079",
    journal = "JHEP",
    volume = "11",
    pages = "079",
    year = "2020"
}

@article{Buchmuller:1985jz,
    author = "Buchmuller, W. and Wyler, D.",
    title = "{Effective Lagrangian Analysis of New Interactions and Flavor Conservation}",
    reportNumber = "CERN-TH-4254/85",
    doi = "10.1016/0550-3213(86)90262-2",
    journal = "Nucl. Phys. B",
    volume = "268",
    pages = "621--653",
    year = "1986"
}

@article{Kuno:1999jp,
    author = "Kuno, Yoshitaka and Okada, Yasuhiro",
    title = "{Muon decay and physics beyond the standard model}",
    eprint = "hep-ph/9909265",
    archivePrefix = "arXiv",
    reportNumber = "KEK-PREPRINT-99-69, KEK-TH-639",
    doi = "10.1103/RevModPhys.73.151",
    journal = "Rev. Mod. Phys.",
    volume = "73",
    pages = "151--202",
    year = "2001"
}

@article{Grzadkowski:2010es,
    author = "Grzadkowski, B. and Iskrzynski, M. and Misiak, M. and Rosiek, J.",
    title = "{Dimension-Six Terms in the Standard Model Lagrangian}",
    eprint = "1008.4884",
    archivePrefix = "arXiv",
    primaryClass = "hep-ph",
    reportNumber = "IFT-9-2010, TTP10-35",
    doi = "10.1007/JHEP10(2010)085",
    journal = "JHEP",
    volume = "10",
    pages = "085",
    year = "2010"
}

@article{Jenkins:2013wua,
    author = "Jenkins, Elizabeth E. and Manohar, Aneesh V. and Trott, Michael",
    title = "{Renormalization Group Evolution of the Standard Model Dimension Six Operators II: Yukawa Dependence}",
    eprint = "1310.4838",
    archivePrefix = "arXiv",
    primaryClass = "hep-ph",
    reportNumber = "CERN-PH-TH/2015-247",
    doi = "10.1007/JHEP01(2014)035",
    journal = "JHEP",
    volume = "01",
    pages = "035",
    year = "2014"
}

@article{Jenkins:2013zja,
    author = "Jenkins, Elizabeth E. and Manohar, Aneesh V. and Trott, Michael",
    title = "{Renormalization Group Evolution of the Standard Model Dimension Six Operators I: Formalism and lambda Dependence}",
    eprint = "1308.2627",
    archivePrefix = "arXiv",
    primaryClass = "hep-ph",
    doi = "10.1007/JHEP10(2013)087",
    journal = "JHEP",
    volume = "10",
    pages = "087",
    year = "2013"
}

@article{Alonso:2013hga,
    author = "Alonso, Rodrigo and Jenkins, Elizabeth E. and Manohar, Aneesh V. and Trott, Michael",
    title = "{Renormalization Group Evolution of the Standard Model Dimension Six Operators III: Gauge Coupling Dependence and Phenomenology}",
    eprint = "1312.2014",
    archivePrefix = "arXiv",
    primaryClass = "hep-ph",
    reportNumber = "CERN-PH-TH-2013-305, CERN-PH-TH/2013-305",
    doi = "10.1007/JHEP04(2014)159",
    journal = "JHEP",
    volume = "04",
    pages = "159",
    year = "2014"
}

@article{Kuno:2013mha,
    author = "Kuno, Yoshitaka",
    collaboration = "COMET",
    title = "{A search for muon-to-electron conversion at J-PARC: The COMET experiment}",
    doi = "10.1093/ptep/pts089",
    journal = "PTEP",
    volume = "2013",
    pages = "022C01",
    year = "2013"
}

@article{BaBar:2006jhm,
    author = "Aubert, Bernard and others",
    collaboration = "BaBar",
    title = "{Search for Lepton Flavor Violating Decays $\tau^\pm \to \ell^\pm \pi^0$, $\ell^\pm \eta$, $\ell^\pm \eta^\prime$}",
    eprint = "hep-ex/0610067",
    archivePrefix = "arXiv",
    reportNumber = "SLAC-PUB-12170, BABAR-PUB-06-061",
    doi = "10.1103/PhysRevLett.98.061803",
    journal = "Phys. Rev. Lett.",
    volume = "98",
    pages = "061803",
    year = "2007"
}

@article{LHCb:2024wve,
    author = "Aaij, Roel and others",
    collaboration = "LHCb",
    title = "{Search for the lepton-flavor violating decay Bs0\textrightarrow{}\ensuremath{\phi}\ensuremath{\mu}\ensuremath{\pm}\ensuremath{\tau}\ensuremath{\mp}}",
    eprint = "2405.13103",
    archivePrefix = "arXiv",
    primaryClass = "hep-ex",
    reportNumber = "LHCb-PAPER-2024-006, CERN-EP-2024-114",
    doi = "10.1103/PhysRevD.110.072014",
    journal = "Phys. Rev. D",
    volume = "110",
    number = "7",
    pages = "072014",
    year = "2024"
}

@article{LHCb:2022wrs,
    collaboration = "LHCb",
    title = "{Search for the lepton-flavour violating decays $B^0 \to K^{*0} \tau^\pm \mu^\mp$}",
    eprint = "2209.09846",
    archivePrefix = "arXiv",
    primaryClass = "hep-ex",
    reportNumber = "LHCb-PAPER-2022-021, CERN-EP-2022-154",
    month = "9",
    year = "2022"
}

@article{LHCb:2022lrd,
    collaboration = "LHCb",
    title = "{Search for the lepton-flavour violating decays $B^0 \to K^{*0} \mu^\pm e^\mp$ and $B_s^0 \to \phi \mu^\pm e^\mp$}",
    eprint = "2207.04005",
    archivePrefix = "arXiv",
    primaryClass = "hep-ex",
    reportNumber = "LHCb-PAPER-2022-008, CERN-EP-2022-097",
    month = "7",
    year = "2022"
}

@article{LHCb:2019bix,
    author = "Aaij, Roel and others",
    collaboration = "LHCb",
    title = "{Search for Lepton-Flavor Violating Decays $B^+ \to K^+ {\mu}^{\pm} e^{\mp}$}",
    eprint = "1909.01010",
    archivePrefix = "arXiv",
    primaryClass = "hep-ex",
    reportNumber = "CERN-EP-2019-172, LHCb-PAPER-2019-022",
    doi = "10.1103/PhysRevLett.123.241802",
    journal = "Phys. Rev. Lett.",
    volume = "123",
    number = "24",
    pages = "241802",
    year = "2019"
}

@article{LHCb:2019ujz,
    author = "Aaij, Roel and others",
    collaboration = "LHCb",
    title = "{Search for the lepton-flavour-violating decays $B^{0}_{s}\to\tau^{\pm}\mu^{\mp}$ and $B^{0}\to\tau^{\pm}\mu^{\mp}$}",
    eprint = "1905.06614",
    archivePrefix = "arXiv",
    primaryClass = "hep-ex",
    reportNumber = "CERN-EP-2019-076, LHCb-PAPER-2019-016",
    doi = "10.1103/PhysRevLett.123.211801",
    journal = "Phys. Rev. Lett.",
    volume = "123",
    number = "21",
    pages = "211801",
    year = "2019"
}

@article{LHCb:2017hag,
    author = "Aaij, Roel and others",
    collaboration = "LHCb",
    title = "{Search for the lepton-flavour violating decays B$_{(s)}^{0} \to e^{\pm}\mu^{\mp}$}",
    eprint = "1710.04111",
    archivePrefix = "arXiv",
    primaryClass = "hep-ex",
    reportNumber = "LHCB-PAPER-2017-031, CERN-EP-2017-242",
    doi = "10.1007/JHEP03(2018)078",
    journal = "JHEP",
    volume = "03",
    pages = "078",
    year = "2018"
}

@article{Kitano:2002mt,
    author = "Kitano, Ryuichiro and Koike, Masafumi and Okada, Yasuhiro",
    title = "{Detailed calculation of lepton flavor violating muon electron conversion rate for various nuclei}",
    eprint = "hep-ph/0203110",
    archivePrefix = "arXiv",
    reportNumber = "KEK-TH-808",
    doi = "10.1103/PhysRevD.76.059902",
    journal = "Phys. Rev. D",
    volume = "66",
    pages = "096002",
    year = "2002",
    note = "[Erratum: Phys.Rev.D 76, 059902 (2007)]"
}

@article{Jenkins:2017jig,
    author = "Jenkins, Elizabeth E. and Manohar, Aneesh V. and Stoffer, Peter",
    title = "{Low-Energy Effective Field Theory below the Electroweak Scale: Operators and Matching}",
    eprint = "1709.04486",
    archivePrefix = "arXiv",
    primaryClass = "hep-ph",
    doi = "10.1007/JHEP03(2018)016",
    journal = "JHEP",
    volume = "03",
    pages = "016",
    year = "2018"
}

@article{Jenkins:2017dyc,
    author = "Jenkins, Elizabeth E. and Manohar, Aneesh V. and Stoffer, Peter",
    title = "{Low-Energy Effective Field Theory below the Electroweak Scale: Anomalous Dimensions}",
    eprint = "1711.05270",
    archivePrefix = "arXiv",
    primaryClass = "hep-ph",
    doi = "10.1007/JHEP01(2018)084",
    journal = "JHEP",
    volume = "01",
    pages = "084",
    year = "2018"
}

@article{Aebischer:2018bkb,
    author = "Aebischer, Jason and Kumar, Jacky and Straub, David M.",
    title = "{Wilson: a Python package for the running and matching of Wilson coefficients above and below the electroweak scale}",
    eprint = "1804.05033",
    archivePrefix = "arXiv",
    primaryClass = "hep-ph",
    doi = "10.1140/epjc/s10052-018-6492-7",
    journal = "Eur. Phys. J. C",
    volume = "78",
    number = "12",
    pages = "1026",
    year = "2018"
}

@article{Straub:2018kue,
    author = "Straub, David M.",
    title = "{flavio: a Python package for flavour and precision phenomenology in the Standard Model and beyond}",
    eprint = "1810.08132",
    archivePrefix = "arXiv",
    primaryClass = "hep-ph",
    month = "10",
    year = "2018"
}

@article{LHCb:2018roe,
    author = "Aaij, Roel and others",
    collaboration = "LHCb",
    title = "{Physics case for an LHCb Upgrade II - Opportunities in flavour physics, and beyond, in the HL-LHC era}",
    eprint = "1808.08865",
    archivePrefix = "arXiv",
    primaryClass = "hep-ex",
    reportNumber = "LHCB-PUB-2018-009, CERN-LHCC-2018-027",
    month = "8",
    year = "2018"
}

@article{Georgi:1993mps,
    author = "Georgi, H.",
    title = "{Effective field theory}",
    doi = "10.1146/annurev.ns.43.120193.001233",
    journal = "Ann. Rev. Nucl. Part. Sci.",
    volume = "43",
    pages = "209--252",
    year = "1993"
}

@article{Davidson:2018kud,
    author = "Davidson, Sacha and Kuno, Yoshitaka and Yamanaka, Masato",
    title = "{Selecting $\mu \to e$ conversion targets to distinguish lepton flavour-changing operators}",
    eprint = "1810.01884",
    archivePrefix = "arXiv",
    primaryClass = "hep-ph",
    doi = "10.1016/j.physletb.2019.01.042",
    journal = "Phys. Lett. B",
    volume = "790",
    pages = "380--388",
    year = "2019"
}

@article{Cirigliano:2017azj,
    author = "Cirigliano, Vincenzo and Davidson, Sacha and Kuno, Yoshitaka",
    title = "{Spin-dependent $\mu \to e$ conversion}",
    eprint = "1703.02057",
    archivePrefix = "arXiv",
    primaryClass = "hep-ph",
    reportNumber = "LA-UR-17-21718, OUHEP-17-1",
    doi = "10.1016/j.physletb.2017.05.053",
    journal = "Phys. Lett. B",
    volume = "771",
    pages = "242--246",
    year = "2017"
}

@article{Davidson:2017nrp,
    author = "Davidson, Sacha and Kuno, Yoshitaka and Saporta, Albert",
    title = "{\textquotedblleft{}Spin-dependent\textquotedblright{} ${\mu \rightarrow e}$ conversion on light nuclei}",
    eprint = "1710.06787",
    archivePrefix = "arXiv",
    primaryClass = "hep-ph",
    doi = "10.1140/epjc/s10052-018-5584-8",
    journal = "Eur. Phys. J. C",
    volume = "78",
    number = "2",
    pages = "109",
    year = "2018"
}

@article{Davidson:2020ord,
    author = "Davidson, S. and Kuno, Y. and Uesaka, Y. and Yamanaka, M.",
    title = "{Probing $\mu e \gamma \gamma$ contact interactions with $\mu \to e$ conversion}",
    eprint = "2007.09612",
    archivePrefix = "arXiv",
    primaryClass = "hep-ph",
    reportNumber = "OCU-PHYS 519, NITEP 72",
    doi = "10.1103/PhysRevD.102.115043",
    journal = "Phys. Rev. D",
    volume = "102",
    number = "11",
    pages = "115043",
    year = "2020"
}

@article{Hoferichter:2022mna,
    author = {Hoferichter, Martin and Men\'endez, Javier and No\"el, Frederic},
    title = "{Improved Limits on Lepton-Flavor-Violating Decays of Light Pseudoscalars via Spin-Dependent \ensuremath{\mu}\textrightarrow{}e Conversion in Nuclei}",
    eprint = "2204.06005",
    archivePrefix = "arXiv",
    primaryClass = "hep-ph",
    doi = "10.1103/PhysRevLett.130.131902",
    journal = "Phys. Rev. Lett.",
    volume = "130",
    number = "13",
    pages = "131902",
    year = "2023"
}

@article{Cirigliano:2021img,
    author = "Cirigliano, Vincenzo and Fuyuto, Kaori and Lee, Christopher and Mereghetti, Emanuele and Yan, Bin",
    title = "{Charged Lepton Flavor Violation at the EIC}",
    eprint = "2102.06176",
    archivePrefix = "arXiv",
    primaryClass = "hep-ph",
    reportNumber = "LA-UR-21-20531",
    doi = "10.1007/JHEP03(2021)256",
    journal = "JHEP",
    volume = "03",
    pages = "256",
    year = "2021"
}

@article{Kumar:2021yod,
    author = "Kumar, Jacky",
    title = "{Renormalization group improved implications of semileptonic operators in SMEFT}",
    eprint = "2107.13005",
    archivePrefix = "arXiv",
    primaryClass = "hep-ph",
    doi = "10.1007/JHEP01(2022)107",
    journal = "JHEP",
    volume = "01",
    pages = "107",
    year = "2022"
}

@article{Davidson:2020hkf,
    author = "Davidson, S.",
    title = "{Completeness and complementarity for $\mu \to e\gamma \mu \to e \bar e e$ and $\mu A \to eA$}",
    eprint = "2010.00317",
    archivePrefix = "arXiv",
    primaryClass = "hep-ph",
    doi = "10.1007/JHEP02(2021)172",
    journal = "JHEP",
    volume = "02",
    pages = "172",
    year = "2021"
}

@article{Belle:2021rod,
    author = "Atmacan, H. and others",
    collaboration = "Belle",
    title = "{Search for $B^{0} \to \tau^\pm \ell^\mp$ ($\ell=e,\mu$) with a hadronic tagging method at Belle}",
    eprint = "2108.11649",
    archivePrefix = "arXiv",
    primaryClass = "hep-ex",
    reportNumber = "Belle Preprint 2021-23; KEK Preprint 2021-27; UC Preprint
  UCHEP-21-04",
    doi = "10.1103/PhysRevD.104.L091105",
    journal = "Phys. Rev. D",
    volume = "104",
    number = "9",
    pages = "L091105",
    year = "2021"
}

@book{Peskin:1995ev,
    author = "Peskin, Michael E. and Schroeder, Daniel V.",
    title = "{An Introduction to quantum field theory}",
    isbn = "978-0-201-50397-5",
    publisher = "Addison-Wesley",
    address = "Reading, USA",
    year = "1995"
}

@article{Crivellin:2023zui,
    author = "Crivellin, Andreas and Mellado, Bruce",
    title = "{Anomalies in Particle Physics}",
    eprint = "2309.03870",
    archivePrefix = "arXiv",
    primaryClass = "hep-ph",
    reportNumber = "PSI-PR-23-34, ZU-TH 53/23, ICPP-73",
    month = "9",
    year = "2023"
}

@article{Myers:1990sk,
    author = "Myers, S. and Picasso, E.",
    title = "{The Design, construction and commissioning of the CERN Large Electron Positron collider}",
    doi = "10.1080/00107519008213789",
    journal = "Contemp. Phys.",
    volume = "31",
    pages = "387--403",
    year = "1990"
}

@article{Wilson:1977nk,
    author = "Wilson, Robert Rathbun",
    title = "{The Tevatron}",
    reportNumber = "FERMILAB-TM-0763",
    doi = "10.1063/1.3037746",
    journal = "Phys. Today",
    volume = "30N10",
    pages = "23--30",
    year = "1977"
}

@article{Evans:2008zzb,
    editor = "Evans, Lyndon and Bryant, Philip",
    title = "{LHC Machine}",
    doi = "10.1088/1748-0221/3/08/S08001",
    journal = "JINST",
    volume = "3",
    pages = "S08001",
    year = "2008"
}

@article{ATLAS:2012yve,
    author = "Aad, Georges and others",
    collaboration = "ATLAS",
    title = "{Observation of a new particle in the search for the Standard Model Higgs boson with the ATLAS detector at the LHC}",
    eprint = "1207.7214",
    archivePrefix = "arXiv",
    primaryClass = "hep-ex",
    reportNumber = "CERN-PH-EP-2012-218",
    doi = "10.1016/j.physletb.2012.08.020",
    journal = "Phys. Lett. B",
    volume = "716",
    pages = "1--29",
    year = "2012"
}

@article{Belle:2022pcr,
    author = "Watanuki, S. and others",
    collaboration = "Belle",
    title = "{Search for the Lepton Flavor Violating Decays B+\textrightarrow{}K+\ensuremath{\tau}\ensuremath{\pm}\ensuremath{\ell}\ensuremath{\mp} (\ensuremath{\ell}=e, \ensuremath{\mu}) at Belle}",
    eprint = "2212.04128",
    archivePrefix = "arXiv",
    primaryClass = "hep-ex",
    doi = "10.1103/PhysRevLett.130.261802",
    journal = "Phys. Rev. Lett.",
    volume = "130",
    number = "26",
    pages = "261802",
    year = "2023"
}

@article{Crivellin:2016vjc,
    author = "Crivellin, Andreas and D'Ambrosio, Giancarlo and Hoferichter, Martin and Tunstall, Lewis C.",
    title = "{Violation of lepton flavor and lepton flavor universality in rare kaon decays}",
    eprint = "1601.00970",
    archivePrefix = "arXiv",
    primaryClass = "hep-ph",
    reportNumber = "CERN-TH-2016-001, INT-PUB-16-001, PSI-PR-16-001",
    doi = "10.1103/PhysRevD.93.074038",
    journal = "Phys. Rev. D",
    volume = "93",
    number = "7",
    pages = "074038",
    year = "2016"
}

@article{Gilman:1979ud,
    author = "Gilman, Frederick J. and Wise, Mark B.",
    title = "{K ---\ensuremath{>} pi e+ e- in the Six Quark Model}",
    reportNumber = "SLAC-PUB-2437",
    doi = "10.1103/PhysRevD.21.3150",
    journal = "Phys. Rev. D",
    volume = "21",
    pages = "3150",
    year = "1980"
}

@article{Inami:1980fz,
    author = "Inami, T. and Lim, C. S.",
    title = "{Effects of Superheavy Quarks and Leptons in Low-Energy Weak Processes k(L) ---\ensuremath{>} mu anti-mu, K+ ---\ensuremath{>} pi+ Neutrino anti-neutrino and K0 \ensuremath{<}---\ensuremath{>} anti-K0}",
    reportNumber = "UT-KOMABA-80-8",
    doi = "10.1143/PTP.65.297",
    journal = "Prog. Theor. Phys.",
    volume = "65",
    pages = "297",
    year = "1981",
    note = "[Erratum: Prog.Theor.Phys. 65, 1772 (1981)]"
}

@article{Dib:1988js,
    author = "Dib, Claudio and Dunietz, Isard and Gilman, Frederick J.",
    title = "{K(L) ---\ensuremath{>} pi0 Lepton+ Lepton- Decays for Large m(t)}",
    reportNumber = "SLAC-PUB-4818",
    doi = "10.1103/PhysRevD.39.2639",
    journal = "Phys. Rev. D",
    volume = "39",
    pages = "2639",
    year = "1989"
}

@article{Fayyazuddin:2018zww,
    author = "Fayyazuddin and Aslam, Muhammad Jamil and Lu, Cai-Dian",
    title = "{Lepton Flavor Violating Decays of $B$ and $K$ Mesons in Models with Extended Gauge Group}",
    eprint = "1805.00177",
    archivePrefix = "arXiv",
    primaryClass = "hep-ph",
    doi = "10.1142/S0217751X18500872",
    journal = "Int. J. Mod. Phys. A",
    volume = "33",
    number = "14n15",
    pages = "1850087",
    year = "2018"
}

@article{deBlas:2017xtg,
    author = "de Blas, J. and Criado, J. C. and Perez-Victoria, M. and Santiago, J.",
    title = "{Effective description of general extensions of the Standard Model: the complete tree-level dictionary}",
    eprint = "1711.10391",
    archivePrefix = "arXiv",
    primaryClass = "hep-ph",
    reportNumber = "CERN-TH-2017-251",
    doi = "10.1007/JHEP03(2018)109",
    journal = "JHEP",
    volume = "03",
    pages = "109",
    year = "2018"
}

@article{Ali:2023kua,
    author = "Ali, Md Isha and Chattopadhyay, Utpal and Rajeev, N. and Roy, Joydeep",
    title = "{SMEFT analysis of charged lepton flavor violating B-meson decays}",
    eprint = "2312.05071",
    archivePrefix = "arXiv",
    primaryClass = "hep-ph",
    doi = "10.1103/PhysRevD.109.075028",
    journal = "Phys. Rev. D",
    volume = "109",
    number = "7",
    pages = "075028",
    year = "2024"
}

@article{Hazard:2016fnc,
    author = "Hazard, Derek E. and Petrov, Alexey A.",
    title = "{Lepton flavor violating quarkonium decays}",
    eprint = "1607.00815",
    archivePrefix = "arXiv",
    primaryClass = "hep-ph",
    reportNumber = "WSU-HEP-1603, SI-HEP-2016-19",
    doi = "10.1103/PhysRevD.94.074023",
    journal = "Phys. Rev. D",
    volume = "94",
    number = "7",
    pages = "074023",
    year = "2016"
}

@article{Abada:2015zea,
    author = "Abada, Asmaa and Be\v{c}irevi\'c, Damir and Lucente, Michele and Sumensari, Olcyr",
    title = "{Lepton flavor violating decays of vector quarkonia and of the $Z$ boson}",
    eprint = "1503.04159",
    archivePrefix = "arXiv",
    primaryClass = "hep-ph",
    reportNumber = "LPT-15-10, SISSA-09-2015-FISI",
    doi = "10.1103/PhysRevD.91.113013",
    journal = "Phys. Rev. D",
    volume = "91",
    number = "11",
    pages = "113013",
    year = "2015"
}

@article{Calibbi:2022ddo,
    author = "Calibbi, Lorenzo and Li, Tong and Marcano, Xabier and Schmidt, Michael A.",
    title = "{Indirect constraints on lepton-flavor-violating quarkonium decays}",
    eprint = "2207.10913",
    archivePrefix = "arXiv",
    primaryClass = "hep-ph",
    reportNumber = "CPPC-2022-08, IFT-UAM/CSIC-22-82",
    doi = "10.1103/PhysRevD.106.115039",
    journal = "Phys. Rev. D",
    volume = "106",
    number = "11",
    pages = "115039",
    year = "2022"
}

@article{Carpentier:2010ue,
    author = "Carpentier, Michael and Davidson, Sacha",
    title = "{Constraints on two-lepton, two quark operators}",
    eprint = "1008.0280",
    archivePrefix = "arXiv",
    primaryClass = "hep-ph",
    doi = "10.1140/epjc/s10052-010-1482-4",
    journal = "Eur. Phys. J. C",
    volume = "70",
    pages = "1071--1090",
    year = "2010"
}

@article{Fernandez-Martinez:2024bxg,
    author = "Fern\'andez-Mart\'\i{}nez, Enrique and Marcano, Xabier and Naredo-Tuero, Daniel",
    title = "{Global Lepton Flavour Violating Constraints on New Physics}",
    eprint = "2403.09772",
    archivePrefix = "arXiv",
    primaryClass = "hep-ph",
    reportNumber = "IFT-UAM/CSIC-24-39",
    month = "3",
    year = "2024"
}

@article{Ali:2025xkw,
    author = "Ali, Md Isha and Chattopadhyay, Utpal and Ghosh, Dilip Kumar and Rajeev, N.",
    title = "{Constraints on lepton flavor universal and non-universal New Physics in $b\, \to\, s\, \ell^+ \ell^-$ decays: a global SMEFT survey}",
    eprint = "2502.20145",
    archivePrefix = "arXiv",
    primaryClass = "hep-ph",
    month = "2",
    year = "2025"
}

@article{Aebischer:2018iyb,
    author = "Aebischer, Jason and Kumar, Jacky and Stangl, Peter and Straub, David M.",
    title = "{A Global Likelihood for Precision Constraints and Flavour Anomalies}",
    eprint = "1810.07698",
    archivePrefix = "arXiv",
    primaryClass = "hep-ph",
    doi = "10.1140/epjc/s10052-019-6977-z",
    journal = "Eur. Phys. J. C",
    volume = "79",
    number = "6",
    pages = "509",
    year = "2019"
}

@article{Belle-II:2025yiq,
    author = "Adachi, I. and others",
    collaboration = "Belle-II, Belle",
    title = "{Search for lepton-flavor-violating $\tau^- \to \ell^- K_s^0$ decays at Belle and Belle II}",
    eprint = "2504.15745",
    archivePrefix = "arXiv",
    primaryClass = "hep-ex",
    month = "4",
    year = "2025"
}

@article{Zyla:2020zbs,
    author = "Zyla, P. A. and others",
    collaboration = "Particle Data Group",
    title = "{Review of Particle Physics}",
    doi = "10.1093/ptep/ptaa104",
    journal = "PTEP",
    volume = "2020",
    number = "8",
    pages = "083C01",
    year = "2020"
}

@article{Belle:2023jwr,
    author = "Nayak, L. and others",
    collaboration = "Belle",
    title = "{Search for $B{}^0_s \rightarrow \ell^{\mp} \tau^{\pm}$ with the Semi-leptonic Tagging Method at Belle}",
    eprint = "2301.10989",
    archivePrefix = "arXiv",
    primaryClass = "hep-ex",
    reportNumber = "Belle Preprint 2023-02; KEK Preprint 2022-49",
    doi = "10.1007/JHEP08(2023)178",
    journal = "JHEP",
    volume = "08",
    pages = "178",
    year = "2023"
}

@article{LHCb:2020car,
    author = "Aaij, Roel and others",
    collaboration = "LHCb",
    title = "{Searches for 25 rare and forbidden decays of $D^{+}$ and $ {D}_s^{+} $ mesons}",
    eprint = "2011.00217",
    archivePrefix = "arXiv",
    primaryClass = "hep-ex",
    reportNumber = "LHCb-PAPER-2020-007, CERN-EP-2020-140",
    doi = "10.1007/JHEP06(2021)044",
    journal = "JHEP",
    volume = "06",
    pages = "044",
    year = "2021"
}

@article{BaBar:2007xeb,
    author = "Aubert, Bernard and others",
    collaboration = "BaBar",
    title = "{Search for the rare decay $B \to  \pi l^+ l^-$}",
    eprint = "hep-ex/0703018",
    archivePrefix = "arXiv",
    reportNumber = "SLAC-PUB-12389, BABAR-PUB-07-002",
    doi = "10.1103/PhysRevLett.99.051801",
    journal = "Phys. Rev. Lett.",
    volume = "99",
    pages = "051801",
    year = "2007"
}

@article{Plakias:2023esq,
    author = "Plakias, I. and Sumensari, O.",
    title = "{Lepton Flavor Violation in Semileptonic Observables}",
    eprint = "2312.14070",
    archivePrefix = "arXiv",
    primaryClass = "hep-ph",
    month = "12",
    year = "2023"
}

@article{Ryd:2009uf,
    author = "Ryd, Anders and Petrov, Alexey A.",
    title = "{Hadronic D and D(s) Meson Decays}",
    eprint = "0910.1265",
    archivePrefix = "arXiv",
    primaryClass = "hep-ph",
    reportNumber = "WSU-HEP-0903",
    doi = "10.1103/RevModPhys.84.65",
    journal = "Rev. Mod. Phys.",
    volume = "84",
    pages = "65--117",
    year = "2012"
}

@article{Hazard:2017udp,
    author = "Hazard, Derek E. and Petrov, Alexey A.",
    title = "{Radiative lepton flavor violating B, D, and K decays}",
    eprint = "1711.05314",
    archivePrefix = "arXiv",
    primaryClass = "hep-ph",
    doi = "10.1103/PhysRevD.98.015027",
    journal = "Phys. Rev. D",
    volume = "98",
    number = "1",
    pages = "015027",
    year = "2018"
}

@article{BaBar:2020faa,
    author = "Lees, J. P. and others",
    collaboration = "BaBar",
    title = "{Search for lepton-flavor-violating decays $D^{0}\rightarrow X^{0}e^{\pm}\mu^{\mp}$}",
    eprint = "2004.09457",
    archivePrefix = "arXiv",
    primaryClass = "hep-ex",
    reportNumber = "BABAR-PUB-20/001, SLAC-PUB-17524",
    doi = "10.1103/PhysRevD.101.112003",
    journal = "Phys. Rev. D",
    volume = "101",
    number = "11",
    pages = "112003",
    year = "2020"
}

@article{Burdman:2001tf,
    author = "Burdman, Gustavo and Golowich, Eugene and Hewett, JoAnne L. and Pakvasa, Sandip",
    title = "{Rare charm decays in the standard model and beyond}",
    eprint = "hep-ph/0112235",
    archivePrefix = "arXiv",
    reportNumber = "SLAC-PUB-9057, LBNL-49074, UH-511-957-01",
    doi = "10.1103/PhysRevD.66.014009",
    journal = "Phys. Rev. D",
    volume = "66",
    pages = "014009",
    year = "2002"
}

@article{deBoer:2015boa,
    author = "de Boer, Stefan and Hiller, Gudrun",
    title = "{Flavor and new physics opportunities with rare charm decays into leptons}",
    eprint = "1510.00311",
    archivePrefix = "arXiv",
    primaryClass = "hep-ph",
    reportNumber = "DO-TH-15-10, QFET-2015-25",
    doi = "10.1103/PhysRevD.93.074001",
    journal = "Phys. Rev. D",
    volume = "93",
    number = "7",
    pages = "074001",
    year = "2016"
}

@article{Ball:2004rg,
    author = "Ball, Patricia and Zwicky, Roman",
    title = "{$B_{d,s} \to  \rho, \omega, K^*, \phi$ decay form-factors from light-cone sum rules revisited}",
    eprint = "hep-ph/0412079",
    archivePrefix = "arXiv",
    reportNumber = "IPPP-04-74, DCPT-04-48, TPI-MINN-04-39",
    doi = "10.1103/PhysRevD.71.014029",
    journal = "Phys. Rev. D",
    volume = "71",
    pages = "014029",
    year = "2005"
}

@article{Khodjamirian:2006st,
    author = "Khodjamirian, Alexander and Mannel, Thomas and Offen, Nils",
    title = "{Form-factors from light-cone sum rules with B-meson distribution amplitudes}",
    eprint = "hep-ph/0611193",
    archivePrefix = "arXiv",
    reportNumber = "SI-HEP-2006-03",
    doi = "10.1103/PhysRevD.75.054013",
    journal = "Phys. Rev. D",
    volume = "75",
    pages = "054013",
    year = "2007"
}

@article{Lu:2011jm,
    author = "Lu, Cai-Dian and Wang, Wei",
    title = "{Analysis of $B\to K^*_J (\to K \pi) \mu^+\mu^-$ in the higher kaon resonance region}",
    eprint = "1111.1513",
    archivePrefix = "arXiv",
    primaryClass = "hep-ph",
    reportNumber = "DESY-11-201",
    doi = "10.1103/PhysRevD.85.034014",
    journal = "Phys. Rev. D",
    volume = "85",
    pages = "034014",
    year = "2012"
}

@article{BESIII:2022exh,
    author = "Ablikim, Medina and others",
    collaboration = "BESIII",
    title = "{Search for the lepton flavor violating decay~$J/\psi\to e\mu$}",
    eprint = "2206.13956",
    archivePrefix = "arXiv",
    primaryClass = "hep-ex",
    doi = "10.1007/s11433-022-1995-0",
    journal = "Sci. China Phys. Mech. Astron.",
    volume = "66",
    number = "2",
    pages = "221011",
    year = "2023"
}

@article{KTeV:2007cvy,
    author = "Abouzaid, E. and others",
    collaboration = "KTeV",
    title = "{Search for lepton flavor violating decays of the neutral kaon}",
    eprint = "0711.3472",
    archivePrefix = "arXiv",
    primaryClass = "hep-ex",
    reportNumber = "FERMILAB-PUB-07-646-E",
    doi = "10.1103/PhysRevLett.100.131803",
    journal = "Phys. Rev. Lett.",
    volume = "100",
    pages = "131803",
    year = "2008"
}

@article{BNL:1998apv,
    author = "Ambrose, D. and others",
    collaboration = "BNL",
    title = "{New limit on muon and electron lepton number violation from K0(L) ---\ensuremath{>} mu+- e-+ decay}",
    eprint = "hep-ex/9811038",
    archivePrefix = "arXiv",
    doi = "10.1103/PhysRevLett.81.5734",
    journal = "Phys. Rev. Lett.",
    volume = "81",
    pages = "5734--5737",
    year = "1998"
}

@article{NA62:2021zxl,
    author = "Cortina Gil, Eduardo and others",
    collaboration = "NA62",
    title = "{Search for Lepton Number and Flavor Violation in $K^+$ and $\pi^0$ Decays}",
    eprint = "2105.06759",
    archivePrefix = "arXiv",
    primaryClass = "hep-ex",
    reportNumber = "CERN-EP-2021-090",
    doi = "10.1103/PhysRevLett.127.131802",
    journal = "Phys. Rev. Lett.",
    volume = "127",
    number = "13",
    pages = "131802",
    year = "2021"
}

@article{Garosi:2023yxg,
    author = "Garosi, Francesco and Marzocca, David and S\'anchez, Antonio Rodr\'\i{}guez and Stanzione, Alfredo",
    title = "{Indirect constraints on top quark operators from a global SMEFT analysis}",
    eprint = "2310.00047",
    archivePrefix = "arXiv",
    primaryClass = "hep-ph",
    doi = "10.1007/JHEP12(2023)129",
    journal = "JHEP",
    volume = "12",
    pages = "129",
    year = "2023"
}

@article{CLEO:1999nsy,
    author = "Briere, Roy A. and others",
    collaboration = "CLEO",
    title = "{Rare decays of the eta-prime}",
    eprint = "hep-ex/9907046",
    archivePrefix = "arXiv",
    reportNumber = "SLAC-REPRINT-1999-101, CLNS-99-1625, CLEO-99-6",
    doi = "10.1103/PhysRevLett.84.26",
    journal = "Phys. Rev. Lett.",
    volume = "84",
    pages = "26--30",
    year = "2000"
}

@conference{JEF,
  author       = {M. Dugger and others},
  title        = {The Jefferson Lab Eta Factory (JEF) experiment},
  doi          = {https://www.jlab.org/exp_prog/proposals/12/PR12-12-003.pdf},
  url          = {https://www.jlab.org/exp_prog/proposals/12/PR12-12-003.pdf},
  place        = {United States},
  year         = {2012},
  month        = {05}
}

@article{REDTOP:2022slw,
    author = "Elam, J. and others",
    collaboration = "REDTOP",
    title = "{The REDTOP experiment: Rare $\eta/\eta^{\prime}$ Decays To Probe New Physics}",
    eprint = "2203.07651",
    archivePrefix = "arXiv",
    primaryClass = "hep-ex",
    reportNumber = "FERMILAB-FN-1153-AD-PPD-T",
    month = "3",
    year = "2022"
}

@article{Chattopadhyay:2019ycs,
    author = "Chattopadhyay, Utpal and Das, Debottam and Mukherjee, Samadrita",
    title = "{Probing Lepton Flavor Violating decays in MSSM with Non-Holomorphic Soft Terms}",
    eprint = "1911.05543",
    archivePrefix = "arXiv",
    primaryClass = "hep-ph",
    doi = "10.1007/JHEP06(2020)015",
    journal = "JHEP",
    volume = "06",
    pages = "015",
    year = "2020"
}

@article{De:2021crr,
    author = "De, Bibhabasu and Das, Debottam and Mitra, Manimala and Sahoo, Nirakar",
    title = "{Magnetic moments of leptons, charged lepton flavor violations and dark matter phenomenology of a minimal radiative Dirac neutrino mass model}",
    eprint = "2106.00979",
    archivePrefix = "arXiv",
    primaryClass = "hep-ph",
    doi = "10.1007/JHEP08(2022)202",
    journal = "JHEP",
    volume = "08",
    pages = "202",
    year = "2022"
}

\end{document}